\documentclass[journal]{IEEEtran} 

\ifCLASSINFOpdf
\usepackage[pdftex]{graphicx}
\else
\fi

\usepackage{balance}
\usepackage{cite}
\usepackage{graphicx}
\usepackage{float}
\usepackage{amsmath}
\usepackage[utf8]{inputenc}
\usepackage[final]{pdfpages}
\usepackage{pdfpages}
\usepackage{lipsum}
\usepackage{textcase}
\usepackage{xurl}
\usepackage{amsmath,esint} 
\usepackage{epstopdf}
\usepackage{array}
\usepackage{color}
\usepackage{subcaption} 

\usepackage[normalem]{ulem}

\usepackage{bm}

\DeclareRobustCommand{\uvec}[1]{{%
  \ifcsname uvec#1\endcsname
     \csname uvec#1\endcsname
   \else
    \bm{\hat{\mathbf{#1}}}%
   \fi
}}

\newcolumntype{P}[1]{>{\centering\arraybackslash}p{#1}}
\newcolumntype{M}[1]{>{\centering\arraybackslash}m{#1}}

\pdfoutput=1

\begin{document}


\title{A Survey on Detection, Tracking, and Classification of Aerial Threats using Radars and Communications Systems}

\author{\IEEEauthorblockN{Wahab Khawaja\IEEEauthorrefmark{1},~Martins Ezuma\IEEEauthorrefmark{4},~\IEEEmembership{Member, IEEE},~Vasilii Semkin\IEEEauthorrefmark{3},~Fatih Erden\IEEEauthorrefmark{4},~Ozgur Ozdemir\IEEEauthorrefmark{4},~\IEEEmembership{Member, IEEE},  and~Ismail Guvenc\IEEEauthorrefmark{4},~\IEEEmembership{Fellow, IEEE}
}

\IEEEauthorblockA{\IEEEauthorrefmark{1}Department of Computer Systems Engineering, Mirpur University of Science and Technology, Mirpur AK, Pakistan}


\IEEEauthorblockA{\IEEEauthorrefmark{3}VTT Technical Research Centre of Finland, Tietotie 3, 02150 Espoo, Finland}

\IEEEauthorblockA{\IEEEauthorrefmark{4} Electrical and Computer Engineering Department, North Carolina State University, Raleigh, NC 27606, USA}

Email: wahab.ali@must.edu.pk, vasilii.semkin@vtt.fi, erdenfatih@gmail.com, \{mcezuma,oozdemi, iguvenc\}@ncsu.edu
}

\maketitle

\begin{abstract}
The use of unmanned aerial vehicles~(UAVs) for different applications has increased many folds in recent years. The UAVs are expected to change the future air operations. However, there are instances where the UAVs can be used for malicious purposes. The detection, tracking, and classification of UAVs is challenging compared to manned aerial vehicles~(MAVs) mainly due to small size, complex shapes, and ability to fly close to the terrain and in autonomous flight patterns in swarms. In this survey, we will discuss current and future aerial threats, and provide an overview of radar systems to counter such threats. We also study the performance parameters of radar systems for the detection, tracking, and classification of UAVs compared to MAVs. In addition to dedicated radar systems, we review the use of joint communication-radar~(JCR) systems, as well as passive monitoring of changes in the common communication signals, e.g., FM, LTE, and any transmissions that may radiate from a UAV, for the detection, tracking, and classification of UAVs are provided. Finally, limitations of radar systems and comparison with other techniques that do not rely on radars for detection, tracking, and classification of aerial threats are provided.


\end{abstract}

\IEEEpeerreviewmaketitle

\section{Introduction}
\begin{figure}[!t] 
    \centering
		\includegraphics[width=0.99\columnwidth]{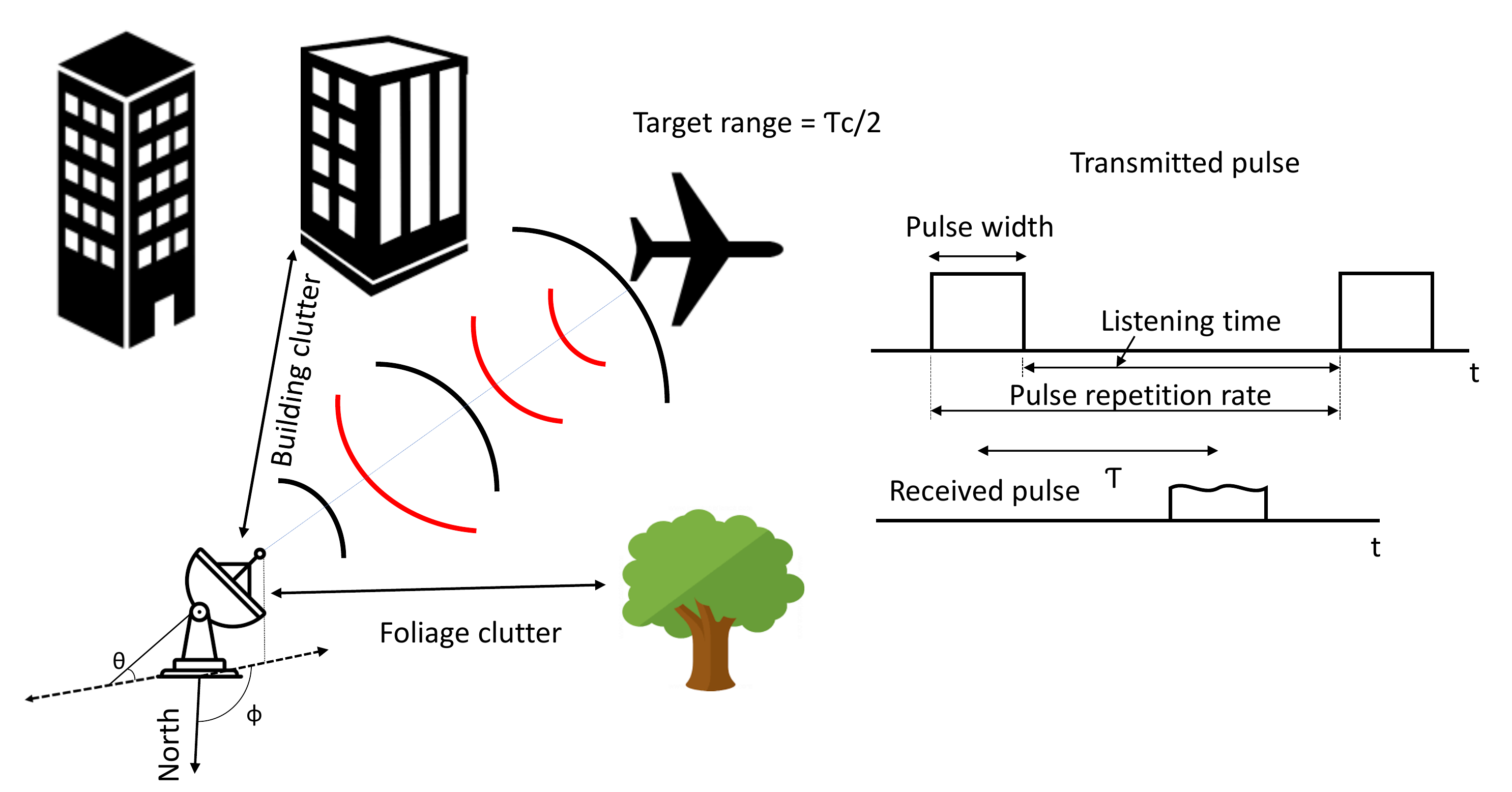}
	   \caption{Basic operation of a pulse-based radar.} \label{Fig:radar_basic1}
\end{figure}

Radar systems are popular and widely used methods for the detection, tracking, and classification of aerial vehicles. Radar technology was first introduced in 1935~\cite{radar_1945}. The research and development of radars were accelerated during the second world war and proved to be extremely effective during the war. Since then, radars have seen decades of improvements in overcoming many challenges. Modern radar systems nowadays use advance electronics, compact antennas, phased arrays, and efficient signal processing to achieve reduced response times, high accuracy, low probability of false alarm~(PFA), unambiguous aerial vehicle detection and tracking at extended ranges, detection and tracking of multiple aerial vehicles simultaneously, integration with multiple sensors~(airborne, ground, and sea-based), and operations in different terrains~\cite{radartypes1,radartypes2}. The extensive training data of different terrains and potential aerial vehicles aided with efficient classification algorithms have helped in the real time classification of different types of aerial vehicles in complex environments. The basic operation of a pulse radar system is shown in Fig.~\ref{Fig:radar_basic1}, where a pulse is transmitted from the radar, and the reflection of the pulse from the aerial vehicle is used to detect and subsequently track the aerial vehicle over time.

Compared to radar systems, the aerial vehicles have also grown in sophistication. The modern aerial vehicles present challenges to conventional and modern radar systems. For example, stealth technology~\cite{ST}, and unmanned aerial vehicles~(UAVs)~\cite{C-UAV2} are difficult to detect by modern radar systems at desired ranges. Compared to stealth technology utilized mainly for manned aerial vehicles~(MAVs) \footnote{While MAV is typically used to refer to micro aerial vehicles in the literature, we use it to refer to manned aerial vehicles in this survey.} which is extremely expensive, complex, and under strict governmental controls, the UAVs offer inherent stealth features. This is mainly due to the small size, complex shapes, and non-metallic construction material of UAVs, and their ability to autonomously fly close to terrain. Moreover, their simple design and ease of manufacturing from off-the-shelf and widely available components, and ease of quick modifications have made countermeasures against UAVs challenging. UAV research and development is one of the fast growing industries in the world. According to \cite{stats}, the overall UAV global market share~(in military, law enforcement, government, commercial, and consumer domains) is estimated at \$27.4 billion in 2021 and is expected to reach \$58.4 billion by 2026.

The affordable prices of readily available components, as well as simple assembly and control of UAVs have allowed the use of UAVs in all the major conflicts of the world. In the last two decades, a large number of different types of UAVs are used in major world conflicts~\cite{UAV_conflicts,NK}. Due to the advantages of UAVs mentioned earlier, UAVs have also attracted the attention of non-state actors, and they have been used for malicious activities in different parts of the world~\cite{NSA,SA}. The UAVs used by amateur users may also introduce major threats if their users do not follow regulatory rules~\cite{UAV_amateur02,UAV_amateur1,UAV_amateur2}. With all these, it is critically important to find ways to counter malicious UAVs as well as unintended threats from amateur UAVs~\cite{wahab_uav_threats,UAV_amateur2}. There are numerous research efforts carried out in the academia and industry to counter the threats and challenges from malicious UAVs. According to a NATO Review Report~\cite{C_UAS}, the global share of UAV countering technologies is expected at $\$6.6$~billion by 2024.

\begin{table*}[!h]
\caption{Types and specifications of UAVs. In the table, L, W, and H represent the length, width, and height, respectively. }
  \centering
\resizebox{\textwidth}{!}{
\begin{tabular}{|p{0.005cm}|p{0.005cm}|p{0.005cm}|p{0.005cm}|p{0.005cm}|}
\hline
		\multicolumn{1}{|c|}{}&\multicolumn{4}{|c|}{\textbf{Specifications of different types of UAVs}}\\		
			\hline
             \multicolumn{1}{|c|}{{\textbf{UAV type}}}&\multicolumn{1}{|c|}{Single/multi-rotor}&\multicolumn{1}{|c|}{Flat/tilt wing}&\multicolumn{1}{|c|}{Hot air Balloons}&\multicolumn{1}{|c|}{Satellites}\\
             \hline
            \multicolumn{1}{|c|}{\textbf{Max. Size (L$\times$ W $\times$ H)}}& \multicolumn{1}{|c|}{1m$\times$1m $\times$0.63m}& \multicolumn{1}{|c|}{14m$\times$1.7m$\times$4m}& \multicolumn{1}{|c|}{17m$\times$17m$\times$24.5m}& \multicolumn{1}{|c|}{73m$\times$109m$\times$20m}\\
             \hline
 \multicolumn{1}{|c|}{\textbf{Airframe material}}& \multicolumn{1}{|c|}{Carbon fiber}& \multicolumn{1}{|c|}{Carbon fiber}& \multicolumn{1}{|c|}{Nylon, polyester}& \multicolumn{1}{|c|}{Aluminum alloys}\\
             \hline
 \multicolumn{1}{|c|}{\textbf{Max. flight Endurance}}& \multicolumn{1}{|c|}{3 hours}& \multicolumn{1}{|c|}{42 hours}& \multicolumn{1}{|c|}{475 hours}& \multicolumn{1}{|c|}{15 years}\\
             \hline
 \multicolumn{1}{|c|}{\textbf{Max. payload capacity}}& \multicolumn{1}{|c|}{8 kg}& \multicolumn{1}{|c|}{100 kg}& \multicolumn{1}{|c|}{600 kg}& \multicolumn{1}{|c|}{29,000 kg}\\
             \hline
 \multicolumn{1}{|c|}{\textbf{Max. flight ceiling}}& \multicolumn{1}{|c|}{6 km}& \multicolumn{1}{|c|}{18 km}& \multicolumn{1}{|c|}{21 km}& \multicolumn{1}{|c|}{35786 km}\\
             \hline
 \multicolumn{1}{|c|}{\textbf{Max. speed}}& \multicolumn{1}{|c|}{29 m/s}& \multicolumn{1}{|c|}{130 m/s}& \multicolumn{1}{|c|}{4.47 m/s}& \multicolumn{1}{|c|}{3138.9 m/s}\\
             \hline
 \multicolumn{1}{|c|}{\textbf{Propulsion System}}& \multicolumn{1}{|c|}{Propeller}& \multicolumn{1}{|c|}{Propeller}& \multicolumn{1}{|c|}{NA}& \multicolumn{1}{|c|}{Chemical thrusters}\\
             \hline
 \multicolumn{1}{|c|}{\textbf{Operating frequencies}}& \multicolumn{1}{|c|}{900 MHz - 5.8 GHz}& \multicolumn{1}{|c|}{900 MHz - 5.8 GHz}& \multicolumn{1}{|c|}{123.3 - 123.5 MHz}& \multicolumn{1}{|c|}{L, C, X, Ku, Ka band} \\
             \hline
 \multicolumn{1}{|c|}{\textbf{Navigation}}& \multicolumn{1}{|c|}{Internal/external}& \multicolumn{1}{|c|}{Internal/external}& \multicolumn{1}{|c|}{NA}& \multicolumn{1}{|c|}{Internal/external}\\
             \hline
 \multicolumn{1}{|c|}{\textbf{Heat signature}}& \multicolumn{1}{|c|}{Small}& \multicolumn{1}{|c|}{Vey small}& \multicolumn{1}{|c|}{Large}& \multicolumn{1}{|c|}{Small}\\
             \hline
 \multicolumn{1}{|c|}{\textbf{RCS}}& \multicolumn{1}{|c|}{Small}& \multicolumn{1}{|c|}{Small/medium}& \multicolumn{1}{|c|}{Large}& \multicolumn{1}{|c|}{Large}\\
             \hline
 		\end{tabular}
\label{Table:UAV_types}
}
\end{table*}

\begin{table*}[!h]
\caption{Types and specifications of manned aerial vehicles.}
  \centering
\resizebox{\textwidth}{!}{
\begin{tabular}{|p{0.005cm}|p{0.005cm}|p{0.005cm}|p{0.005cm}|p{0.005cm}|p{0.005cm}|p{0.005cm}|p{0.005cm}|}
\hline
		\multicolumn{1}{|c|}{}&\multicolumn{4}{|c|}{\textbf{Specifications of different types of manned aerial vehicles}}\\		
			\hline
             \multicolumn{1}{|c|}{{\textbf{Manned aerial vehicle type}}}&\multicolumn{1}{|c|}{Jet airliners}&\multicolumn{1}{|c|}{Fighter Jets}&\multicolumn{1}{|c|}{Turbofan/turboprop/piston engine}&\multicolumn{1}{|c|}{Helicopters}\\
             \hline
            \multicolumn{1}{|c|}{\textbf{Max. Size (L$\times$W$\times$H)}}& \multicolumn{1}{|c|}{71m$\times$76m$\times$20m}& \multicolumn{1}{|c|}{26m$\times$16m$\times$20m}& \multicolumn{1}{|c|}{27m$\times$27m$\times$7.5m}& \multicolumn{1}{|c|}{37.5m$\times$36m$\times$7.8m}\\
             \hline
\multicolumn{1}{|c|}{\textbf{Airframe material}}& \multicolumn{1}{|c|}{Aluminum alloys}& \multicolumn{1}{|c|}{Carbon fiber}& \multicolumn{1}{|c|}{Aluminum alloys}& \multicolumn{1}{|c|}{Aluminum alloys/steel} \\
             \hline
 \multicolumn{1}{|c|}{\textbf{Max. flight endurance}}& \multicolumn{1}{|c|}{23 hours}& \multicolumn{1}{|c|}{8.5 hours}& \multicolumn{1}{|c|}{10 hours}& \multicolumn{1}{|c|}{5 hours} \\
             \hline
 \multicolumn{1}{|c|}{\textbf{Max. Payload capacity}}& \multicolumn{1}{|c|}{38000 kg}& \multicolumn{1}{|c|}{12000 kg}& \multicolumn{1}{|c|}{6800 kg}& \multicolumn{1}{|c|}{20000 kg} \\
             \hline
 \multicolumn{1}{|c|}{\textbf{Max. flight ceiling}}& \multicolumn{1}{|c|}{15.2 km}& \multicolumn{1}{|c|}{18.5 km}& \multicolumn{1}{|c|}{7.62 km}& \multicolumn{1}{|c|}{7.6 km}\\
             \hline
 \multicolumn{1}{|c|}{\textbf{Max. Speed}}& \multicolumn{1}{|c|}{283 m/s}& \multicolumn{1}{|c|}{736 m/s}& \multicolumn{1}{|c|}{144 m/s}& \multicolumn{1}{|c|}{93 m/s}\\
             \hline
 \multicolumn{1}{|c|}{\textbf{Propulsion type}}& \multicolumn{1}{|c|}{Jet engine}& \multicolumn{1}{|c|}{Jet engine}& \multicolumn{1}{|c|}{Propeller}& \multicolumn{1}{|c|}{Propeller}\\
             \hline
 \multicolumn{1}{|c|}{\textbf{Operating frequencies}}& \multicolumn{1}{|c|}{118 MHz - 136.9 MHz}& \multicolumn{1}{|c|}{225 MHz to 399.9 MHz}& \multicolumn{1}{|c|}{118 MHz - 136.9 MHz}& \multicolumn{1}{|c|}{123.02 MHz}\\
             \hline
 \multicolumn{1}{|c|}{\textbf{Navigation}}& \multicolumn{1}{|c|}{Internal/external}& \multicolumn{1}{|c|}{Internal/external}& \multicolumn{1}{|c|}{Internal/external}& \multicolumn{1}{|c|}{Internal/external}\\
             \hline
\multicolumn{1}{|c|}{\textbf{Heat signature}}& \multicolumn{1}{|c|}{Large}& \multicolumn{1}{|c|}{Large}& \multicolumn{1}{|c|}{Medium}& \multicolumn{1}{|c|}{Medium} \\
             \hline
\multicolumn{1}{|c|}{\textbf{RCS}}& \multicolumn{1}{|c|}{Very large}& \multicolumn{1}{|c|}{Large}& \multicolumn{1}{|c|}{Very large}& \multicolumn{1}{|c|}{Large} \\
             \hline
 		\end{tabular}
\label{Table:MAV_types}
}
\end{table*}

There are different techniques available in the literature used for the detection, tracking, and classification of malicious UAVs and related aerial threats. Popular techniques include using radar systems, electro-optical/infra-red~(EO/IR) imaging, radio frequency (RF) analysis techniques, and sound/noise analysis of aerial vehicles~\cite{wahab_uav_threats}. Radar systems are dominantly used compared to other techniques. Sophisticated and multiple radar systems can be integrated to detect and track UAVs~\cite{modern_detect, drone_detect_techs, UAV_stealth_airborne}. However, the probability of miss detection for UAVs is high mainly due to the small radar cross-section~(RCS), and high maneuverability close to the terrain. Moreover, as the UAVs are used for different applications and recreational purposes, it is sometimes difficult to learn about the intent of the UAV flight.

In a complex environment, e.g., a dense urban area, radar systems face different challenges for the detection, tracking, and classification of UAVs mainly due to: 1) high-rise buildings that often obstruct the field of view of the radar for low-flying UAVs; 2) dynamic clutter conditions due to moving ground vehicles and pedestrians, and commercial and private air traffic~(during take-off and landing); and 3) shape and flight features of UAVs similar to birds makes the classification difficult for radars. Due to the limitations of the radar systems in a dense urban environment, a radar-based solution may not always be sufficient. Alternatively, or as a complementary approach, communication systems that have generally good coverage in densely populated urban areas can be used for the detection, tracking, and classification of UAVs. The transmissions from popular communication systems, e.g., long-term evolution~(LTE) and frequency modulation~(FM) broadcasts can be analyzed passively to detect, track and classify UAVs~\cite{Comm_gen}. Similarly, communications~(control and onboard sensor data transfer) between the UAVs or between the UAV and the ground station~(GS) can be analyzed passively for the detection and classification of UAVs~\cite{RF_analysis}. Moreover, joint communication-radar~(JCR) systems or joint radar-communication~(RadCom) systems can be used for the detection, tracking, and classification of UAVs in active mode, efficiently utilizing the available spectrum~\cite{JCR2}.  

The purpose of this paper is to provide a comprehensive overview of radar systems and communication systems for detection, tracking, and classification of emerging aerial threats and challenges. A comparison of radar systems to other techniques for detection, tracking, and classification of UAVs is also provided. The rest of the paper is organized as follows.  Section~\ref{Section:future_threats} discusses the current and future aerial threats and challenging features of UAVs,  Section~\ref{Section:EM_surroundings} provides the interaction of the electromagnetic~(EM) waves with the surroundings, and Section~\ref{Section:TX_RX} provides key parameters or metrics of the radar systems. The detection and ranging of aerial vehicles is provided in Section~\ref{Section:detection_ranging}, Section~\ref{Section:radar_track_clas} provides tracking and classification of aerial vehicles, Section~\ref{Section:Radar_types} reviews different types of radar systems for countering aerial threats, the miscellaneous factors affecting the radar performance are discussed in Section~\ref{Section:Misc_factors}. The use of communication systems for the detection, tracking, and classification of UAVs is provided in Section~\ref{Section:Comm_Sys}. We discuss methods other than radar systems for detection, tracking, and classification of aerial vehicles in Section~\ref{Section:others_than_radar}, and finally, Section~\ref{Section:conclusions} concludes the paper.

\section{Current and Future Aerial Threats and their Challenging Features} \label{Section:future_threats}

In this section, current and future aerial threats from aerial vehicles and in particular UAVs are discussed. The challenging features of UAVs are also provided.  

\subsection{Current and Future Aerial Threats}
The aerial vehicles can be broadly classified as MAVs and UAVs. Table~\ref{Table:UAV_types} and Table~\ref{Table:MAV_types} show the major types and specifications of typical UAVs and MAVs, respectively. In Table~\ref{Table:UAV_types} and Table~\ref{Table:MAV_types}, maximum values of size, flight endurance, payload capacity, flight ceiling, and speed for UAVs and MAVs, respectively, sorted from the internet are provided. From Table~\ref{Table:UAV_types} and Table~\ref{Table:MAV_types}, it can be observed that the UAVs have in general smaller size, weight, payload capacity, maximum flight ceiling, speed, power consumption, heat signature, and RCS compared to MAVs. Moreover, UAVs offer longer flight endurance, and simple measuring sensors compared to MAVs. The cost of UAVs is also significantly small compared to MAVs. The MAVs are owned by large public or private sector entities and their flying is strictly controlled by national and international laws. Therefore, there is less chance of MAVs being used for malicious activities. The UAVs on the other hand can be easily produced using simple design and readily available off-the-shelf components. Therefore, UAVs are more prone to be used by non-state actors for malicious activities.

Examples of major current and future threats from malicious UAVs can be listed as follows: 1) use by non-state actors during conflicts; 2) conventional,
biological, and chemical threats carried by UAVs; 3) threats to sensitive infrastructure, e.g., chemical and nuclear facilities; 4) threats to important personalities, vehicles, and locations; 5) threats to crowded areas; 6) malicious activities using UAVs; 7) starting a fire, and identity theft; 8) smuggling of contraband articles using UAVs; 9) threats to the aviation industry; 10) planting improvised explosive devices~(IEDs), and mines on ground and at sea; and 11) hacking UAVs and flying them for malicious purposes. Moreover, different UAV flight scenarios that can present a challenge for the detection, tracking, and classification of UAVs are as follows:
\begin{itemize}
    \item During the autonomous, and mechanically controlled UAV flight scenarios, the absence of an external RF control link and external navigation reference can provide immunity against majority of electronic countermeasures~(ECM) such as RF jamming, external navigation spoofing and the UAV will not be detected by RF analysis of the control link between the UAV and the controller. Also, during the mechanical flight mode, the UAV is immune to high energy EM radiation burst countermeasure. 
    \item A scenario where the autonomous UAV performs close terrain and infrastructure hugging makes the detection and tracking difficult. 
    \item A hybrid scenario over land, over/underwater, and in the air can provide flexibility and achieve long operational ranges.
    \item The UAVs are generally small, low, and slow, flying aerial vehicles similar to birds. A scenario where the design of the UAV closely resembles the birds, makes the detection, classification, and tracking difficult. 
    \item Smart modular design scenario where the structural modifications of UAVs can be made in real-time using artificial intelligence~(AI) and 3D printing, which can make the detection, tracking, and classification to become challenging. 
    \item Swarms of small and miniaturized UAVs are difficult to track~\cite{UAV_swarms_new}. 
\end{itemize}

\subsection{Challenges from Physical and Motion Characteristics of UAVs}

UAVs have small RCS compared to MAVs. The small RCS of a UAV is mainly due to small size, and non-metallic and small number of flat surfaces. The complex geometry of the multi-rotor UAVs generally results in reflections in particular directions~\cite{RCS_vehicles2}. The scattering of radio waves in directions other than the desired one in the presence of clutter results in weak radar returns. Moreover, there are certain frequency bands, e.g., K-band where the detection works relatively better compared to other frequency bands for small UAVs~\cite{UAV_kband}. The conventional radar systems are generally configured/calibrated to measure conventional targets. However, UAVs come in different shapes and sizes, and therefore, calibration of the radar systems are required to detect different types of UAVs. 

The ability to fly at low altitudes makes it difficult for a radar system to differentiate the slow moving UAV from the dynamic clutter in urban environments and highways. UAVs can also fly autonomously and can perform terrain/infrastructure hugging making the detection and tracking for a radar system difficult. Moreover, UAVs have high maneuverability compared to MAVs and can perform sharp pitch, roll and yaw movements. The high maneuverability makes the tracking and trajectory estimation challenging. UAVs can also be equipped with a number of different sensors that can help to evade countermeasures against UAVs.

\subsection{Challenges from Multiple UAVs} \label{Section:multipleUAVs}
The detection and tracking of multiple aerial vehicles simultaneously is a challenging task. In particular, multiple closely flying aerial vehicles are difficult to differentiate from one another (if the range/angular resolution is small) and can introduce range ambiguity. For example, the detection and tracking of UAVs flying in swarms is challenging by conventional radar systems~\cite{multipleUAVs_difficulty,swarm_detect}, and specially designed algorithms are required for detection and tracking such swarms~\cite{UAVSwarms_Threat,UAV_swarms_new}.

\section{Electromagnetic Waves Interaction with the Aerial Vehicles  and Surroundings} \label{Section:EM_surroundings}
In this section, physical and motion characteristics of the aerial vehicles due to interaction of EM waves with the aerial vehicles are discussed. The effect of the objects in the surroundings, terrain, and atmosphere on the detection of aerial vehicles by a radar system is also reviewed. 

\subsection{Physical Characteristics}
The physical characteristics of an aerial vehicle are estimated using the RCS of the aerial vehicle. The RCS, also called the EM signature of an aerial vehicle, is a measure of detectability of an object  using a radar. The RCS of an aerial vehicle $\sigma$ is given as $\sigma = \lim_{R \to \infty} 4\pi R^2 \frac{|E_{\rm s}|^2}{|E_{\rm i}|^2}$, where $R$ is the slant range, while $E_{\rm i}$ and $E_{\rm s}$ are the incident and scattered electric fields from an aerial vehicle, respectively. There are three regions of the RCS when measured for a conducting sphere~\cite{RCS_new}. The three regions are Rayleigh~($\lambda \gg a$), resonance/Mie, and optical~($\lambda \ll a$), where $\lambda$ is the wavelength, and $a$ is the radius of the sphere. 

The RCS of an aerial vehicle depends on the frequency, polarization, angle of illumination, and geometry and electrical properties of the material of the aerial vehicle. The effect of the frequency on the RCS measurements is provided in \cite{class_new9,RCS_freq2, semkin_rcs_database}. The RCS can be changed by changing the polarization as given in \cite{polarization_RCS}. The RCS of an aerial vehicle also depends on the angle of the illumination. If the beamwidth of the radar is smaller than the size of the aerial vehicle, then due to the motion of the aerial vehicle and steering of the radar beam, the angle of illumination (of different parts of the aerial vehicle) varies, and RCS of the aerial vehicle fluctuates during measurements. 

The RCS also depends on the aerial vehicle's geometry. For example, the RCS of a flat plate is different compared to a sphere~\cite{RCS_freq2}. The shape of the exterior of the aerial vehicle can be designed to reflect the radar energy in directions other than the radar. This can introduce stealth to an aerial vehicle. However, the stealth due to shape features only (no EM waves absorption) will depend on the angle of illumination of the aerial vehicle. For example, the backscattered radar energy may not be collected by a given radar but can be collected by a radar/RX at a different location.

An obvious dependence of the RCS is on the material of the aerial vehicle. It is well-known that metals are better reflectors of EM energy compared to dielectrics. Therefore, another approach to produce stealth is by selecting materials for the surface of the aerial vehicles that have specific permeability and permittivity, e.g., radar-absorbent materials~(RAM)~\cite{RAM_stealth}. RAM absorbs the majority of the incident radar energy and therefore, provides stealth capabilities. RAM can be applied either in the form of sheets or paint layers. The basic principle of RAM is to match the intrinsic impedance of the RAM equal to the impedance of the incoming waves for maximum power transfer. The RAM should be able to cover wide frequency band impedance matching.

Meta-materials can also provide attenuation and dispersion of the radar EM energy incident on the aerial vehicles~\cite{metamaterial_stealth1,metamaterial_stealth2,Metamaterials_new} by: 1) changing the reflection/diffraction of EM waves from the surface of an aerial vehicle; 2) changing the polarization of the reflected waves; and 3) as a RAM. Meta-material can operate at wide frequency bands. Moreover, the shape of the aerial vehicle is not required to be changed, therefore, the aerodynamics of the aerial vehicle is not compromised using meta-material surfaces. However, aerial vehicles with meta-material surfaces are not always invisible from a radar system. There are number of ways to detect scattered energy from a meta-material surface, e.g., using passive radars/receivers~\cite{Passive_stealth,Passive_stealth2}.

\begin{figure}[!t] 
    \centering
	\begin{subfigure}{\columnwidth}
    \centering
	\includegraphics[width=0.8\textwidth]{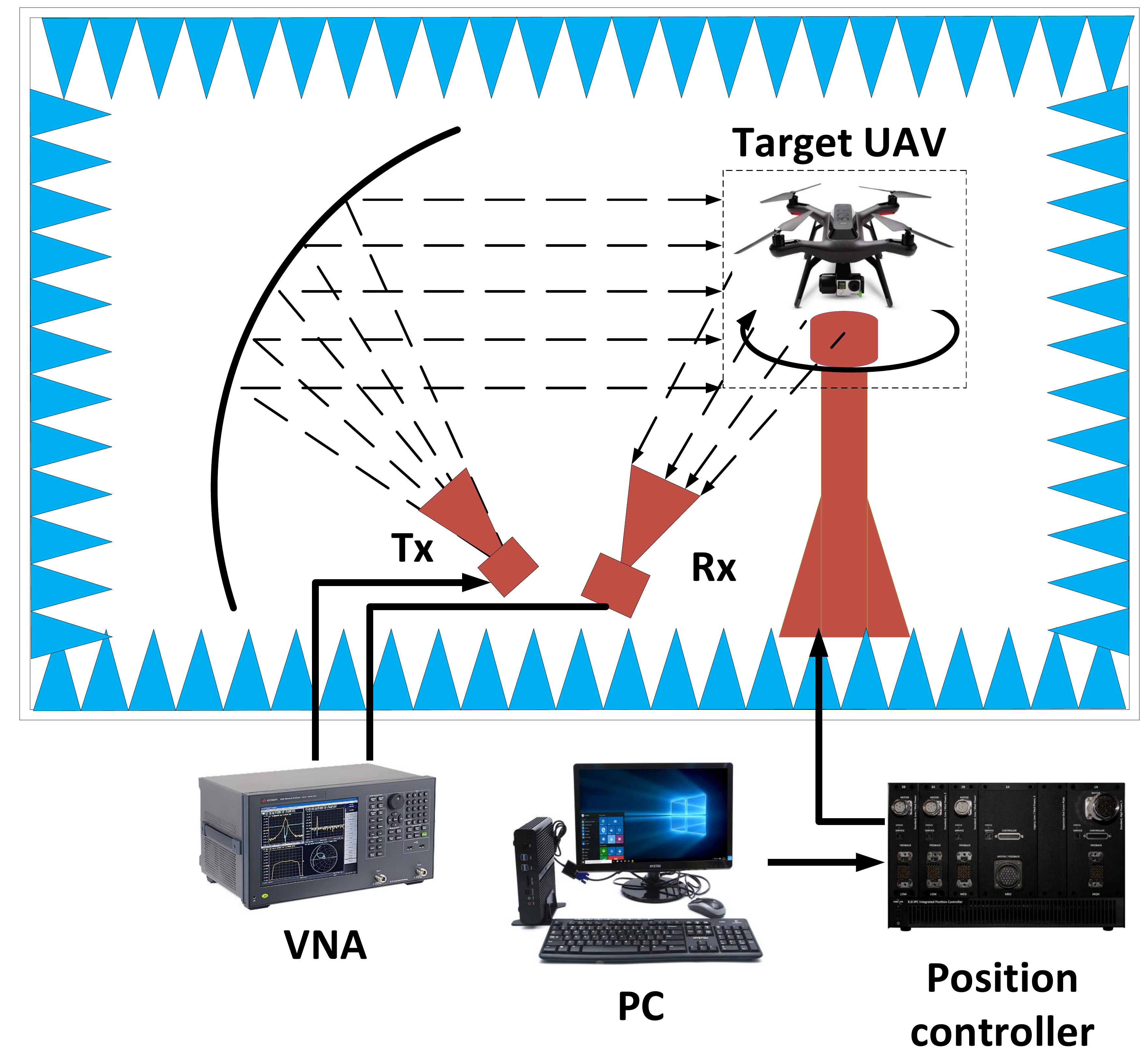}
	  \caption{}  
    \end{subfigure}    
    \begin{subfigure}{\columnwidth}
    \centering
	\includegraphics[width=0.8\textwidth]{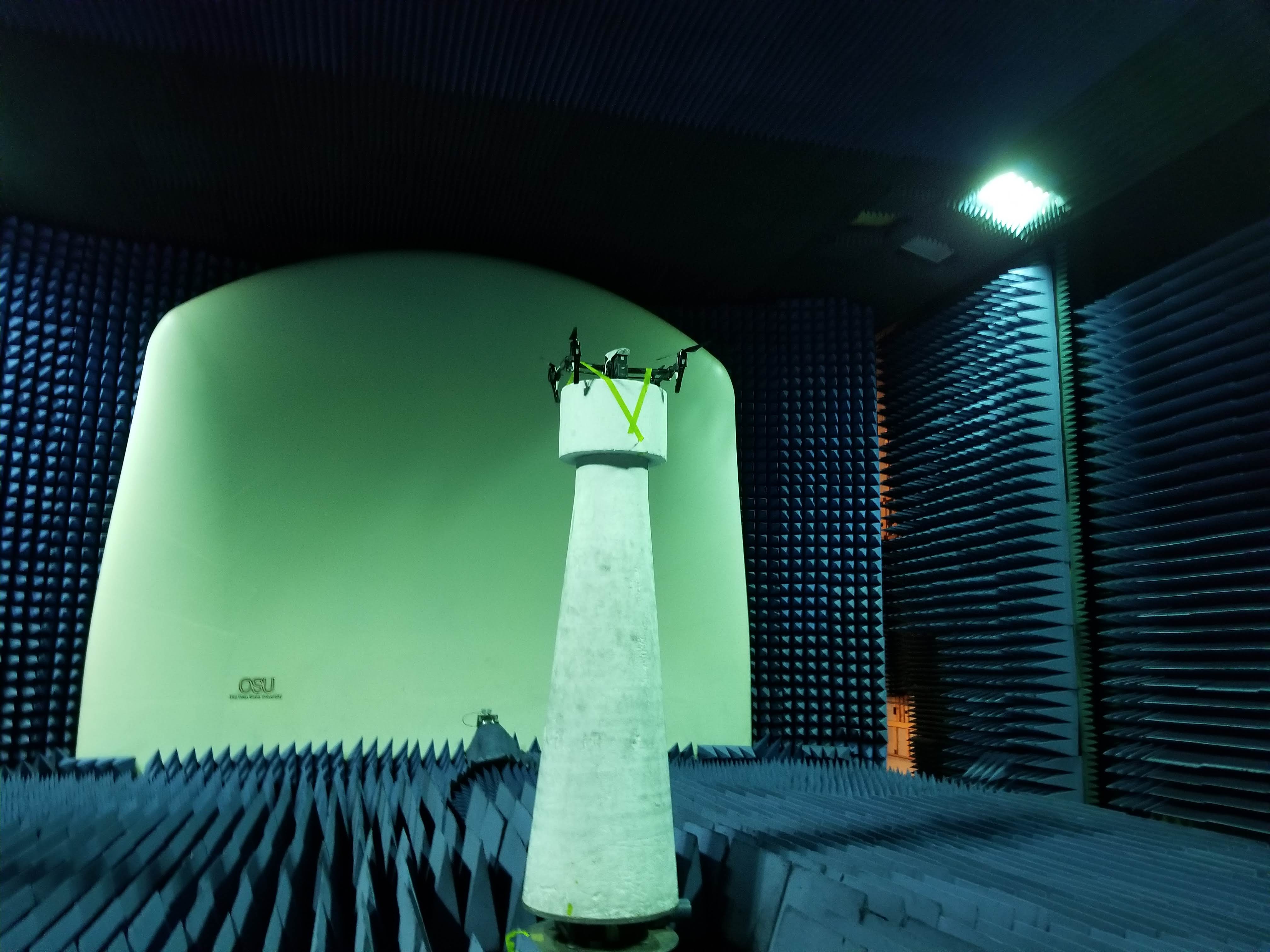}
	  \caption{}  
    \end{subfigure}
    \caption{The measurement setup in \cite{class_new9} is shown. The measurements were conducted in the anechoic chamber} \label{Fig:Martins_RCSsetup}
\end{figure}

\begin{figure}[!t] 
    \centering
	\begin{subfigure}{\columnwidth}
    \centering
	\includegraphics[width=0.63\textwidth]{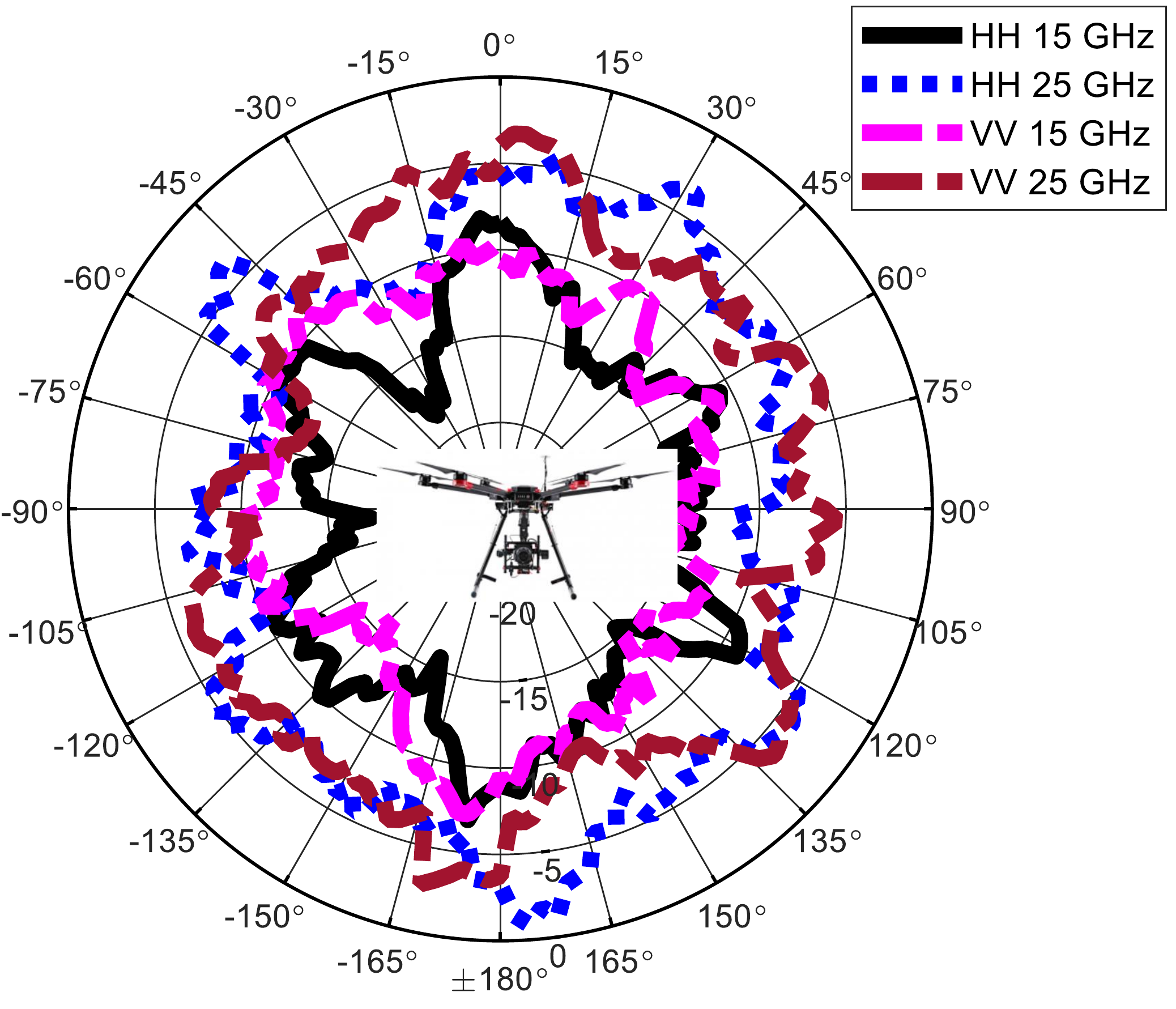}
	  \caption{}  
    \end{subfigure}    
    \begin{subfigure}{\columnwidth}
    \centering
	\includegraphics[width=0.63\textwidth]{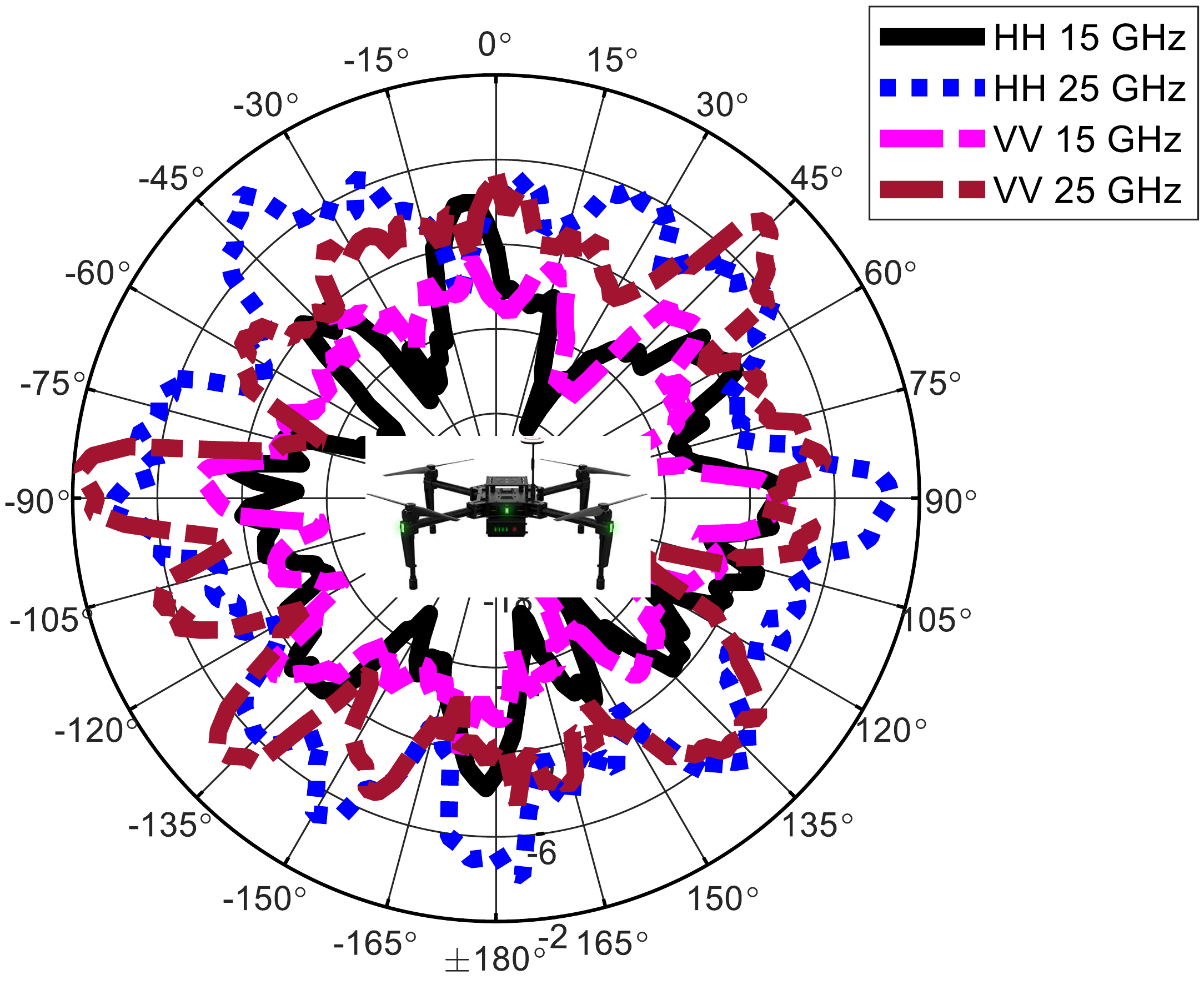}
	  \caption{}  
    \end{subfigure}
    \begin{subfigure}{\columnwidth}
    \centering
	\includegraphics[width=0.63\textwidth]{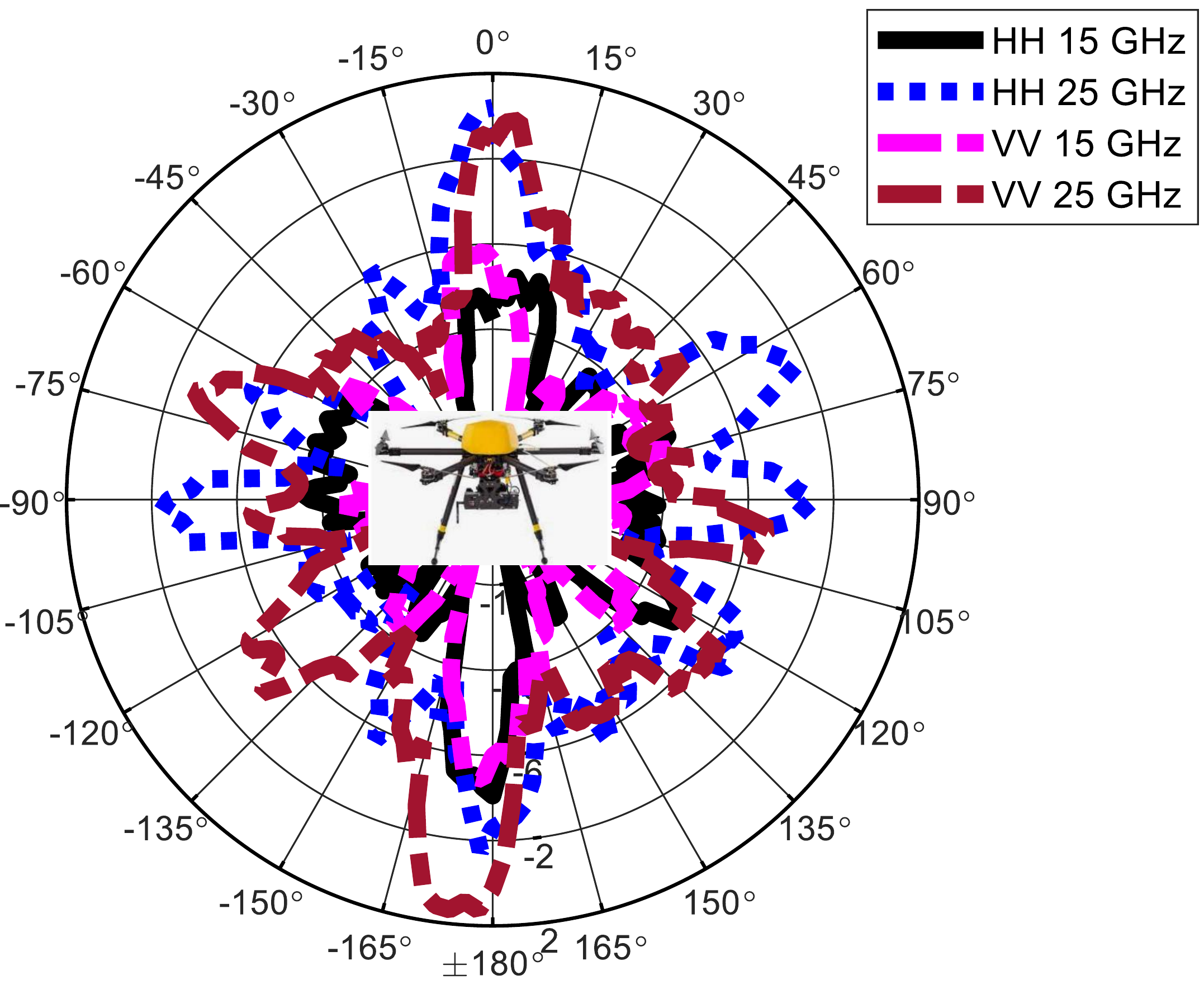}
	  \caption{}  
    \end{subfigure}
    \begin{subfigure}{\columnwidth}
    \centering
	\includegraphics[width=0.63\textwidth]{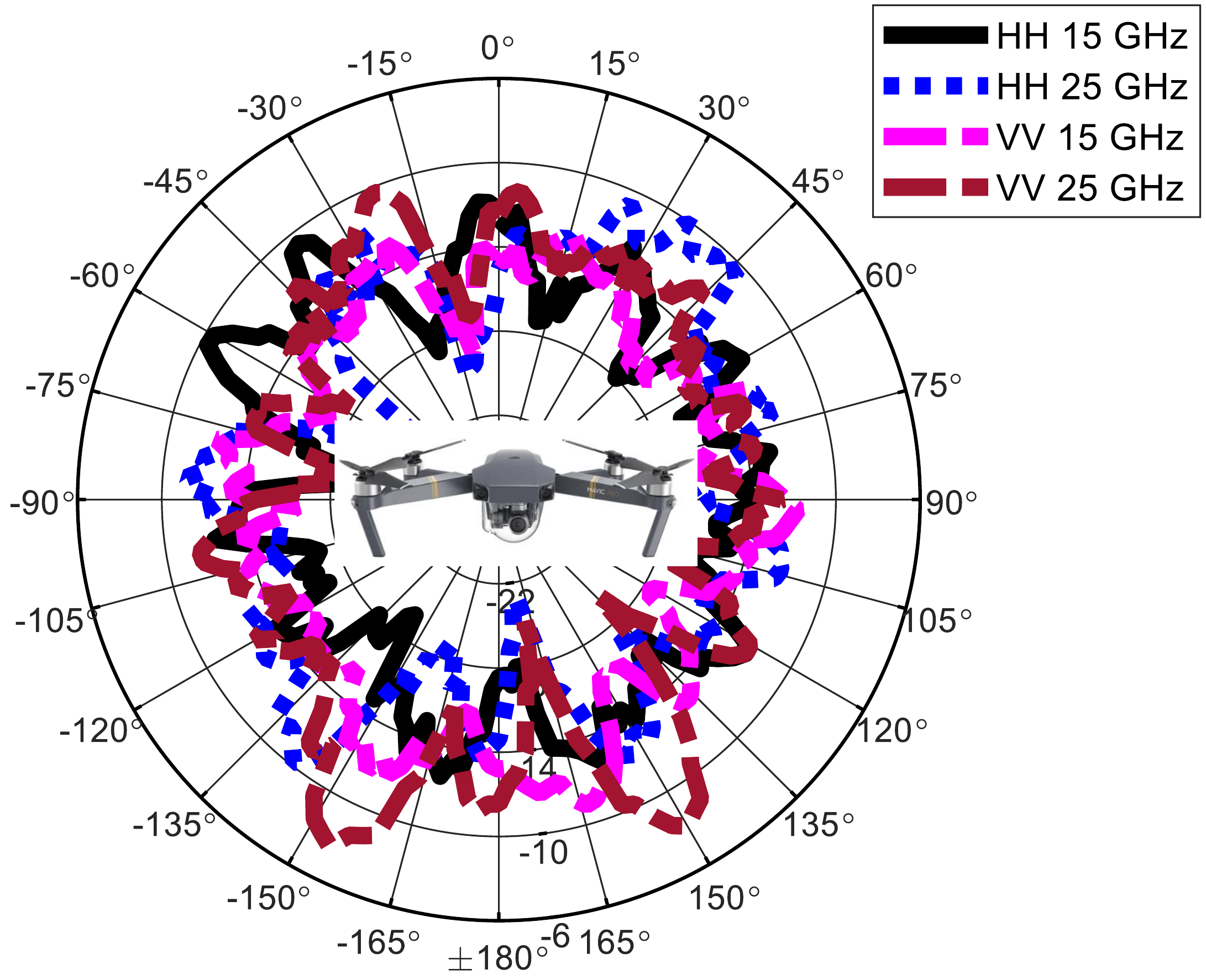}
	  \caption{}  
    \end{subfigure}
    \caption{The RCS of four different UAVs measured at 15~GHz and 25~GHz using vertical-vertical and horizontal-horizontal polarization~(regenerated from \cite{class_new9}). The measured RCS (in dBsm) is plotted against the azimuth angle for the range [0$^\circ$~360$^\circ$]. The four UAVs are (a) DJI Matrice 600 Pro, (b) DJI Matrice 100, (c) Trimble
zx5, and (d) DJI Mavic Pro 1.} \label{Fig:RCS_anechoic_martins}
\end{figure}


The RCS of an aerial vehicle can be used for classification. In \cite{class_new9}, RCS of six different types of UAVs were measured at $15$~GHz, and $25$~GHz. The measurements were carried out in the anechoic chamber. Vertical-vertical and horizontal-horizontal polarization were used in the measurements. Fifteen different classification algorithms from statistical learning, machine learning, and deep learning were used for UAV identification. The measurement setup is shown in Fig.~\ref{Fig:Martins_RCSsetup}. The RCS of four UAVs measured at $15$~GHz and $25$~GHz and at respective polarization pairs of vertical-vertical and horizontal-horizontal are provided in Fig.~\ref{Fig:RCS_anechoic_martins}. The confusion matrix of the best and worst-performing classifiers is also provided in \cite{class_new9}. The RCS of different types of aerial vehicles obtained by different radar systems are provided in Table~\ref{Table:RCS_radarsystems}, along with central frequency and sounding signal used in the specific studies reported in the literature. 

\begin{table*}[t]
	\begin{center}
     \footnotesize
		\caption{RCS of different types of aerial vehicles using different types of radar systems.} \label{Table:RCS_radarsystems}
\begin{tabular}{@{}|P{ 5.1cm}|P{3.4cm}|P{2.8cm}|P{3.8cm}|P{0.6cm}|@{}}
 \hline
\textbf{Aerial vehicle type}&\textbf{RCS}&\textbf{Center frequency}&\textbf{Sounding signal}&\textbf{Ref.}\\
\hline
DJI Matrice 600 Pro, DJI Matrice 100, Trimble zx5, DJI Mavic Pro 1, DJI Inspire 1 Pro, DJI Phantom 4 Pro& lognormal, generalized extreme
value, and gamma distributions&15 GHz, 25 GHz&Continuous wave &\cite{RCS_clutter1}\\
\hline
DJI Phantom 4 Pro& 0.01~m$^2$&25~GHz&FMCW &\cite{RCS_clutter2}\\
\hline
DJI Phantom 3& 0.01~m$^2$&Ku band&Pulse based phased array radar &\cite{clutter3}\\
\hline
Mavic Pro& 0.03~m$^2$&2.4~GHz&Continuous wave, linear frequency modulated &\cite{clutter4}\\
\hline
Iris+, X8, and High one& See Table~II in \cite{clutter5}&3~GHz, 9.7~GHz, 15~GHz, and 24.3~GHz&Pulse Ku-band
short range battlefield radar&\cite{clutter5}\\
\hline
Phantom 3, fixed-wing buzzard& Maximum value 0.09~m$^2$&X-band&FMCW&\cite{clutter6}\\
\hline
Ground targets observed from airborne platform& See Fig.~6, and Fig.~7 in \cite{clutter8}&215~MHz-730~MHz&UWB, SAR&\cite{clutter8}\\
\hline
Aircraft's weak scattering source& Ability to measure aerial targets with RCS of $1\times10^{-4}$~m$^2$&9~GHz-11~GHz&CW signal&\cite{clutter9}\\
\hline
UAV& 0.08~m$^2$&S-band&Pulses&\cite{clutter11}\\
\hline
Point targets&Rayleigh, and Rician distributions&C,L,P, and X bands&Pulses&\cite{clutter13}\\
\hline
\end{tabular}
		\end{center}
			\end{table*}

\subsection{Motion Characteristics} \label{Section:Doppler_ambiguity}
The motion characteristics of an aerial vehicle include velocity, pitch, yaw, roll angles, and rate of climb. The velocity of an aerial vehicle can be obtained by measuring the Doppler shift in frequency of the received radar signal. The Doppler shift in the frequency $f_{\rm d}$ of an echo signal due to motion of the aerial vehicle is given as $f_{\rm d} = \frac{2v\cos\alpha}{\lambda}$, where $v$ is the velocity of the aerial vehicle, and $\alpha$ is the angle between the radar's line of sight towards the aerial vehicle and direction of travel of the aerial vehicle. The Doppler shift is used to estimate the velocity of the aerial vehicle. Similar to range ambiguity, there is also Doppler ambiguity~\cite{range_doppler_ambiguity}. Due to Doppler ambiguity, an aerial vehicle will appear stationary at multiples of pulse repetition frequency~(PRF). The relative velocity with ambiguity is given as 
\begin{equation}
    V_{\rm r} = \frac{\lambda \big(f_{\rm b}\pm y f_{\rm r}\big)}{2},~~~y=0,1,2,\hdots, \label{Eq:doppler_amb}
\end{equation}
where $f_{\rm r}$ is the PRF and $f_{\rm b}$ represents the Doppler frequency bins. It can be observed that for maximum unambiguous Doppler/velocity we require large $f_{\rm r}$. On the other hand, for the maximum unambiguous range, we require the PRF to be small. Therefore, there is a trade-off between Doppler ambiguity and range ambiguity. The categorization of range and Doppler ambiguities in measurements based on PRF is given in Table~\ref{Table:Ambiguities}. 

\begin{table*}[!t]
	\begin{center}
     \footnotesize
		\caption{Ambiguities in range and Doppler~(velocity) measurements and PRF.} \label{Table:Ambiguities}
\begin{tabular}{@{}|P{4.5cm}|P{ 2.2cm}|P{2.2cm}|P{2.2cm}|@{}}
 \hline
\textbf{Measurement type}&\textbf{Low PRF}&\textbf{Medium PRF}&\textbf{High PRF}\\
\hline
Range measurements&Unambiguous&Ambiguous&Very ambiguous\\
\hline
Velocity~(Doppler) measurements&Very ambiguous&Ambiguous&Unambiguous\\
\hline
\end{tabular}
		\end{center}
			\end{table*}
			
Some popular radar systems that can provide velocity estimates based on Doppler measurements are continuous-wave~(CW) and pulse-Doppler radars. CW radars are the simplest radars and transmit a CW at a given frequency. Any shift in the frequency of the CW due to reflection from a moving aerial vehicle is translated to corresponding radial velocity~\cite{cw}. The CW radar is further divided into other types, e.g., frequency-modulated continuous-wave~(FMCW) radar that can provide both range and velocity estimation of an aerial vehicle. The pulse-Doppler radar also provides velocity and range estimates~\cite{pd}. The pulse-Doppler offers a combination of the features of the pulse radar and CW radar. Pulse-Doppler radar is often used on airborne platforms for the detection of moving aerial vehicles in air and stationary/slow moving objects on ground. 

In addition to the main Doppler shift, there are additional Doppler shifts due to motion/rotation of sub-parts of the aerial vehicle. The additional Doppler shifts due to motion/rotation of sub-parts of an aerial vehicle at the micro level are categorized as micro-Doppler, which is a helpful feature often used in the classification of an aerial vehicle. The micro-Doppler also helps to determine the phases of the flight of the aerial vehicle based on the rotation of the propellers. The motion of blades of a helicopter or propellers of a plane, flapping of wings by birds are some of the examples that produce noticeable micro-Doppler. Representative mico-Doppler measurements using radar systems are  provided in \cite{micro_doppler}. In \cite{micro_doppler}, Doppler modulations modeling, and analysis of micro-Doppler phenomenon in different scenarios are provided using simulation and real-time radar data. In \cite{UAV_microDoppler_new}, the unique micro-Doppler signatures of the UAVs were identified and the UAV was categorized based on the micro-Doppler signature. The speed and subsequently length of the propeller was obtained from the returned radar echoes in \cite{UAV_microDoppler_new}. The measurements were taken using CW radar operating at $94$~GHz. 

\subsection{Clutter}
The reflection of the EM waves is the major phenomenon observed by any radar. Dependent on the geometry/shape of the object and the wavelength of the incident radar wave, specular reflection or scattering is observed. The reflection of EM waves depends on the incidence angle, operating frequency, shape, size, and material of the reflector. The collection of radar reflections~(specular and scattered) from objects that are not of interest such as buildings, cars, and other man-made structures, hills, and forests, is called clutter~\cite{radar_clutter_basics}. The clutter can be static or dynamic. A major source of static clutter is the ground/sea surface reflection~\cite{GRC}. Dynamic clutter can be from moving vehicles/objects in the environment, e.g., rotation of wind turbines, and movement of the foliage.

 The clutter is related to the angular and range resolution of a radar system. If the angular resolution is small (i.e., large beamwidth) large clutter is observed and vice versa. If the range resolution is high, the clutter can be differentiated from the aerial vehicle, and the signal to clutter ratio increases. The clutter cross-section can be thousands of times larger than the cross-section of the aerial vehicle observed by radar. Looking down from an airborne radar towards the ground produces large clutter returns from the ground compared to look up radars.
 
There are three main types of clutter. The point, surface, and volume clutter. The point clutter is due to radar beam reflections mainly from tall buildings, birds, and other mobile objects in the environment. The surface clutter is the radar returns from the sea or ground surface. The volume clutter is due to radar reflections from hail, snow, and rain. The surface clutter is easier to cancel compared to point clutter as the reflections from the surface are always present and their statistics do not change significantly. The point clutter can be removed by using different types of motion filtering and clutter rejection algorithms discussed in Section~\ref{Section:clutter_reject}. The volume clutter can be detected and tracked using specialized weather radars, e.g., dual-polarized radars. Other precipitation sensors can also be used to detect volume clutter. The different clutter distributions observed by radar systems in different environments and sounding signals are provided in Table~\ref{Table:Clutter_dist}. From this table, it can be observed that the clutter distributions depend mainly on the type of the environment.

\begin{table*}[htbp]
	\begin{center}
     \footnotesize
		\caption{Clutter distributions obtained from different types of radar systems and environment. } \label{Table:Clutter_dist}
\begin{tabular}{@{}|P{ 4.4cm}|P{1.5cm}|P{2.2cm}|P{5.5cm}|P{0.5cm}|@{}}
 \hline
\textbf{Radar/antenna type}&\textbf{Center frequency}&\textbf{Sounding signal}&\textbf{Environment \& clutter distribution}&\textbf{Ref.}\\
\hline
Ancortek radar (now Luswave Technology), horn antenna&25~GHz&FMCW &Rocky terrain, and Weibull distribution&\cite{RCS_clutter2}\\
\hline
Airborne wideband radar&3.5~GHz&Pulses &Not specified, Non-uniform distribution&\cite{clutter2}\\
\hline
Non-coherent airborne pulse radar&S-band&Pulses &Farmland and sea clutter, and Weibull and lognormal distributions&\cite{clutter7}\\
\hline
Bi-static radar&S-band&Pulses &Farmland and sea clutter, and Weibull and lognormal distributions&\cite{clutter7}\\
\hline
Bi-static radar&2.4~GHz&Linear frequency modulated chirp&Sea clutter, and K-distributions&\cite{clutter10}\\
\hline
Airborne surface surveillance radar&X-band&Linear frequency modulated chirp&Sea clutter, and K, Pareto, Chi-squared, and exponential distributions&\cite{clutter12}\\
\hline
Space-time adaptive radar
&L-band&Pulses&Heterogeneous clutter, and Gamma distribution&\cite{clutter14}\\
\hline
Pulse radar
&15.5~GHz&Pulses&Uniform terrain, and Rayleigh distribution&\cite{clutter15}\\
\hline
\end{tabular}
		\end{center}
			\end{table*}

\subsection{Terrain Effects}
The propagation of radio waves depends on the terrain and atmospheric effects. There are four main types of terrain: urban, suburban, hilly/mountainous, and rural/open area. Each terrain can have a different number, and type of scatterers, e.g., buildings, cars, and trees. The propagation effects~(reflection, diffraction and scattering) on radio waves will be different in different terrain. Therefore, a radar system working in a rural terrain will require calibration and training before operation in an urban terrain. The atmospheric effects, e.g., wind in a forest area or strong air currents at sea can also affect the propagation characteristics of radio waves.

The terrain strongly affects the performance of any radar system. In addition to the free space loss, the terrain can introduce additional losses mainly due to obstruction. The terrain obstructions can result in the reduction of the signal-to-noise ratio~(SNR) and hence reduce the overall probability of detection. Therefore, the probability of detection or PFA can vary with the terrain. However, if the aerial vehicle flies at high altitudes above the terrain, the effects of the terrain are negligible on the detection. The effect of the mountainous terrain on the detection of an aircraft by a surveillance radar is provided in~\cite{Matlab_terrain}. The radar location is at Rocky Mountain Metropolitan Airport in Broomfield, Colorado, USA. The radar station is at a height of $10$~m above the ground. The aircraft is flying in a corkscrew trajectory from the radar location. The radar operates at $6$~GHz and the input power is $1$~kW. The rest of the simulation parameters are provided in~\cite{Matlab_terrain}. The line-of-sight~(LOS) between the radar and the aircraft is obstructed by the mountains. The detection of an aircraft by the surveillance radar in a hilly terrain can be further improved by increasing the transmit power~(peak power), increasing the pulse width~(PW), increasing the physical size of the antenna aperture of the radar, and using coherent integration of a large number of pulses~\cite{Matlab_terrain}.

\subsection{Atmospheric Effects}
The radar energy suffers attenuation as it travels through the atmosphere due to the presence of gases and water vapors. The attenuation increases in the presence of rain, fog, dust, and clouds. This attenuation is also called atmospheric attenuation. The attenuation in free space also depends on the frequency used. The peak received power from a monostatic radar $P_{\rm R}$ in free space is given as
\begin{equation}
    P_{\rm R} = \frac{P_{\rm T}G^2\lambda^2 \sigma}{(4\pi)^3R^4 L}, \label{Eq:FSPL_radar}
\end{equation}
where $P_{\rm T}$ is the transmitted power, $G$ is the gain of the radar antenna, $\lambda$ is the wavelength, $\sigma$ is the RCS of the interacting object, and $L$ is the losses from the hardware (RF losses in transmitter~(TX)/receiver~(RX)). In addition to free space attenuation in \ref{Eq:FSPL_radar}, there are additional losses due to antenna polarization mismatch~\cite{ant_pol_loss}, and atmospheric absorption~\cite{atmos_loss}. 

The region close to the earth's surface and below the horizon is called the diffraction region. In the diffraction region, the diffraction can be categorized as knife-edge or cylinder edge diffraction~\cite{refraction_reflection}. Above the diffraction region is the troposphere. The EM waves are refracted from the troposphere and the refraction characteristics depend on the dielectric constant, which itself depends on the temperature, pressure, and gaseous/water vapor content. Above the troposphere, small refraction occurs only and the region is called the interference zone. Above the interference zone, there is the ionosphere layer that includes ionized free electrons. The effect of the free electrons~(and positive ions) on the EM waves depends on the operating frequency. The free electrons can affect the EM waves through absorption, refraction, polarization change, and noise emission~\cite{ionosphere_electrons}. 

Atmospheric conditions vary with altitude, e.g., air density and humidity, rain rate, fog, and cloud water contents. The rain, hail, snow, and upper atmospheric conditions can directly affect the detection by radar~\cite{atmospheric_new}. The rain, hail, snow, fog, and other precipitation conditions can result in echoes from these particles that result in masking the actual echoes. The attenuation of radar signals due to rain, hail, snow, fog, and other precipitation phenomenon depends on the frequency of operation and droplet size distribution. The presence of charged particles, e.g., dust and sand can also change the ratio of received power to transmit power (pathloss) and is dependent on the distance between the charged particles and the RX~\cite{charged_particles}. 

\section{Radar Transmission and Reception}  \label{Section:TX_RX}
In this section, we will discuss the factors related to the transmission and reception of a radar system. The effect of these factors on the detection, tracking, and classification of aerial vehicles will also be discussed.  

\subsection{Transmit Power}
The transmit power of radars can vary from a few milliwatts to megawatts. The transmit power is mainly dependent on the application and constrained by the platform. The high transmit power allows a greater range for radar systems by overcoming the free space attenuation~(\ref{Eq:radar_range}). In the past, magnetron was used in low to high power radar applications~\cite{Magnetron}. The magnetron acts as a power oscillator when voltage is applied. Other radar system applications use power amplifiers for amplification of transmit power. A major benefit of the power amplifier is that stable high power signals can be produced from precise low powered signals. Instead of using a single power amplifier, a cascade of multiple power amplification stages is generally used~\cite{radar_transmitter2}. Vacuum tubes and solid-state amplifiers are mainly used for power amplification~\cite{radar_transmitter}. The size and amplification power of vacuum tube amplifiers are larger compared to solid-state amplifiers. However, the cost of vacuum tube amplifiers is significantly higher compared to solid-state amplifiers. The detection and tracking of medium to small RCS aerial vehicles at long ranges can be accomplished with a medium radar transmit power. However, for UAVs with small RCS, large transmit power is required. From (\ref{Eq:radar_range}), a large transmit power can compensate for the small RCS of a UAV. Similarly, for tracking from (\ref{Eq:track_eq}) a higher transmit power results in better SNR. Actively steered phased arrays that use solid-state power amplifiers can be used at medium to short ranges for UAV tracking. 

\subsection{Types of Sounding Signals} \label{Section:SS}
\begin{figure}[!t]
	\centering
	\includegraphics[width=\columnwidth]{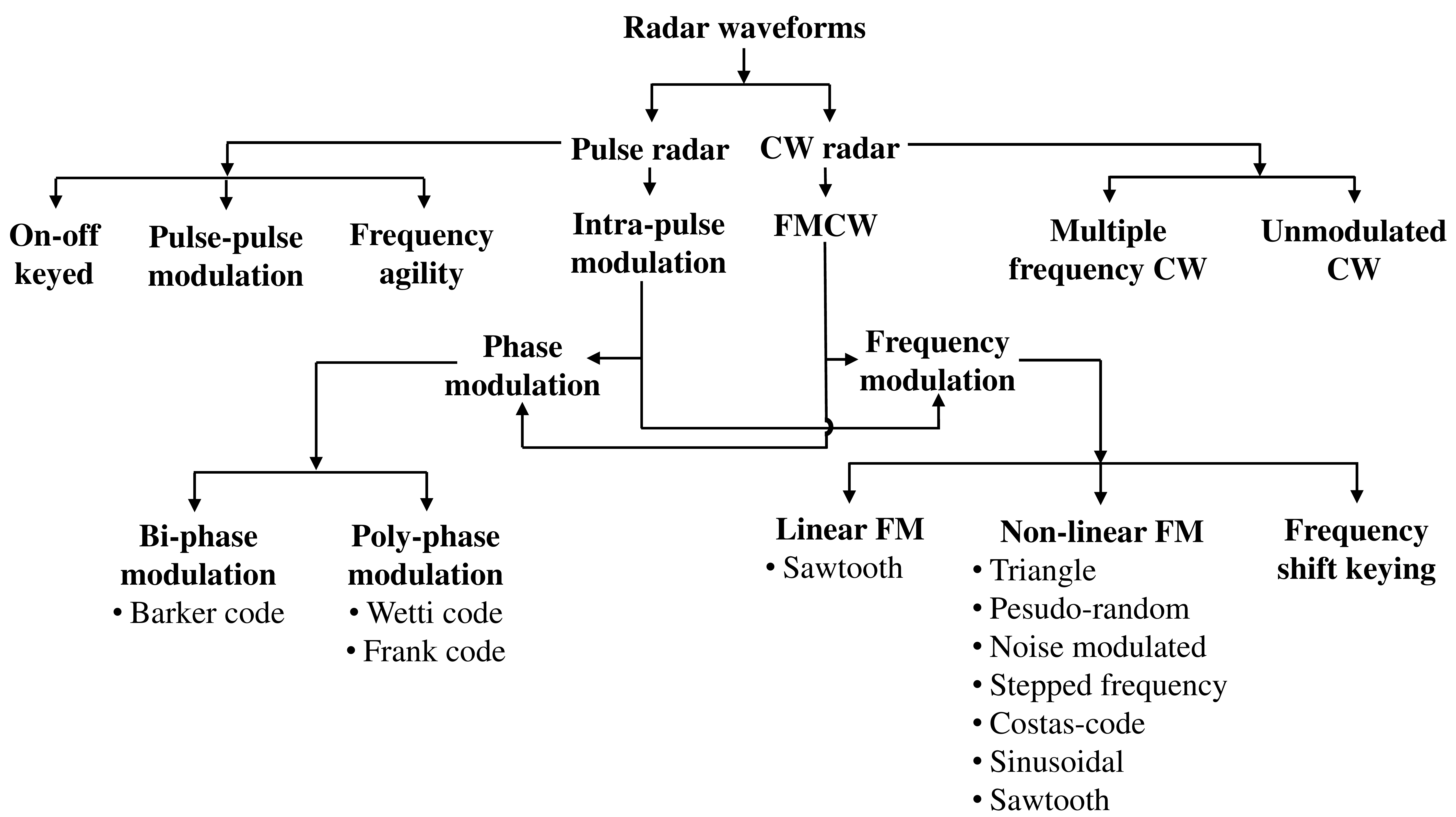}
	\caption{Different types of radar waveforms.}\label{Fig:Waveforms_radar}
\end{figure}

There are mainly two types of sounding signals used for radar transmissions: CW, and pulse transmissions. In addition to CW and pulse, other signal waveforms can be used. Different signal waveforms used by radar systems are provided in Fig.~\ref{Fig:Waveforms_radar}. CW radar systems are simpler compared to pulse radar systems. The transmission and reception takes place at the same time for the CW radar system. The peak power for CW radar is the same as the average power. Therefore, CW radar systems use low-power solid-state TXs. Major limitations of CW radar systems are: 1) CW radars cannot provide the range of the aerial vehicle directly~(as there is no basis for time delay measurement of the target for range calculation); 2) CW radars cannot differentiate between aerial vehicles when they are in the same direction and travel at the same speed~(because CW radars measure the Doppler shift at a single frequency and can only provide the direction and speed of the target); 3) CW radars cannot detect stationary or slow-moving aerial vehicles~(as the CW radar depends mainly on the Doppler shift that arises due to the motion of the target); and 4) the range of the CW radar systems is small compared to pulse radar systems (due to small but continuous power transmission). The CW radar systems can be amplitude, phase, or frequency modulated. The modulated CW radars can perform additional tasks, e.g., an FMCW radar can measure range and velocity simultaneously. 

\begin{figure}[!t]
	\centering
	\includegraphics[width=0.999\columnwidth]{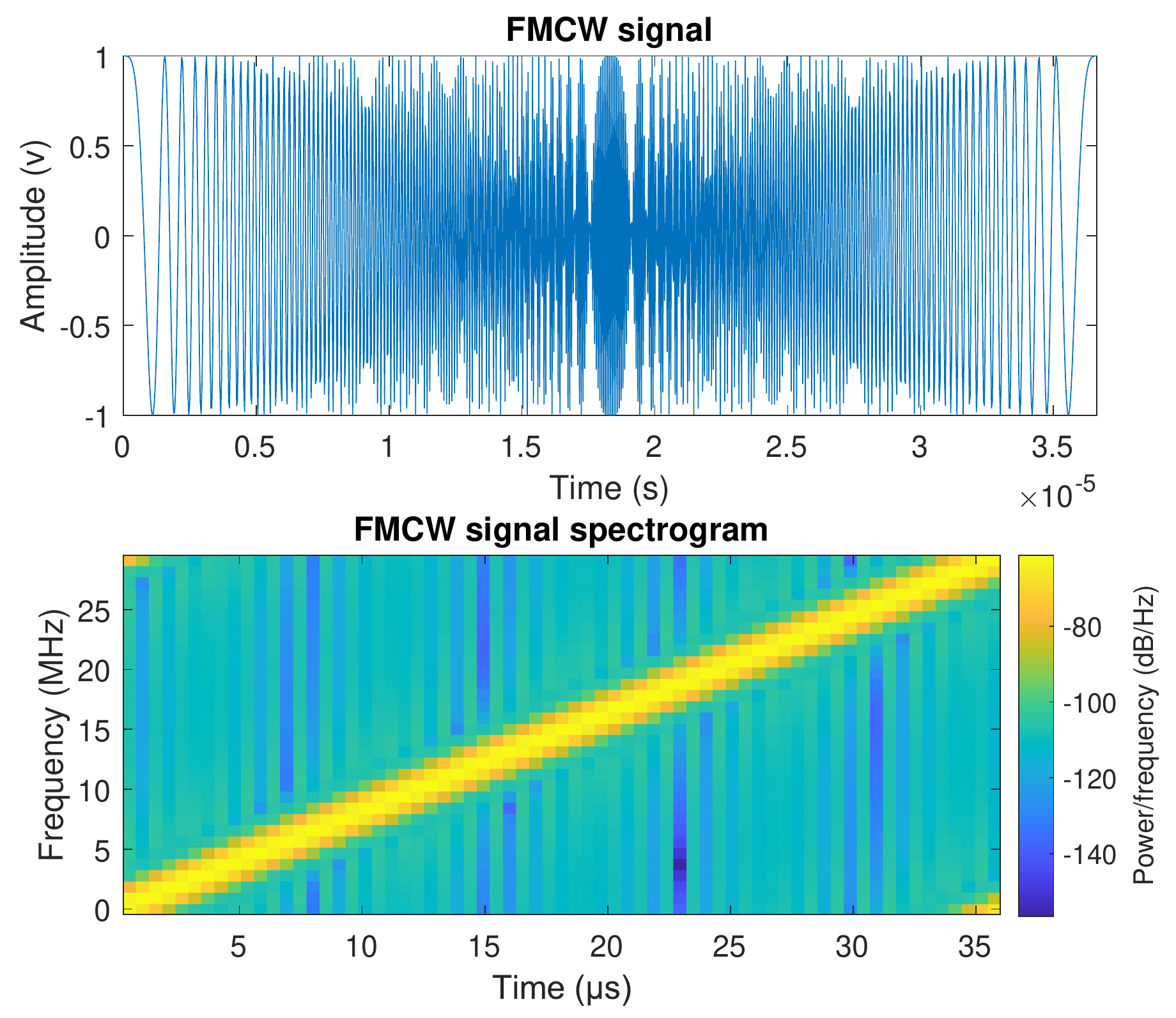}
	\caption{FMCW time and frequency domain signal representation. The center frequency is $18$~GHz, the range resolution is $5$~m, the maximum ambiguous range is $1$~km, maximum Doppler shift is $13$~kHz, and maximum beat frequency is $6.5$~MHz. }\label{Fig:FMCW_signal}
\end{figure}

The FMCW radars are the most popular among the CW radar's family for UAV detection. The FMCW radar uses two frequencies to obtain the phase difference information. The range measurement with the FMCW radar is given as $R = \frac{c\Delta \phi}{4\pi \Delta f}$, where $\Delta \phi$ is the phase difference, and $\Delta f$ is the frequency difference between the two frequencies of the FMCW. For FMCW, the unambiguous range is limited by $\frac{\lambda}{2}$, and therefore, it offers a limited range. An advantage of FMCW radar is that the FMCW radar can be used to detect stationary aerial vehicles. A time and frequency domain representation of the FMCW radar signal is shown in Fig.~\ref{Fig:FMCW_signal}. The center frequency is $18$~GHz, and the range resolution is $5$~m. The radar can detect aerial vehicles at a maximum unambiguous range of $1$~km, and the maximum Doppler shift and maximum beat frequency are $13$~kHz and $6.5$~MHz, respectively. A $35$~GHz FMCW radar is used for the detection and tracking of small UAVs in \cite{UAV_FMCW_new35}. An FMCW radar is built using software-defined radio~(SDR) USRP B210, and GNU radio in \cite{UAV_FMCW_new}. The radar is capable of detecting small UAVs of RCS $0.1$~m$^2$ at a range of $150$~m. The radar in \cite{UAV_FMCW_new} uses pulse compression and coherent integration. In \cite{UAV_SFCW}, a CW step frequency radar is used for the detection of small quadcopters at short ranges. 

Pulse radar is the most common type of a radar system. A major advantage of pulse radar compared to CW radar is the high dynamic range due to the isolation of TX and RX. The high dynamic range allows long-range detection capability. However, the range resolution of pulse radar is lower compared to FMCW radar. There is also a range ambiguity problem for pulse radar systems. A modification of pulse radar is the pulse-Doppler radar that combines the capabilities of pulse and CW radar systems~\cite{pulse_radar}. The pulse-Doppler can determine the range and velocity of an aerial vehicle simultaneously and has good look-down clutter rejection capabilities when used on aerial vehicles. Pulse-Doppler radar systems are used for long-range aerial vehicle~(manned or UAV) detection and tracking. In the literature different types of pulse radars are available. The FMCW and ultra-wideband~(UWB) pulse radars can be used for short-range detection and tracking of UAVs~\cite{fmcw1,uwb1}. 

\subsection{Pulse Width and Duty Cycle} \label{Section:PW_dutycycle}
\begin{figure}[!t]
	\centering
	\includegraphics[width=\columnwidth]{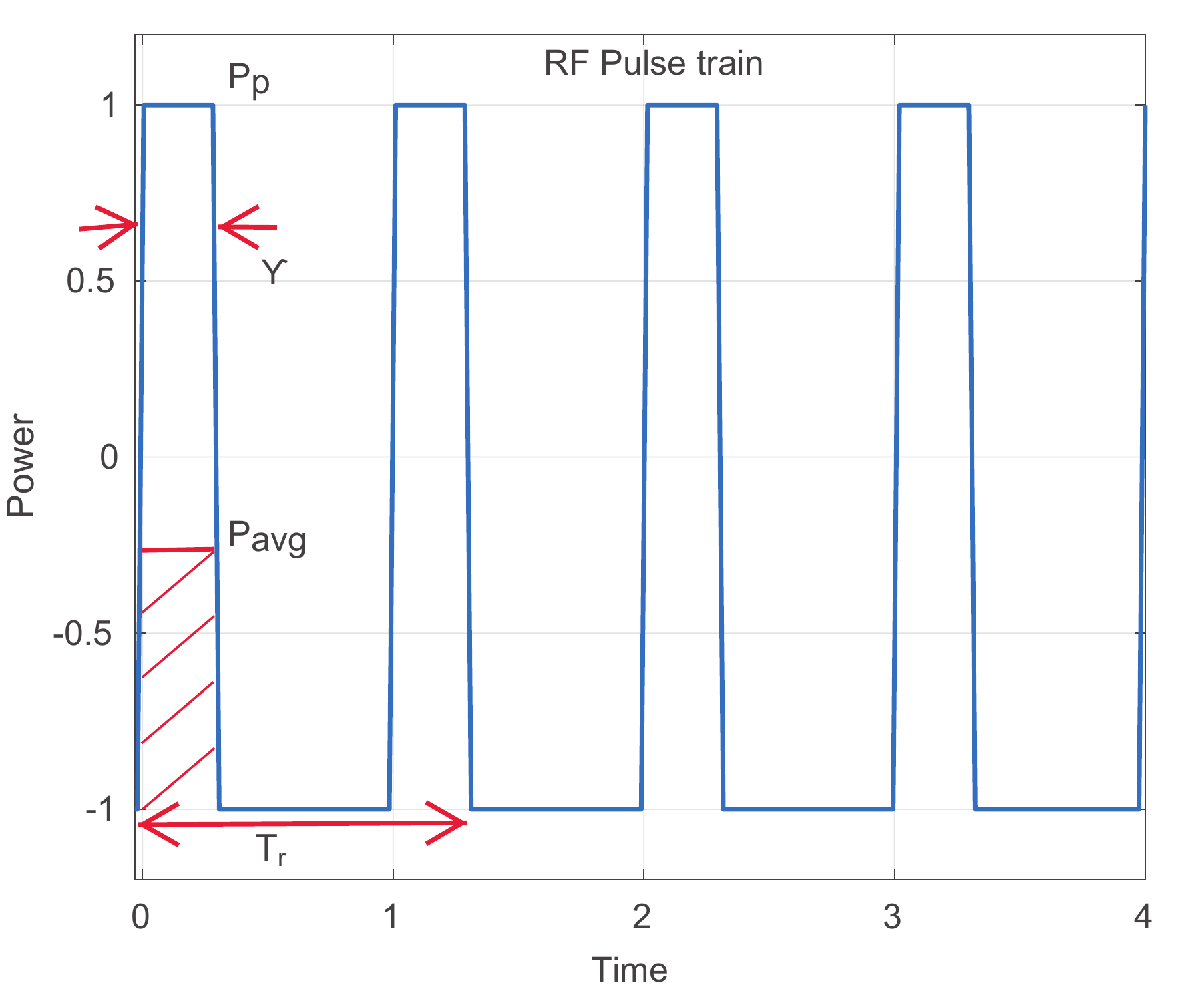}
	\caption{An RF pulse train showing the pulse width, pulse repetition interval, and pulse peak and average power. }\label{Fig:pulsetrain}
\end{figure}

Continuous high energy transmissions are not possible from a single TX/RX antenna~(monostatic radar) and it can damage the RX if not stopped by a Duplexer. Therefore, a listening time is required for pulse radars. The listening time is constrained by the duty cycle, PW, and pulse repetition interval~(PRI). The peak power, average power, PW, PRI, and duty cycle for a radar pulse train shown in Fig.~\ref{Fig:pulsetrain} are related to each other as follows~\cite{pulse_parameters}:
\begin{equation}
    \rm{duty~cycle}  = P_{\rm avg}/P_{\rm p} = \rm{\gamma}/\rm{t_{\rm r}} = \rm{\gamma} \times \rm{f_{\rm r}}, \label{Eq:dutycycle}
\end{equation}
where $P_{\rm p}$ is the peak power, and $\gamma$ is the PW. 

From (\ref{Eq:dutycycle}), the maximum energy that can be transmitted using a pulse radar depends on the PW, PRI, duty cycle, and listening time. The PRI is also directly related to the maximum unambiguous range of a radar~\cite{radarbook}. For a given duty cycle, increasing the PW increases the maximum energy carried by the pulse, however, increasing the PW can reduce the PRI, range resolution~(with no pulse compression), and unambiguous range. To overcome this trade-off either a dead interval can be added to the PRI or pulses at random intervals can be transmitted. PRF staggering where pulses at different intervals are transmitted for each scan can also be used to trace the ambiguous range returns. 

The waveform parameters can be changed adaptively based on the real-time returns of a radar. An example is waveform scheduling, where the waveform parameters are adjusted adaptively by the radar. In waveform scheduling, the PRF is changed based on the detection of the aerial vehicle/s. For example, if an aerial vehicle at $2$~km is detected at an unambiguous range of $5$~km then, the radar can switch to a higher PRF corresponding to $2$~km. Waveform scheduling can also help to efficiently detect changes in the speed of an aerial vehicle by adjusting the PRF. 

According to the Federal Communications Commission document~\cite{ITUwaveforms}, which handles devices operating at $5$~GHz, the radar test waveforms are divided into two major types. The two major types are short pulse and long pulse test waveforms. The categorization of short and long pulse types is based mainly on PRI, PW, and the number of pulses. Table~\ref{Table:shortwaveforms}, and Table~\ref{Table:longwaveforms} show the characteristics of short and long pulse radar test waveforms, respectively. 

\begin{table*}[htbp]
	\begin{center}
     \footnotesize
		\caption{Short pulse radar test waveform~(adapted from \cite{ITUwaveforms}).} \label{Table:shortwaveforms}
\begin{tabular}{@{}|P{ 0.7cm}|P{ 1.5cm}|P{2.5cm}|P{4.0cm}|P{2.7cm}|P{2.6cm}|@{}}
 \hline
\textbf{Type of Radar}&\textbf{PW ($\mu$s)} &\textbf{PRI ($\mu$s)} &\textbf{Number of pulses} &\textbf{Minimum
percentage of successful detection}&\textbf{Minimum number of trials}\\
\hline
0&1&1428&18&See note~1 in \cite{ITUwaveforms}&See note~1 in \cite{ITUwaveforms}\\
\hline
1&1&Test~A, and Test~B, in \cite{ITUwaveforms}&$\rm{Roundup}\Big( \big(\frac{1}{360}\big).\big(\frac{19.10^6}{PRI}\big)\Big)$&60\%&30\\
\hline
2&1-5&150-230&23-29&60\%&30\\
\hline
3&6-10&200-500&16-18&60\%&30\\
\hline
4&11-20&200-500&12-16&60\%&30\\
\hline
\end{tabular}
		\end{center}
			\end{table*}

\begin{table*}[htbp]
	\begin{center}
     \footnotesize
		\caption{Long pulse radar test waveform~(adapted from \cite{ITUwaveforms}).} \label{Table:longwaveforms}
\begin{tabular}{@{}|P{ 0.7cm}|P{ 1.3cm}|P{1.5cm}|P{1.5cm}|P{2.2cm}|P{2.0cm}|P{2cm}|P{2cm}|@{}}
 \hline
\textbf{Type of radar}&\textbf{PW ($\mu$s)} &\textbf{Chirp width (MHz)}&\textbf{PRI ($\mu$s)} &\textbf{Number of pulses per burst}&\textbf{Number of bursts}&\textbf{Minimum percentage of successful detection}&\textbf{Minimum number of trials}\\
\hline
5&50-100&5-20&1000-2000&1-3&8-20&80\%&30\\
\hline
\end{tabular}
		\end{center}
			\end{table*}

\subsection{Signal Processing Techniques} \label{Section:Signal_processing}
\begin{figure}[!t]
	\centering
	\includegraphics[width=\columnwidth]{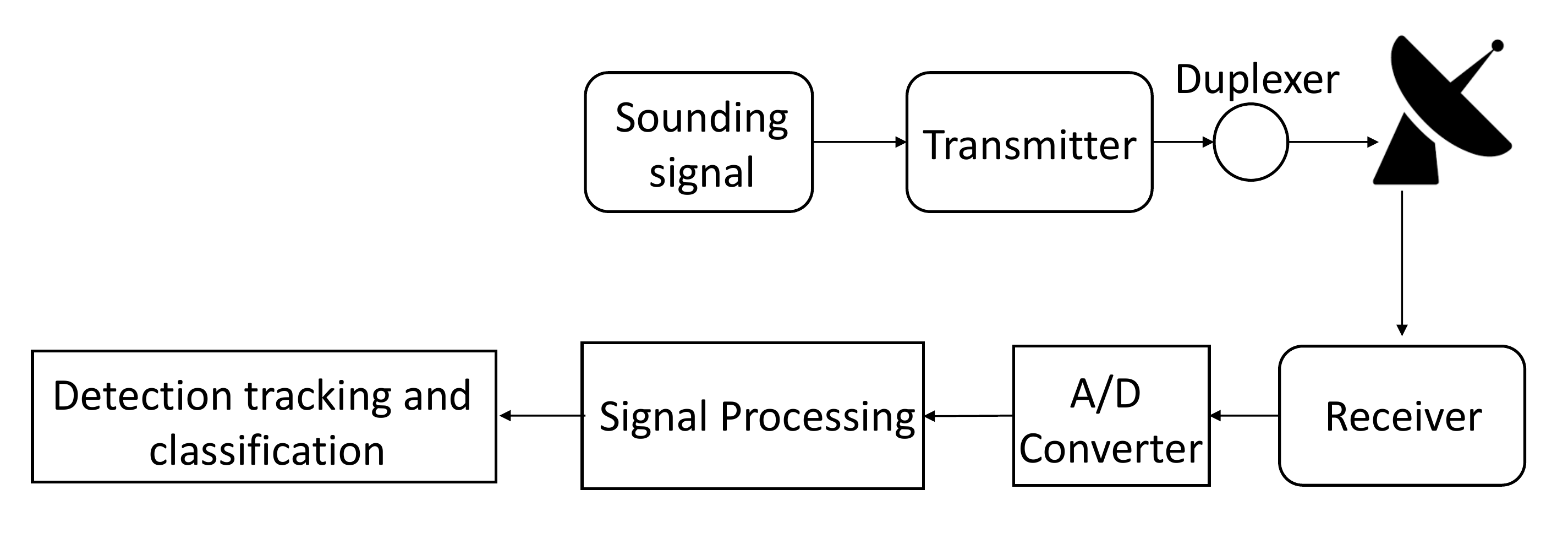}
	\caption{Basic signal processing of a radar system.}\label{Fig:radar_signal_processing}
\end{figure}

\begin{figure}[!t]
	\centering
	\includegraphics[width=\columnwidth]{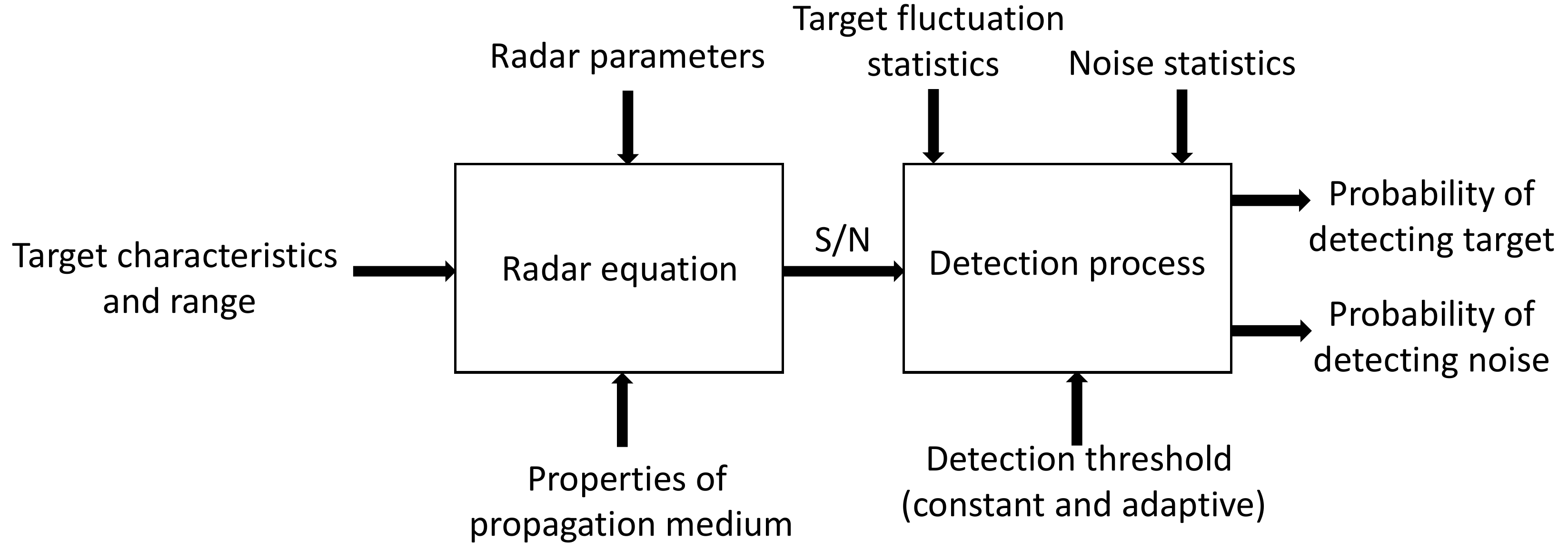}
	\caption{Block diagram of the radar processing for detection of an aerial vehicle.}\label{Fig:radar_processing}
\end{figure}

The signal processing is used mainly at three stages of a radar system. The first stage is at the TX side for signal generation, waveform shaping, modulation, and preparation of the signal for the RF front end. The second stage is at the RX side for processing of the received pulses, e.g., analog to digital conversion, noise removal, matched filtering, and pulse compression. The third stage is post-processing after the signal is recovered at the RX. Post-processing includes detection, ranging, tracking, Doppler processing, and classification using different algorithms. Fig.~\ref{Fig:radar_signal_processing} and Fig.~\ref{Fig:radar_processing} show the basic signal processing blocks for a radar system. The signal processing for all three stages will be dependent on the platform over which radar is mounted. If the radar is mounted on a moving platform, e.g., a ship or an aerial vehicle, then the real-time motion characteristics of the platform are included in the signal processing. Following are some of the popular signal processing techniques used in radar systems. 

\subsubsection{Detection and Tracking}
Detection and tracking of a single conventional target is performed using typical radar signal processing techniques. However, detection and tracking of multiple aerial vehicles is a challenging task. Modern signal processing techniques~\cite{multiple_targets1, multiple_targets2, multiple_targets3} can be used for detection and tracking of multiple aerial vehicles simultaneously. Similarly, detection and tracking of modern aerial threats, e.g., UAVs and stealth aerial vehicles requires additional signal processing. Due to absence of direct or weak reflection from a modern aerial threat, high order reflections and multiple diffractions from an aerial target are processed using complex algorithms. For example, non-directive signals from TV and radio broadcasting, and mobile communications can be processed in the passive mode for detection of UAVs~\cite{dvb_uav}.

\subsubsection{Range and Velocity Calculation}
The range resolution is important for estimating the features of an aerial vehicle. The range resolution is inversely proportional to the PW, whereas, the range is directly proportional to the PW for CW pulse radar. Therefore, to achieve high-range resolution and long-range simultaneously, pulse compression is used. In pulse compression, a long~(wide) pulse is frequency or phase-modulated to have a range resolution similar to a narrow pulse. Linear frequency modulation and binary phase coding are popular modulation methods for pulse compression. The pulse compression also helps to achieve high SNR. The radar systems can accurately measure the range of the aerial vehicle, however, finding the accurate direction is challenging. There are different ways to find the direction of the radar echo, such as interferometry~\cite{interferometry}. Similarly, monopulse radar can be used for obtaining accurate directional information of the aerial vehicle~\cite{monpulse_radar}.

Doppler processing is essential for estimating the velocity of an aerial vehicle in all the radar systems. The frequency shift of the received signal from the center frequency is used to obtain the Doppler shift and subsequently the velocity estimate of the aerial vehicle. Micro-Doppler processing can also be used to obtain the micro-Doppler signature of a moving aerial vehicle. The micro-Doppler signature helps in the classification of the aerial vehicle~\cite{class_new6}. 

\subsubsection{Modulation and Coding}
Different types of modulations are used by different types of radar systems~\cite{modulation1}. Popular modulations for radar systems are 1) amplitude modulation, 2) pulse-amplitude modulation, 3) linear frequency modulation, 4) pulse linear frequency modulation, 5) CW linear frequency modulation, 6) stepped frequency modulation, and 7) FMCW. Different modulation patterns available for measurements are sawtooth, triangular, square-wave, staircase, and sinusoidal. Coding is also used to make radar communications reliable and efficient. Different coding schemes can be used for different types of radar systems~\cite{coding1,coding2}.  

\subsubsection{Coherent Integration}
In pulse radar systems, integration of received pulses is used to increase the SNR. The pulse integration can be coherent or non-coherent. The coherent integration requires both the in-phase and quadrature components of the received signal. The in-phase and quadrature components are used to obtain the phase of the received signal. A coherent processing interval~(CPI) can be used to obtain higher SNR. In CPI, multiple pulses~(generally in groups) using the same PRF and frequency are coherently integrated. Different CPIs can be used to extract additional information about an aerial vehicle~\cite{UAV_CPI_diff}, leading to better detection and tracking. For perfectly coherent integration of $N$ pulses, the SNR is $N\times SNR$. In non-coherent integration, the phase information is not available. The gain of the non-coherent integration is significantly small compared to coherent integration. 

\subsubsection{Clutter Rejection}
The filtering of received pulses polluted with the various copies of the transmitted pulses~(multipath) is generally performed before the matched filtering. Moreover, the signal obtained from the antenna sidelobes and unintended scatterers is generally considered clutter and subsequently rejected. Motion filtering/clutter rejection is used to remove clutter from the aerial vehicle echoes. The clutter rejection is simpler for ground-based radars and relatively stationary surroundings. A simple clutter rejection/motion filter subtracts either two consecutive channel impulse responses~(CIRs) or subtracts the mean CIR from the instantaneous CIR~\cite{uwb_wahab1}. However, the clutter rejection becomes complicated when the radar is on a moving platform and the channel is fast varying. For moving platforms and fast varying channels, adaptive clutter rejection is required~\cite{adaptive_clutter}. Moving target indicator~(MTI) and pulse-Doppler processing are popular motion filters. MTI and pulse-Doppler processing use the Doppler principle to reject clutter and enhance the detection of moving aerial vehicles. 

The threshold is critically important to properly reject clutter without suppressing the echoes from the target. If a threshold is not properly adjusted, it can either lead to false alarms or miss aerial vehicle detection. The threshold for the received echoes in a radar system depends on the noise floor, which is not constant and changes with temperature and atmospheric events e.g. rain, and clutter. Therefore, adaptive thresholding where the threshold is adjusted adaptively based on the noise floor is used. The advantage of adaptive thresholding compared to the fixed threshold is that the PFA and miss detection are significantly reduced.

\subsubsection{Cognitive Radar and AI Techniques}
Cognitive radar systems require higher signal processing compared to conventional radar systems. The processing for cognitive radar includes sensing the environment and adjusting the transmission and reception parameters of the radar accordingly~\cite{cognitive_adaptive}. The adjustment of transmission and reception parameters of the radar systems can be carried out cognitively using AI~\cite{cognitive_AI}. Furthermore, different AI algorithms are available for the detection and classification of an aerial vehicle~\cite{AI_class1,AI_class2}. The stored database of features of different types of aerial vehicles and real-time features of a potential aerial vehicle are compared using AI algorithms for detection and classification. The AI methods for the detection and classification of aerial vehicles are expected to replace traditional filtering techniques used by radar systems. 

\subsubsection{Others}
\begin{itemize}
\item Compressed sensing is used for solving problems related to detection, tracking, and classification of aerial vehicles by radar systems~\cite{compressed2,compressed1}. Other radar estimation problems, e.g., the high-resolution direction of arrival of the aerial vehicle is also addressed using the compressed sensing techniques~\cite{compress_sensing2}. Compressed sensing is popular for the detection of small UAVs~\cite{cs_detect} by passive radars~\cite{passive_compress}.
\item Jam-resistant signals that use a large bandwidth and robust frequency-hopping rates are resilient against intentional jamming. Additionally, encryption and authentication are used to avoid signal spoofing.
\item Majority of the radar systems nowadays use digital signal processing. Analog to digital converter~(ADC) is a basic component of a digital system. The sampling rate of the ADC will determine the overall rate of the signal processing of a radar system. 
\end{itemize}

\subsection{Motion Filtering} \label{Section:clutter_reject}
The reflection of transmitted radar energy from objects other than the aerial vehicle in an environment is categorized as clutter. The clutter can be from static or moving objects in an environment. Clutter is a random process and power spectral density~(PSD) is concentrated around $f=0$~(zero mean in frequency content). Ideally, the PSD from a static clutter is a delta function, but it has a small spread in practical cases. The mean and spread of the PSD of static clutter are significantly smaller than the aerial vehicle. The clutter mainly depends on the terrain, atmospheric conditions, seasonal changes, and radar platform. The clutter rejection becomes challenging for mobile radar platforms in unknown terrain and for fast varying channels. In \cite{track_new2}, both time domain and micro-Doppler radar return signatures are used to differentiate the UAV from the ground clutter. Plot of Doppler spectrum of clutter and UAVs is provided in~\cite{track_new2}. The spectrum of micro-Doppler signature of the clutter is significantly narrow compared to the UAV that helps in the differentiating the two. In \cite{clutter15}, detection of a small RCS UAV in a cluttered environment is provided. The RCS of the target UAV is $0.01-0.1$m$^2$ at $15.5$~GHz and different distributions of the amplitude of radar clutter are provided in different terrains, e.g., a Rayleigh distribution in a uniform terrain.  

The simplest form of motion filtering or clutter rejection for pulse-based radar systems is obtained by subtracting the instantaneous CIR from the mean CIR~\cite{uwb_wahab1}. In \cite{clutter_filter}, a machine learning approach is used for the classification and modeling of clutter in the presence of interference and noise. Orthogonal frequency-division multiple access is used for the collection of clutter signals and joint radar sensing and communications, and machine learning is applied to obtain a classification accuracy of $79\%$. In \cite{clutter_reject1}, the performance of the adaptive array processing technique is compared with the displaced phase center aperture processing technique in a non-homogeneous cluttered environment. The adaptive array processing technique using the joint angle-Doppler domain localized generalized likelihood ratio test performs better than the displaced phase center aperture processing technique. In \cite{clutter_reject2}, clutter mitigation techniques are discussed for airborne radars. The space-time adaptive processing~(STAP) provides the best performance for airborne radars in a spatially large coverage and cluttered environment. 

\begin{figure}[!t]
	\centering
	\includegraphics[width=0.9\columnwidth]{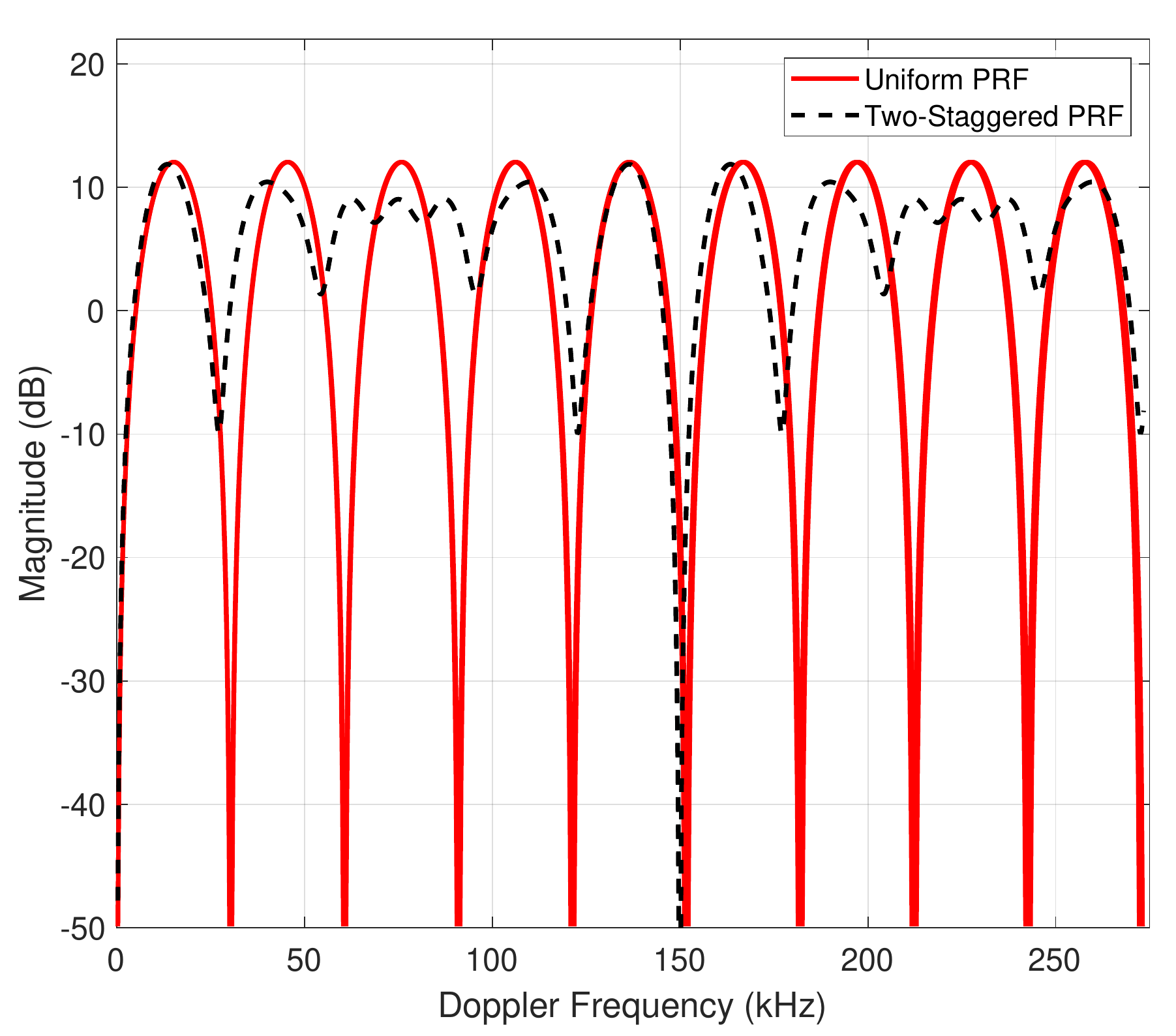}
	\caption{MTI filter plots for three pulse canceller using uniform PRF and 2-staggered PRFs at $30$~kHz and $25$~kHz, respectively~(regenerated from \cite{Matlab_MTI}). }\label{Fig:Uniform_staggered_PRF}
\end{figure}

MTI is a popular clutter rejection filter. MTI can be used to discriminate moving aerial vehicles, e.g., UAVs from clutter~\cite{mti_uav}. In~\cite{mti_uav, mti_uav2} MTI is used for the detection and classification of UAVs. The main principle of the MTI is that the phase of a moving aerial vehicle changes with time, whereas, the phase of a static aerial vehicle remains constant. The amplitude of transfer function of MTI radar is $|H(f)| = 2\sin\big(\pi f_{\rm d}T\big)$, where $f_{\rm d}$ is the Doppler frequency, and $T = \frac{1}{f_{\rm r}}$. In MTI radar, there are blind speeds due to $f_{\rm r}$ that results in velocity/Doppler ambiguity shown in Fig.~\ref{Fig:Uniform_staggered_PRF}. In Fig.~\ref{Fig:Uniform_staggered_PRF}, a periodic null is obtained for the uniform PRF, whereas, for the 2-staggered PRF~(at $30$~kHz and $25$~kHz), the null is present only at five times the uniform PRF, hence resulting in fewer blind speeds. Overall, MTI only separates the aerial vehicle returns from the clutter, uses short waveforms~(usually 2 to 3 pulses), and does not provide velocity estimates of the aerial vehicle. Due to Doppler ambiguities and inability to detect the blind speed of a moving target, MTI filters are not preferred for airborne radars. 

\begin{figure}[!t]
	\centering
	\includegraphics[width=0.9\columnwidth]{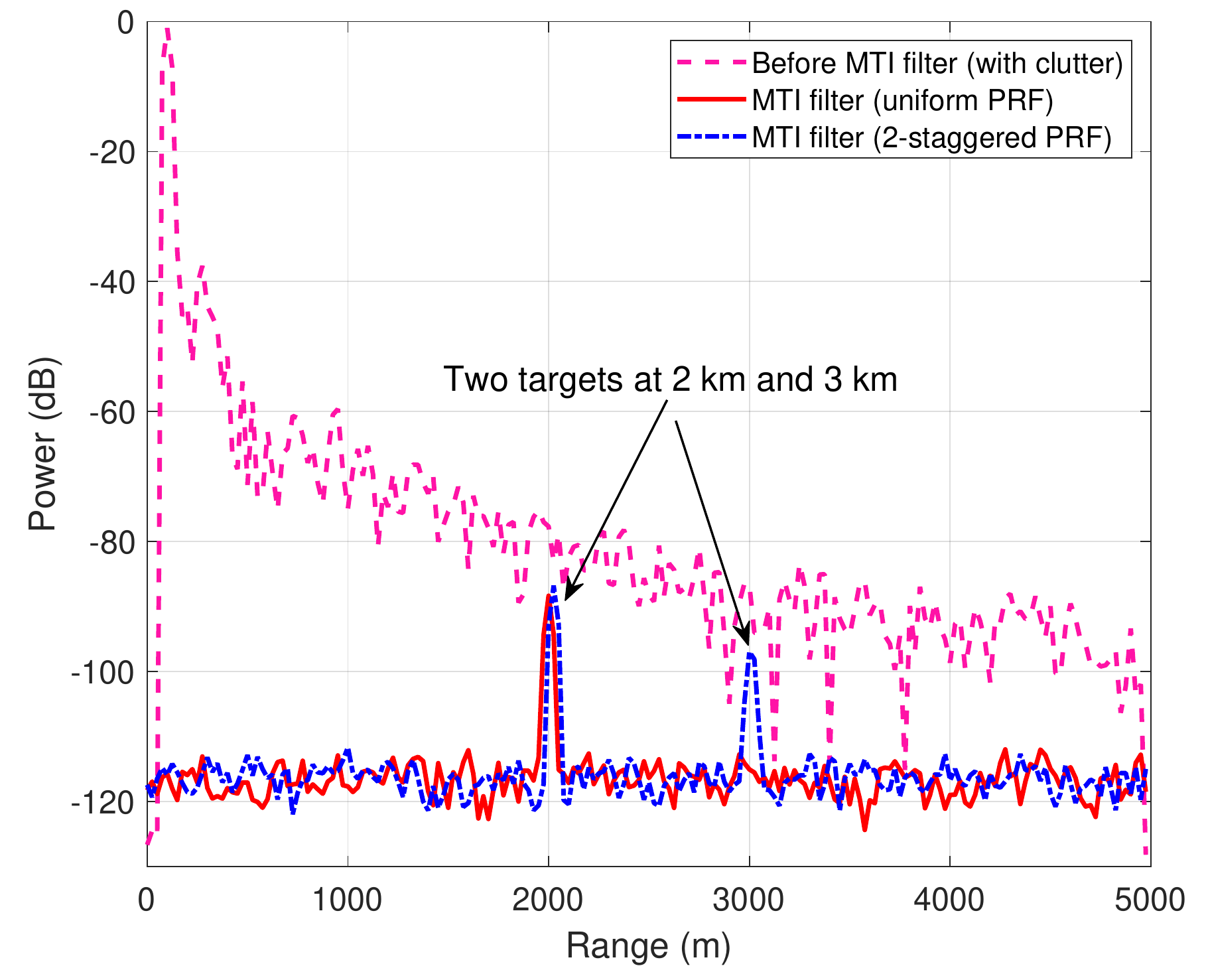}
	\caption{A scenario for a monostatic radar returns in the presence of two moving aerial vehicles and clutter~(regenerated from \cite{Matlab_MTI}). The first aerial vehicle present at $2$~km range from the radar is moving at a speed of $80$~m/s. The second aerial vehicle at $3$~km from the radar is moving at a \textit{blind speed} of $449.4$~m/s. MTI filtering with uniform PRF and 2-staggered PRF is used. The uniform PRF is able to remove the clutter and detect the first aerial vehicle but not able to detect the second aerial vehicle moving at a blind speed. On the other hand, MTI filtering using 2-staggered PRF is able to detect both aerial vehicles and remove clutter. }\label{Fig:MTI_filtering}
\end{figure}

An example of monostatic radar returns and MTI filtering~\cite{Matlab_MTI} is shown in Fig.~\ref{Fig:MTI_filtering}. The details of the setup are provided in~\cite{Matlab_MTI}. There are two aerial vehicles at $2$~km, and $3$~km range from the radar. The speed of the first aerial vehicle is $80$~m/s, whereas the speed of the second aerial vehicle is intentionally set to the \textit{blind speed} of $449.4$~m/s. The PRF is set at $30$~kHz. In Fig.~\ref{Fig:MTI_filtering}, the radar returns are shown with and without the MTI filtering. When no MTI filtering is applied, the clutter results in a higher noise floor, and the two aerial vehicles are not visible. Applying an MTI filter with uniform PRF~(three pulse canceller), the clutter is removed and the first aerial vehicle can be detected as shown in Fig.~\ref{Fig:MTI_filtering}. However, due to the repetition of nulls at blind speeds of the Doppler frequencies shown in Fig.~\ref{Fig:Uniform_staggered_PRF}, the second aerial vehicle cannot be detected. Using MTI filter and 2-staggered PRF, the nulls at blind frequency are significantly far off as shown in Fig.~\ref{Fig:Uniform_staggered_PRF}. This makes it possible to detect the second aerial vehicle using 2-staggered PRF as shown in Fig.~\ref{Fig:MTI_filtering}. 

Pulse-Doppler is also a popular approach for clutter rejection. A major advantage of pulse-Doppler radar compared to MTI radar is that the pulse-Doppler radar can provide velocity estimates of the aerial vehicle. Pulse-Doppler separates the aerial vehicle into different velocity regimes, provides good clutter rejection and velocity estimates, and uses long waveforms (usually $10$ to $1000$ pulses)~\cite{pulsedoppler}. Both MTI and pulse-Doppler can measure the range and velocity. However, the range is unambiguous for MTI, whereas, the velocity/Doppler measurement is ambiguous. For the puse-Doppler, the range is ambiguous, whereas, the velocity is unambiguous. Pulse-Doppler is widely used on airborne platforms. Pulse-Doppler can filter the strong ground returns on airborne platforms when the radar is looking towards the ground providing look-down capability. 

\subsection{Antenna Type and Polarization}
Directional antennas are mainly used by radar systems for the detection and tracking of aerial vehicles. The high gain from a directional antenna helps to achieve long-range~\big(see (\ref{Eq:radar_range})\big). In addition, the position of the narrow beamwidth directional beam can be used to roughly estimate the position of the aerial vehicle. The size of a radar antenna depends on the radar type, frequency, and platform. The size of search radar antennas is larger compared to track and guidance radar antennas due to the respective frequencies used. Moreover, airborne antennas are compact compared to ground and ship-mounted radar antennas.

\begin{figure}[!t]
	\centering
	\includegraphics[width=\columnwidth]{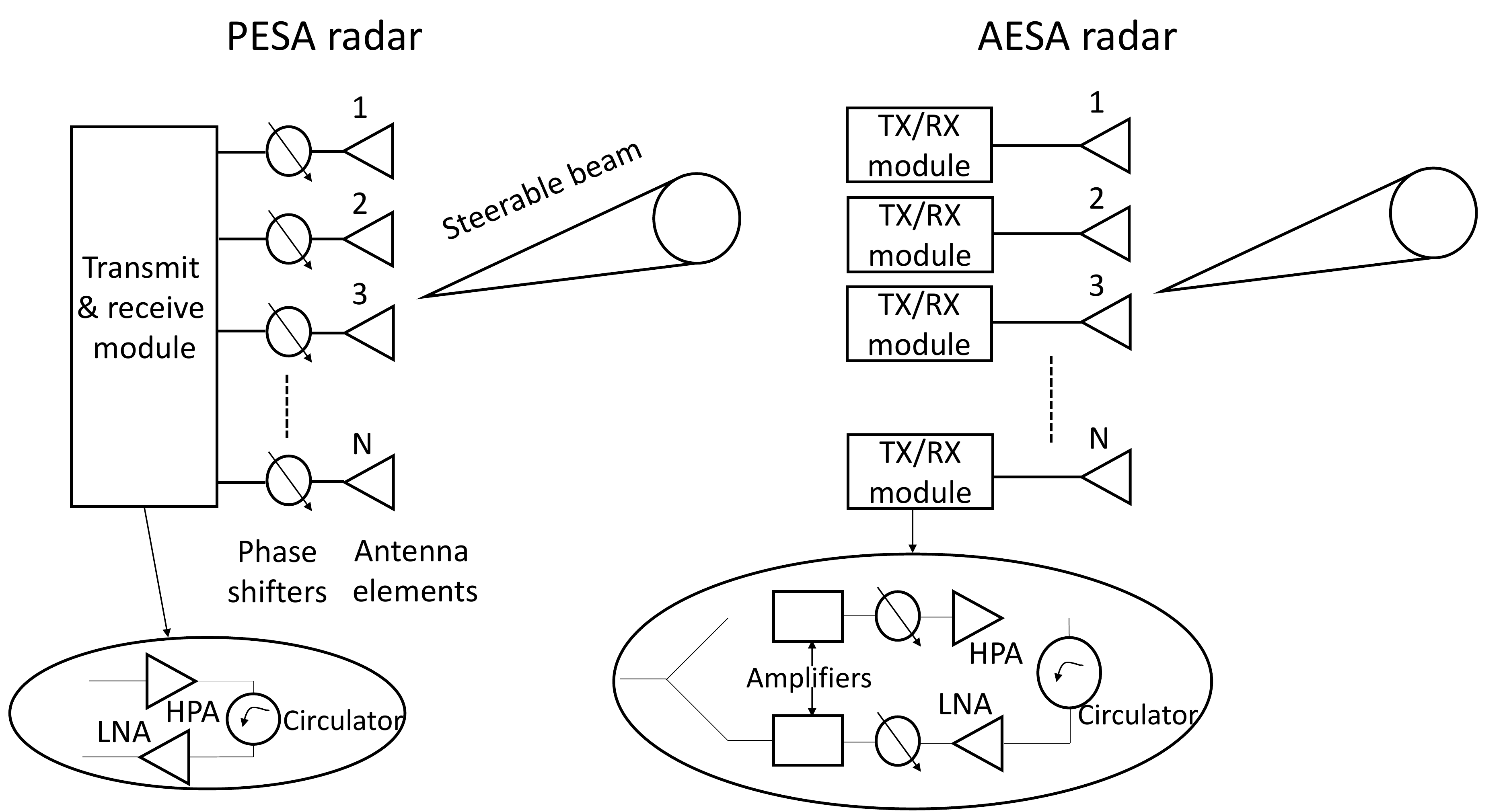}
	\caption{Passive and active electronically scanned array radar systems. }\label{Fig:AESA_PESA}
\end{figure}

Popular radar antennas were simple parabolic reflectors and Cassegrain feed parabolic reflectors that were mechanically rotated. Nowadays, electronically scanned phased arrays are used that require minimum mechanical assembly for rotation. The electronically scanned phased arrays are further divided into active and passive types~\cite{aesa_pesa} shown in Fig.~\ref{Fig:AESA_PESA}. The size and shape of the active electronically scanned array~(AESA) and passive electronically scanned array~(PESA) are similar, however, the number of TXs and RXs are different. Reconfigurable radar antennas can also be used for the detection and tracking of aerial vehicles~\cite{reconfig_ant2}. In \cite{reconfig_ant2} a multi-functional reconfigurable antenna is used for a cognitive radar. The re-configurable antenna is adaptively controlled to work within the framework of the cognitive radar in~\cite{reconfig_ant2}. 

Radars can use different types of polarization-dependent on the application. The RX of the radar is generally capable of receiving more than one component of polarization. Vertical or horizontal polarization is mainly used by radar systems. Circular polarization is suitable for aerial platforms. Adaptive antenna polarization can also be used dependent on the aerial vehicle and environmental scenario~\cite{reconfig_pol}. Horizontal-horizontal polarization is found to be better for UAV detection at very high frequency~(VHF)/ultra high frequency~(UHF) bands and low altitude angles in~\cite{HH_detect}.

\subsection{Frequency and Bandwidth}
The majority of the radar systems operate between $400$~MHz to $36$~GHz frequency range. The use of a particular frequency band depends on: 1) the role of the radar, i.e., search, tracking, or guidance; 2) the type of aerial vehicles expected to detect and track; 3) the terrain; 4) the platform over which radar is mounted; and 5) range of the radar. The long-range search radar uses low frequencies generally VHF and UHF bands. The low frequencies allow long-range detection capabilities. The close-range tracking, and guidance radar systems require higher resolution and narrow beams compared to search radars and therefore use higher frequencies compared to search radars. The aerial vehicle type can also influence the frequency selection. For example, for stealth aerial vehicles, low frequencies are effective mainly due to scattering from the surface of the aerial vehicle~\cite{stealth1}. Similarly, frequency sweeping across a large band increases the probability of detection of stealth and UAVs. Frequency-hopping is used by all modern radar systems. The major benefit of frequency hopping is resistance against jamming.

The terrain is an important factor that is taken into account for the frequency selection of a radar system. For example, in a mountainous area, low frequencies (wavelength comparable to the size of the mountains peaks) can avoid shadow zones due to diffraction and can detect/track terrain hugging aerial vehicles in the shadow zones. Other factors that influence the selection of the radar frequency are radio interferers in the area, clutter, and weather conditions. The platform over which radar is installed will also determine the wavelength (or size of radar antenna)~\cite{frequency_sel}. The popular frequency bands used for different radar applications are provided in Table~\ref{Table:Freq_target}. 

\begin{table*}[t]
	\begin{center}
     \footnotesize
		\caption{Different frequency bands for radars.} \label{Table:Freq_target}
\begin{tabular}{@{}|P{ 1.5cm}|P{2.2cm}|P{10.8cm}|@{}}
\hline
 \textbf{Radar band}& \textbf{Frequency (GHz)}& \textbf{Popular applications}\\
\hline
Millimeter&40-100&UAV detection, tracking, and navigation, airborne radar, spaceborne radar, SAR\\
\hline
Ka&26.5–40&Airborne close range targeting, airport surveillance, traffic speed detection, SAR\\
\hline
K&18–26.5&Small UAV detection, airborne close range targeting, traffic speed detection\\
\hline
Ku&12.5–18&High resolution mapping, satellite altimetry, air-traffic control, air-borne radar\\
\hline
X&8–12.5&Short range tracking,  guidance, UAV detection and tracking, marine radar, air-traffic control\\
\hline
C&4–8&Long range tracking, weather monitoring, SAR\\
\hline
S&2–4&Moderate range surveillance, air-traffic control, weather monitoring, surface ship radar \\
\hline
L&1–2&Long range surveillance, small UAV detection, atmospheric studies, air-traffic control, SAR\\
\hline
UHF&0.3–1&Very long range early warning against aerial threats, anti-stealth\\
\hline
VHF&0.03 to 0.3&Very long range early warning against aerial threats, anti-stealth\\
\hline
\end{tabular}
		\end{center}
			\end{table*}

The bandwidth of a pulse radar RX is larger than the reciprocal of the transmitted PW. The range resolution of a radar is dependent on the bandwidth or PW. Greater the bandwidth, the better the resolution. The bandwidth and the signal power~(or range) are exchangeable due to the dependence of both on the PW. UWB radars are popular due to their high resolution~\cite{UWB_highresol}. The multipath components~(MPCs) from a UWB radar can be resolved in the order of centimeters, therefore, fine details of the aerial vehicle or image can be obtained. The large bandwidth is also helpful to distinguish an aerial vehicle from the clutter. However, the range of UWB radars is limited. Another major limitation of large bandwidth is that the noise will increase with the bandwidth, resulting in a lowering of the SNR. In some radar systems, e.g., cognitive radars~\cite{cog_band}, the bandwidth can be tuned in real-time depending on the scenario. An UWB radar and multiple aerial vehicles tracking algorithms are used in \cite{UAV_UWBradar}. Range and position-based multiple aerial vehicle tracking are carried out using linear multi-target integrated probabilistic data association~(LM-IPDA), and multisensor~(LM-IPDA) in \cite{UAV_UWBradar}.    

\begin{figure}[!t] 
    \centering
	\begin{subfigure}{\columnwidth}
    \centering
	\includegraphics[width=\columnwidth]{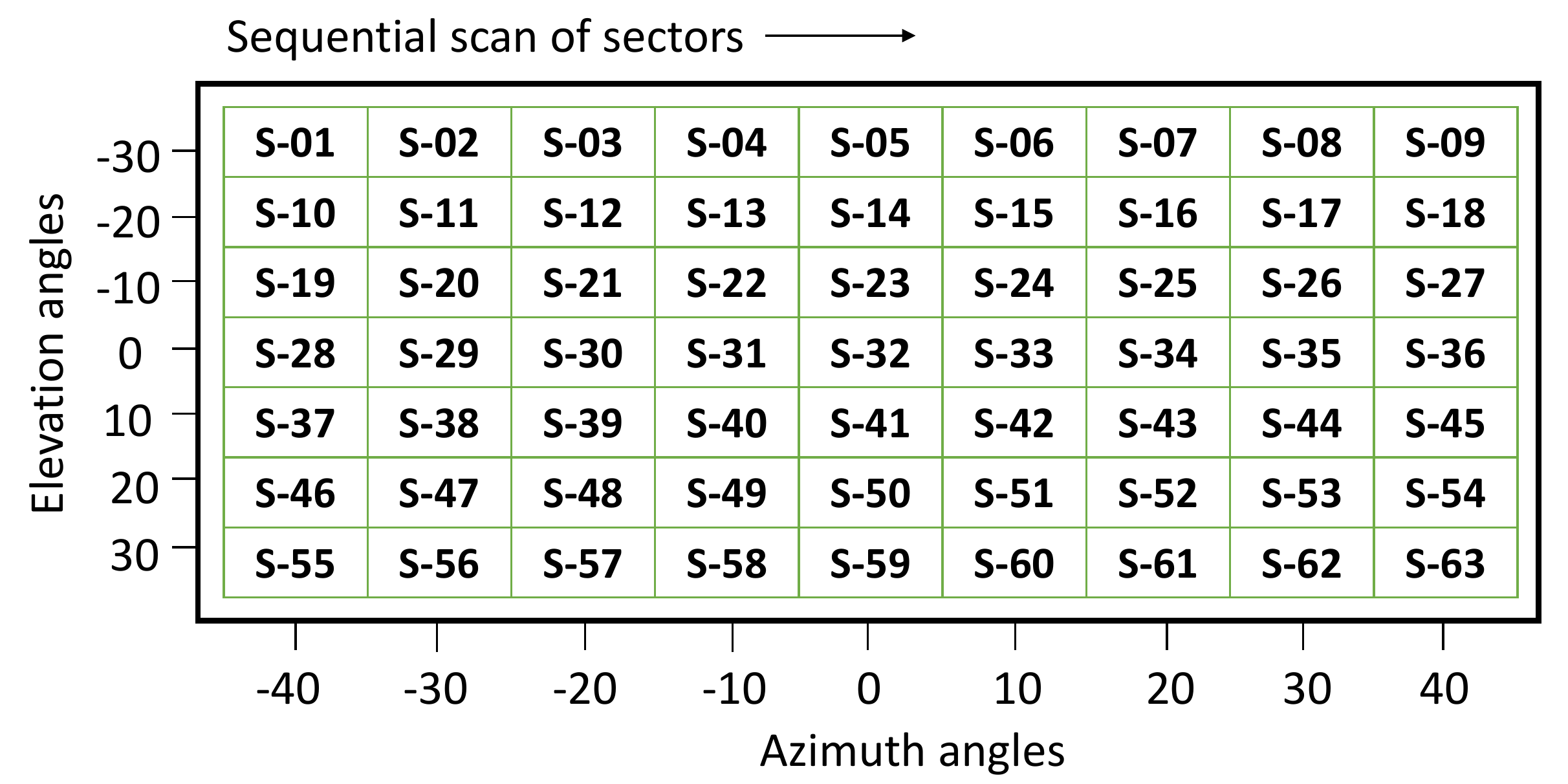}
	  \caption{}  
    \end{subfigure}    
    \begin{subfigure}{\columnwidth}
    \centering
	\includegraphics[width=\columnwidth]{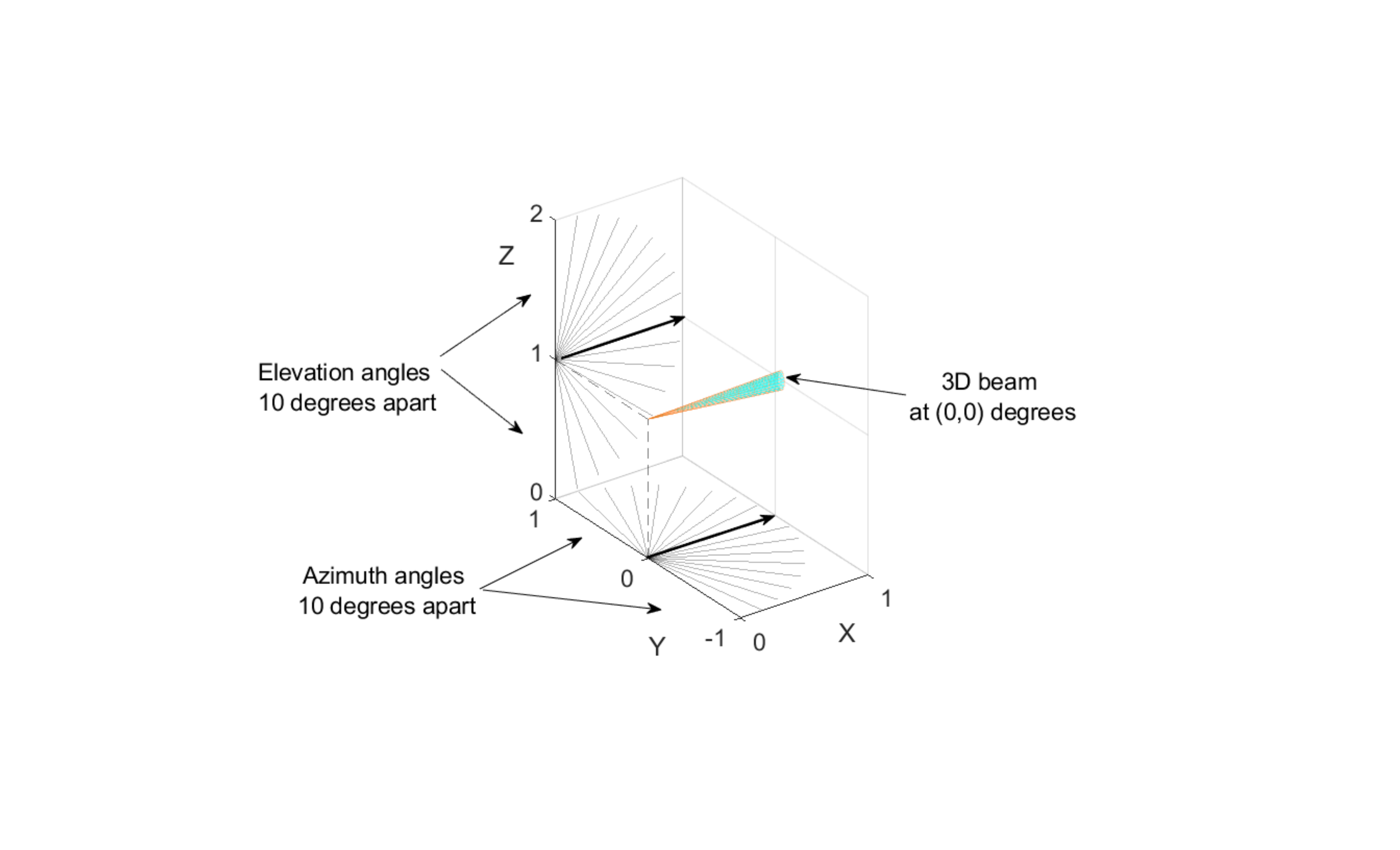}
	  \caption{}  
    \end{subfigure}
    \caption{(a) A subset of the scan angles in the elevation and azimuth planes. Each pair of azimuth and elevation angle is labeled as a sector. The sectors are generally scanned in a sequential manner, (b) beam scanning at different sectors~(angle pairs in the azimuth and elevation planes shown in Fig.~\ref{Fig:Scan_beams}(a)). The current location of the beam is at $0^{\circ}$ in the azimuth and elevation plane.} \label{Fig:Scan_beams}
\end{figure}

\subsection{Range and Angular Resolution} \label{Section:Range_Angular_resolution}
There are two main types of resolutions for the radars. The first is the range resolution and the second is the angular resolution. The range resolution of a radar system is its ability to distinguish between multiple aerial vehicles that are close or to distinguish between different parts of a single aerial vehicle. The range resolution and PW of radar are related by: $S_{\rm r} \geq \frac{c\rm{\gamma}}{2}$, where $S_{\rm r}$ is the range resolution, and $c$ is the speed of light. The smaller the PW, the greater the range resolution, and vice versa. However, the range resolution of a radar system can be increased using pulse compression~\cite{pulsecompressionradar} without changing the PW. The SNR using pulse compression is given as 
\begin{equation}
    S/N = \frac{P_{\rm t}G^2\lambda^2\sigma(\gamma = n\gamma_{\rm c})}{(4\pi)^3R^4kT_{\rm e}FL}, \label{Eq:pulse_compression}
\end{equation}
where $\gamma_{\rm c}$ is the PW of the compressed subpulse, $n$ is the number of the subpulses, $T_{\rm e}$ is the effective noise temperature, and $F$ is the noise figure. From (\ref{Eq:pulse_compression}), due to pulse compression, the bandwidth from compressed subpulses increases by a factor $n = \frac{\gamma}{\gamma_{\rm c}}$ without affecting the PW~(or range) $\gamma$.

The angular resolution of a radar system depends on the antenna beamwidth generally defined using the half-power~(-$3$~dB) beamwidth~\cite{radar_beamwidth}. It is always good to have narrow beamwidths so that TX energy is concentrated in a given direction and not wasted around the aerial vehicle. Narrow beamwidths can provide high angular resolution. The narrow horizontal beamwidths help in the high resolution of the bearing of the aerial vehicle. On the other hand, the wide vertical beamwidths allow compensation of pitch and roll angles of own aircraft/ship and help in detection of low flying aerial vehicles close to the terrain, e.g., UAVs.  The angular resolution between any two aerial vehicles is given as $S_{\rm a} \leq 2R\sin\frac{\theta}{2}$, where $S_{\rm a}$ is the angular resolution, $R$ is the slant range, and $\theta$ is the antenna beamwidth.

Fig.~\ref{Fig:Scan_beams}(a) and Fig.~\ref{Fig:Scan_beams}(b) show the angular sectors in the 3D plane where the beam is steered sequentially in a given duration of time~\cite{matlab_scan_angles}. If the angular resolution is high, there are a large number of scanning sectors. An obvious limitation of the high angular resolution and a large number of scanning sectors from Fig.~\ref{Fig:Scan_beams}(a) and Fig.~\ref{Fig:Scan_beams}(b) is the overall long scan duration. The location of the beam from Fig.~\ref{Fig:Scan_beams}(b) can be used to roughly estimate the position of the aerial vehicle in 3D coordinates. Different algorithms are used to track and forecast the trajectory of a detected aerial vehicle corresponding to sector positions.

\section{Detection and Ranging Using Radar Systems} \label{Section:detection_ranging}
In this section, the detection and ranging of aerial vehicles using radar systems is provided. 

\subsection{Detection Using Radar Systems}
Radars are mainly used for the detection of moving aerial vehicles. The echoes from a moving aerial vehicle are motion filtered to remove the clutter. The amplitude, delay (between the transmitted and received pulse), and phase shift (between the transmitted and received pulses) are used to estimate the range, velocity, and type of the aerial vehicle. Fig.~\ref{Fig:radar_basic1} earlier in this survey shows the basic operation of a radar to measure the elevation angle~$\theta$ and the azimuth angle~$\phi$ of an aerial vehicle. 

The detection of an aerial vehicle can be carried out by a radar system in active and passive modes. A radar system in an active mode provides accurate detection of the aerial vehicle due to active illumination of the aerial vehicle compared to passive radar. However, the transmissions from the active radar can be used to locate the position of the radar. The operation of the passive radar is similar to bi-static/multi-static radars, however, the source of illumination of the aerial vehicle is non-cooperative, e.g., TV broadcast signals, FM signals, and cellular phone signals or signals from other radars in the vicinity. The echoes collected by the passive radar are dependent on the terrain and wireless channel conditions. A major advantage of a passive radar is that the detection of the radar unit can be avoided due to no active transmissions. In literature, passive radars are used to detect small UAVs. In \cite{UAV_passive_radar}, a passive Global System for Mobile Communication~(GSM) radar is used to collect weak reflections from a UAV. A track-before-detection approach in \cite{UAV_passive_radar} helps in better detection and tracking of small UAVs in a passive mode. Experiments are conducted in \cite{UAV_passive_radar} to detect and track a quadcopter using passive GSM radar achieving a high detection rate. To combine the advantages of both active and passive radars, a hybrid radar system can also be used~\cite{hybrid_active_passive}. The operation of both active and passive radars for the detection of UAVs should follow the joint advisory issued by the Federal Aviation Administration, Federal Communications
Commission (FCC), Department of Justice (DOJ), and Department of Homeland Security.~\cite{advisory_CUAS}. 


The detection range of radar can be extended by: 1) increasing the transmit power; 2) using long-duration pulses; 3) using low-frequency signals; and 4) increasing the gain of the radar antennas. The detection ranges are also dependent on the geometry and the material of the aerial vehicle. The geometry and material of aerial vehicles can be modified using stealth techniques to reduce the detection ranges~\cite{stealth_tech}. Generally, detection ranges are provided by radar systems for different types of aerial vehicles. For example in \cite{radar_type}, an improved \& enhanced multi-mission hemispheric radar~(IEMHR) is able to detect different types of aerial vehicles at different detection ranges.

\subsection{Search Radars for Aerial Vehicle Detection}
There are three main types of radar systems for aerial vehicles detection and subsequent tracking. The three main types are air search radar, tracking radar, and guidance radar. The air search (or simply, search) radars are mainly used for early warning and can be placed on the ground, airborne platforms, and ships~\cite{early_warning1,early_warning2}. The main purpose of search radars is to detect and provide range and bearing information of aerial vehicles at long ranges by scanning in the $360^\circ$ azimuth plane around the radar. The search radars can detect aerial vehicles ranging from large planes to small UAVs. Aerial vehicles with large RCS can be detected by the majority of the search radars using conventional settings. However, small aerial vehicles can be detected with search radars, e.g., using wide PWs and high transmit power. 

Search radars sweep across a band of low frequencies. The low frequencies and high transmit power used by the search radars can cover a large area extending to hundreds of miles. The PRF is also low to allow long-range aerial vehicle detection. The placement of search radars on the aerial platforms can further increase the detection range and allow the use of high frequencies and offers better resolution compared to ground-based search radars. There can be 2D and 3D search radars. The 2D search radars only provide the range and bearing information, whereas 3D radars can also provide the height information of the aerial vehicle. The 2D search radar generally uses a single lobe scanning in the 360$^\circ$. Examples of long-range 2D radars include AN/SPS-49 that operates on L-band~\cite{SPS49} and the detection range is around $475$~km. The beamwidth is narrow at $3.3^\circ$ and helps against jamming. 

The 3D long-range primary search radars include AN/SPS-48~\cite{SPS48}, RAT-31DL~\cite{31DL}, and SMART-L~(uses an AESA and has a range of 2000~km)~\cite{smartL}. The pencil beams and high data rates allow effective processing for clutter. The reduction in the peak power provides defense against anti-radiation and other ECM. Over the horizon radar~(OTHR) is used as an early warning search radar for detection of aerial vehicles beyond the horizon. Over the horizon long-range is achieved by refraction of EM waves~(in the high frequency band) from the ionosphere~\cite{OTH}. OTHR are categorized into skywave and groundwave systems. Compared to airborne long-range early warning radar systems, the operational cost of the OTHR is significantly small. Moreover, advancements in the signal processing has allowed to overcome the range resolution issues for the OTHR.

\begin{figure}[!t] 
    \centering
			\includegraphics[width=0.99\columnwidth]{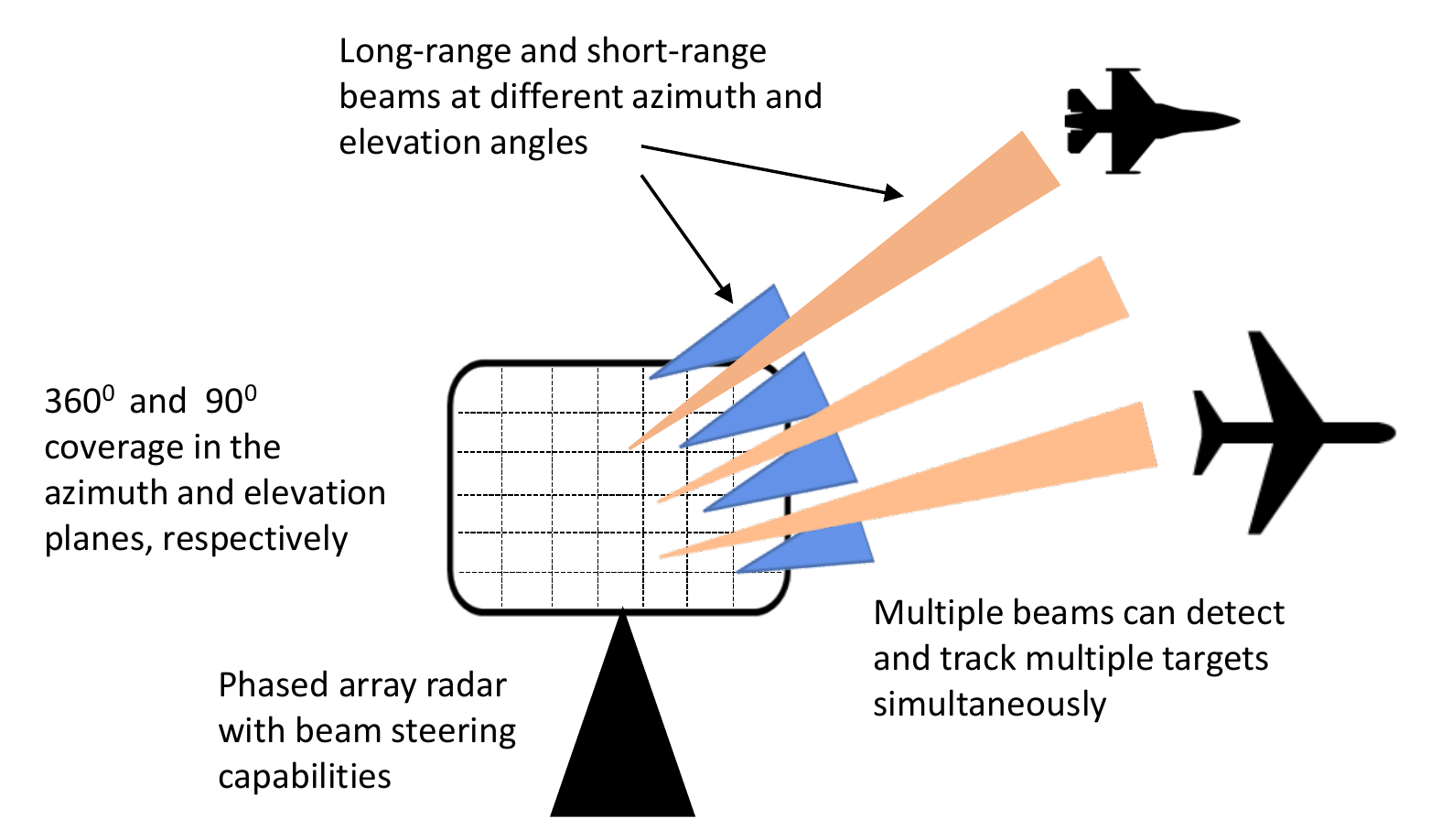}
	   \caption{3D beam steering using a phased array radar. The beams can be short or long-range and are electronically steered at different azimuth and elevation angles. The multiple radar beams help to detect and track multiple aerial targets simultaneously.} \label{Fig:radar_basic2}
\end{figure}

Beams of typical search radars are shown in Fig.~\ref{Fig:radar_basic2}, where 3D beams at different spatial positions are generated and steered in the $360^{\circ}$ azimuth plane and $90^\circ$ elevation plane. The steering can be performed using electronically scanned phased arrays aided by the mechanical rotation of the antenna assembly to cover the $360^\circ$ azimuth plane. An example of PESA radar system is AN/SPY-1~\cite{SPY-1}, whereas, AN/SPY-6 provides active electronic steering~\cite{SPY-6}. The 3D active and passive beam scanning can be used to detect and track multiple aerial vehicles simultaneously, which is possible mainly due to spatially separated narrow radar beams. In search radars, adjustment of sensitivity time control is important for the detection of small aerial vehicles, e.g., small UAVs.

\subsection{Detection Using Mechanically, and Electronically Scanned Platforms}
Mechanical scanning is conventional and has many disadvantages compared to electronic scanning, e.g., scanning delay and equipment size. Both AESA and PESA can be used for ECM, passive scanning, beamforming, etc., using the narrowband or wideband signals. In PESA the phase-shifting elements only work together to create a beam of different shapes and steer, whereas, in AESA these phase-shifting elements can be TX or RX themselves as shown in Fig.~\ref{Fig:AESA_PESA}. Using the different frequencies at each dwell, a low probability of intercept~(LPI) can be achieved easily with AESA compared to PESA mainly due to the use of modern solid-state transmit/receive modules~\cite{LPI_AESA}. Moreover, with AESA, we can form multiple beams at different frequencies that can be helpful for the detection of stealth aerial vehicles. However, if different beams have different frequencies, then the monopulse measurement for finding the precise location of the aerial vehicle cannot be achieved. A major drawback of the PESA compared to AESA is that PESA has a single-point failure. 

In \cite{UAV_array}, simulations are performed to detect a quadcopter UAV at a range of $5$~km using uniform rectangular phased array radar and electronic steering. Pulse Doppler radar principle and electronic and mechanical scanned radar are used in \cite{UAV_scan_elect_mech}. The mechanical scanning strategy  is adopted in the azimuth plane, while the electronic scanning strategy is used in the elevation plane for detection and tracking of UAVs. In \cite{UAV_phasedarray}, a phased array X-band radar based on AD$9361$ is used. The X-band small phased array radar is capable of detecting small UAVs of RCS $0.01$ at azimuth range and height of $5$~km and $250$~m, respectively, in urban environments. In \cite{UAV_fixed_radar}, a fixed wideband choke horn antenna is used for the detection of small UAVs. The sector beams are synthesized in a choke horn antenna that helps to obtain a wide beamwidth, which eliminates the need for mechanical rotation.

\subsection{Ranging Using Radar Systems}
The ranging of an aerial vehicle is generally performed using the basic radio wave echo principle. Similar to (\ref{Eq:FSPL_radar}), the range equation for the monostatic radar can be written as
\begin{equation}
E_{\rm r} = P_{\rm T}\frac{4\pi A_{\rm t}}{\lambda^2}\frac{1}{4\pi R^2}\frac{1}{L}\sigma\frac{1}{4\pi R^2}A_{\rm r}T_{\rm d},
\end{equation}
where $E_{\rm r}$ is the received energy, $A_{\rm t}$ is the TX antenna aperture, $A_{\rm r}$ is the RX antenna aperture, and $T_{\rm d}$ is the dwell time. The ranging of an aerial vehicle takes place once detected. The echoes from the aerial vehicle are used to calculate the range of the aerial vehicle from the radar. The frequently used maximum range equation for radar from (\ref{Eq:FSPL_radar}) is given as
\begin{equation}
    R_{\rm max} = \sqrt[4]{\frac{P_{\rm T}G^2\lambda^2\sigma}{P_{\rm R, min}(4\pi)^3L}}, \label{Eq:radar_range}
\end{equation}
where $R_{\rm max}$ is the maximum range for the radar, corresponding to minimum received power $P_{\rm R, min}$ and other radar parameters. It can be observed that all the parameters on the right-hand side of equation (\ref{Eq:radar_range}) are controlled by the radar operator except the RCS. The radar range ambiguity equation similar to (\ref{Eq:doppler_amb}) is given as: 
\begin{equation}
    R = \frac{c\big( t \pm x t_{\rm r}\big)}{2},~~~x=0,1,2,\hdots,
\end{equation}
where $t$ is the delay corresponding to the range and $t_{\rm r}$ is the PRI. The range and Doppler ambiguities in the measurements based on PRF is given in Table~\ref{Table:Ambiguities}.

The range equation for AESA radar is modified compared to the conventional radar equation~\cite{AESA_rangeeq}. In particular, the contribution of individual TX/RX modules is considered in the AESA range equation given as
\begin{equation}
    R = \sqrt[4]{\frac{N^3p\pi^2\lambda^2\sigma T_{\rm d}}{(4\pi)^3kT_{\rm s}D_{\rm x}(n')L_{\rm t}L_{\rm a}}}, \label{Eq:AESA}
\end{equation}
where $N$ is the number of TX and RX modules, $p$ is the mean power of each TX/RX module, $G_{\rm t} = G_{\rm r} = \pi N$ for the broadside direction and $N$ TX/RX modules, $T_{\rm d}$ is the dwell time, $k$ is the Boltzmann constant, $T_{\rm s}$ is the total system temperature, $D_{\rm x}(n')$ is the effective detectability factor~(see \cite{AESA_rangeeq}), instead of SNR, $L_{\rm t}$ is the transmission line loss, and $L_{\rm a}$ is the atmospheric loss.  

\section{Tracking and Classification of Aerial Threats Using Radar Systems}\label{Section:radar_track_clas}
In this section tracking and classification of different types of aerial vehicles using radar systems is provided.  

\subsection{Tracking of Aerial Vehicles Using Radar Systems}
Using radar we can continuously obtain the position information and velocity of an aerial vehicle in space that can help in the aerial vehicle's tracking. The search and track radar equations are given, respectively, as
\begin{equation}
    S/N = \frac{P_{\rm avg}A t_{\rm s}\sigma}{4\pi \Omega R^4 k T_{\rm s}L}, \label{Eq:search_eq}
\end{equation}
\begin{equation}
    S/N = \frac{P_{\rm t}G^2\lambda^2\sigma}{(4\pi)^3 R^4 k T_{\rm s}B_{\rm n}L}, \label{Eq:track_eq}
\end{equation}
where $S/N$ is the SNR, $P_{\rm avg}$ is the average power, $A$ is the antenna aperture, $t_{\rm s}$ is the scan time for the solid angle of search $\Omega$ in (\ref{Eq:search_eq}). In (\ref{Eq:track_eq}), $B_{\rm n}$ is the noise bandwidth of the RX, and the gain of the TX and RX antennas are same. Typically, $S/N\geq 10$ is required for both search and track operations. 

There are different techniques available in the literature for the localization of an aerial vehicle~\cite{local1,local2,local3}. Kalman and Particle filters are popular for the localization of aerial vehicles~\cite{localization_kalman_particle}. Imaging techniques are also used for the localization of aerial vehicles~\cite{localization_image1,localization_image2}. The localization of aerial vehicles for non-LOS~(NLOS) scenarios~\cite{local_NLOS1} and multiple moving aerial vehicles~\cite{multi_target1,multi_target2} is often challenging. Therefore, NLOS scenarios for an aerial vehicle are avoided by using multiple radars at different locations to ensure that at least one radar has a LOS path to the aerial vehicle. Moreover, advanced algorithms are used to localize an aerial vehicle in NLOS scenarios~\cite{local_NLOS2}.

Aerial vehicle tracking parameters observed by radar are position~(range, azimuth, and elevation angles of the aerial vehicle from the radar reference), geometry, and speed. If there are multiple aerial vehicles, the parameters of each individual aerial vehicle are tracked. To perform tracking either special resources are allocated towards the aerial vehicle using manual or computer-controlled sensors, e.g., directing antenna beam towards the estimated aerial vehicle's trajectory for tracking or a dedicated tracking radar is used~\cite{tracking_radar}. A dedicated tracking radar provides continuous positioning information of an aerial vehicle typically using a narrow circular beam. 
 
There are multiple techniques available in the literature for the tracking of aerial vehicles~\cite{tracking1,tracking2}. Machine learning and AI algorithms are also used to forecast the trajectory of the aerial vehicle~\cite{tracking_AI} that can assist in tracking. However, tracking of an aerial vehicle becomes complicated if there are multiple aerial vehicles~\cite{trakcing_multiple}. Tracking of a UAV using a radar and Bernoulli filter is performed in \cite{UAV_Bernoulli}. The detection of trajectories of highly maneuverable UAVs is provided in \cite{UAV_trajectory_new}. A filtration algorithm is applied for tracking UAV trajectories. The proposed approach in \cite{UAV_trajectory_new} provides better tracking of highly maneuverable UAVs in noisy environments compared to other tracking techniques for aerial vehicles. The different algorithms available in the literature for tracking aerial vehicles by radar systems are provided in Table~\ref{Table:Tracking_algos}. In Table~\ref{Table:Tracking_algos}, type of aerial vehicle, radar type, tracking features, and tracking algorithm from different literature references are covered.  

\begin{table*}[htbp]
	\begin{center}
     \footnotesize
		\caption{Tracking of aerial vehicles by radar systems using different tracking algorithms.} \label{Table:Tracking_algos}
\begin{tabular}{@{}|P{ 2.6cm}|P{ 2.8cm}|P{3.5cm}|P{4.5cm}|P{0.8cm}|@{}}
 \hline
\textbf{Aerial vehicle type}&\textbf{Radar/sensor type} &\textbf{Tracking features}&\textbf{Tracking algorithm}&\textbf{Ref.}\\
\hline
Motoar Sky MS-670&FMCW Interferometric radar&Radial velocity&Kalman filter&\cite{track_new1}\\
\hline
Micro-UAV&Multistatic radar NetRAD&Micro-Doppler, 2D aerial vehicle state, bi-static range&Extended Kalman filter&\cite{track_new2}\\
\hline
Multi-rotor UAV&Networked radar systems&Detections from multiple radars&Recursive random sample consensus algorithm, tuned Kalman filter&\cite{track_new3}\\
\hline
Multi-rotor UAVs&A dynamic radar network onboard UAVs&State space model&Local Bayesian estimator, extended Kalman filter&\cite{track_new4}\\
\hline
Multi-rotor UAVs&Camera&Camera images of an aerial vehicle and environment&CSRT, MIL, MOSSE, and KCF tracker algorithms discussed in \cite{track_new5}&\cite{track_new5}\\
\hline
DJI Phantom 3, Parrot Bebop 2, Parrot Disco, DJI MAVIC Pro.&Velodyne VLP-16 lidar&Speed, overall motion, laser scanned data&Kalman filter&\cite{track_new6}\\
\hline
UAVs&Mobile radar&3D position coordinates and velocity of the UAV&Kalman filter variants with east-north-up improvements&\cite{track_new7}\\
\hline
F450 and phantom 3&FMCW radar, acoustic and optical sensors&Data from three sensors&Kalman filter, single source detection algorithm, multiple drone detection algorithm, MUSIC algorithm&\cite{track_new8}\\
\hline
Small quadri-motor UAVs&Acoustic antenna, and microphones&Acoustic signals from UAV&Beamforming and direction of arrival estimation using time-difference-of-arrival&\cite{track_new9}\\
\hline
Miniature rotor UAV&Ku-band radar&Micro-Doppler features&Hough transform&\cite{track_new8}\\
\hline
\end{tabular}
		\end{center}
			\end{table*}

The tracking of aerial vehicles can also be carried out using guidance radars~\cite{guidance_radar}. The guidance radars emit a very narrow and intense beam of EM energy for accurate localization of the aerial vehicle. The guidance radars are also called targeting radars and are used to obtain the spatial~(range, azimuth, and elevation angles) and motion information~(velocity) of the aerial vehicle. The guidance radar can either be placed on the ground, mounted on ships, and aircraft, or ride a projectile. There are three phases (modes) of operation of the guidance radar system. The first is the designation, followed by acquisition and tracking. During the track phase, once the radar follows/track all the motions of the aerial vehicle, the aerial vehicle is said to be locked on.

\subsection{Trajectory Estimation of Aerial Vehicles Using Radar Systems}
The trajectory estimation is an important segment of the overall aerial vehicle's tracking. Trajectory estimation depends on many factors, e.g., measurement setup, terrain, and estimation algorithm. Also, 2D and 3D maps can assist in estimating an aerial vehicle's trajectory. The azimuth angle of the beam, and ranging information of the aerial vehicle is used to plot the 2D map of the aerial vehicle's trajectory. In order to obtain the 3D map of the aerial vehicle's movement, a separate radar that measures the elevation of the aerial vehicle was used. Nowadays, the elevation angles of the steered beams~(in the elevation plane) are used to estimate the height of the aerial vehicle~\cite{localization_3d}. Multiple radar units can also be used to obtain the position information of the aerial vehicle using triangulation~\cite{localization1}.

Trajectory estimation is important in estimating the 1) origin coordinates of the aerial vehicle, 2) terminal point of impact, 3) guidance towards an aerial vehicle, 4) classification of the aerial vehicle based on the trajectory features, and 5) predicting the intent of the aerial vehicle~\cite{intent_trajectory}. In \cite{trajectory_wahab}, channel parameters of ground-to-air UAV propagation channel following a given trajectory are forecasted during the blockage. The MPCs are arranged in individual path bins based on the minimum Euclidean distance among the channel parameters of the MPCs and forecasted using the Autoregressive approach. In \cite{trajectory_UAV1}, radar data is used to estimate the UAV trajectory parameters. Bayesian filters are used for UAV trajectory estimation and tracking in \cite{trajectory_UAV2}. A dynamic Bayesian model is used for multi-target threat value and A-Star algorithm is used to capture the optimal trajectory of the aerial vehicle in \cite{trajectory_UAV3}.

\subsection{Classification of Aerial Vehicles Using Radar Systems}
There are different methods for the classification of an aerial vehicle~\cite{classification1,classification2}. The main idea is to compare the features of the aerial vehicle obtained from measurements with the database of stored features of possible aerial vehicles. Some of the important classification features are size, shape, velocity, maneuverability, and RF control link characteristics~(if any) of the aerial vehicle. In order to accurately classify an aerial vehicle, the following conditions should be met: 1) the scattered signal should be independent of the operating frequency, aspect angle, and polarization; 2) the noise, clutter, and interference from different sources should be minimum; 3) there should be sufficient features to classify an aerial vehicle; and 4) the duration of the classification process should also be small. Nearly all types of modern aerial vehicle classification methods use machine learning and AI algorithms. The machine learning and AI algorithms learn the given features of an aerial vehicle and use different prediction algorithms to classify a given aerial vehicle~\cite{classification_AI,classification_AI2}. A drawback of aerial vehicle classification using machine learning and AI algorithms is large processing delays.   

Radar automatic aerial vehicle recognition~(RATR) is a popular research area for modern radar systems~\cite{RATR}. RATR algorithms help to accurately and quickly classify aerial vehicles without human feed. The features of the aerial vehicle are extracted from the radar data and modeled. Moreover, a classifier and training database is required for an RATR algorithm to work. Convolutional neural network~(CNN) based RATR is known to outperform conventional classification methods. High range resolution profile~(HRRP) obtains detailed features of the aerial vehicle for RATR~\cite{RATR_HRRP1,RATR_HRRP2}. Compared to synthetic aperture radar~(SAR) and inverse SAR images, obtaining and storing aerial vehicle data using HRRP is easier. Using HRRP, structural characteristics and geometrical shape of an aerial vehicle can be obtained. However, significant training data is required by RATR algorithm to correlate with the extensive data from the HRRP.  

Distinguishing small RCS aerial vehicles, e.g., UAVs from birds is challenging. The similarity between birds and small UAVs is that both are low and slow flying, and have small RCS. Maritime radar systems with additional post-processing can be used for birds detection. In \cite{UAV_bird_UAV}, motion characteristics of the birds and UAVs collected by a surveillance radar are used for their classification using a random forest classifier. The radar system used in \cite{UAV_bird_UAV} is a bird surveillance radar system installed at airports and operates at S-band. The measurements for the classification of UAVs and birds using L-band staring radar are provided in \cite{UAV_Lband_staring} in an urban area.

In \cite{UAV_classification_new}, a staring radar is used for distinguishing UAVs from birds. Staring radar operating at an Land and symmetrical peaks are extracted from the micro-Doppler signatures for UAV classification. The symmetrical peak extraction algorithm can distinguish between UAVs and birds by focusing on the relationship between the components producing micro-Dopper and the main body of the aerial vehicle in \cite{UAV_classification_new}. In \cite{UAV_classification_new2}, AI enabled detection and classification of birds and UAVs is provided. The classification is performed using an interactive multiple models and recurrent neural network. Classification of small aerial vehicles using range-Doppler diagrams is provided in \cite{UAV_classification_RD}. The non-kinematic features e.g. micro-Doppler signatures are extracted from the radar signal for better tracking and situational awareness. 

In \cite{UAV_tracks_classify} the problem of assigning correct labels to different radar tracks is taken up. Classification of UAV and non-UAV tracks~(e.g., birds, or manned aircraft) is carried out in \cite{UAV_tracks_classify}. Simulations are carried out to compare the statistical features of the UAV track with bird and aircraft track. It was shown 
that UAV tracks can be correctly labeled with an accuracy of $99\%$ using a given set of track features. The paper also presents an counter-UAV system based on RF detection and CNN. CNN is also used for the classification of UAVs in \cite{UAV_CNN2}. Micro-Doppler radar signatures and CNN are used for the classification of UAVs from birds. In \cite{UAV_classification_new3}, six CNN architectures used for computer vision applications are compared for the classification of UAVs. The classification is obtained for low SNR detection. In \cite{class_new9}, a comparative analysis of UAV classification techniques based on the RCS is provided. The RCS of six different types of UAVs is measured at $15$~GHz and $25$~GHz in an anechoic chamber. Different classification algorithms from machine learning, statistical learning, and deep learning were used. For example, assuming equal a priori probabilities for each of the UAV type or class, the statistical rule for UAV classification given in \cite{class_new9} as
\begin{eqnarray}
\begin{aligned}
\widehat{C}  &=  \arg\max_{C=1,2,\cdots, M}{\ln P({C}=j|\boldsymbol{\sigma})},\\
   &=  \arg\max_{C=1,2,\cdots, M} {\ln P(\boldsymbol{\sigma}|C=j)},
   \label{decision_rule}
\end{aligned}
\end{eqnarray}
where $P(C = j|\sigma)$ is the posterior probability for a $j^{\rm th}$ class, and $P(\sigma|C=j)$ is the conditional class density, and there are $M$ number of different UAV classes .

\begin{figure}[!t]
	\centering
	\includegraphics[width=0.999\columnwidth]{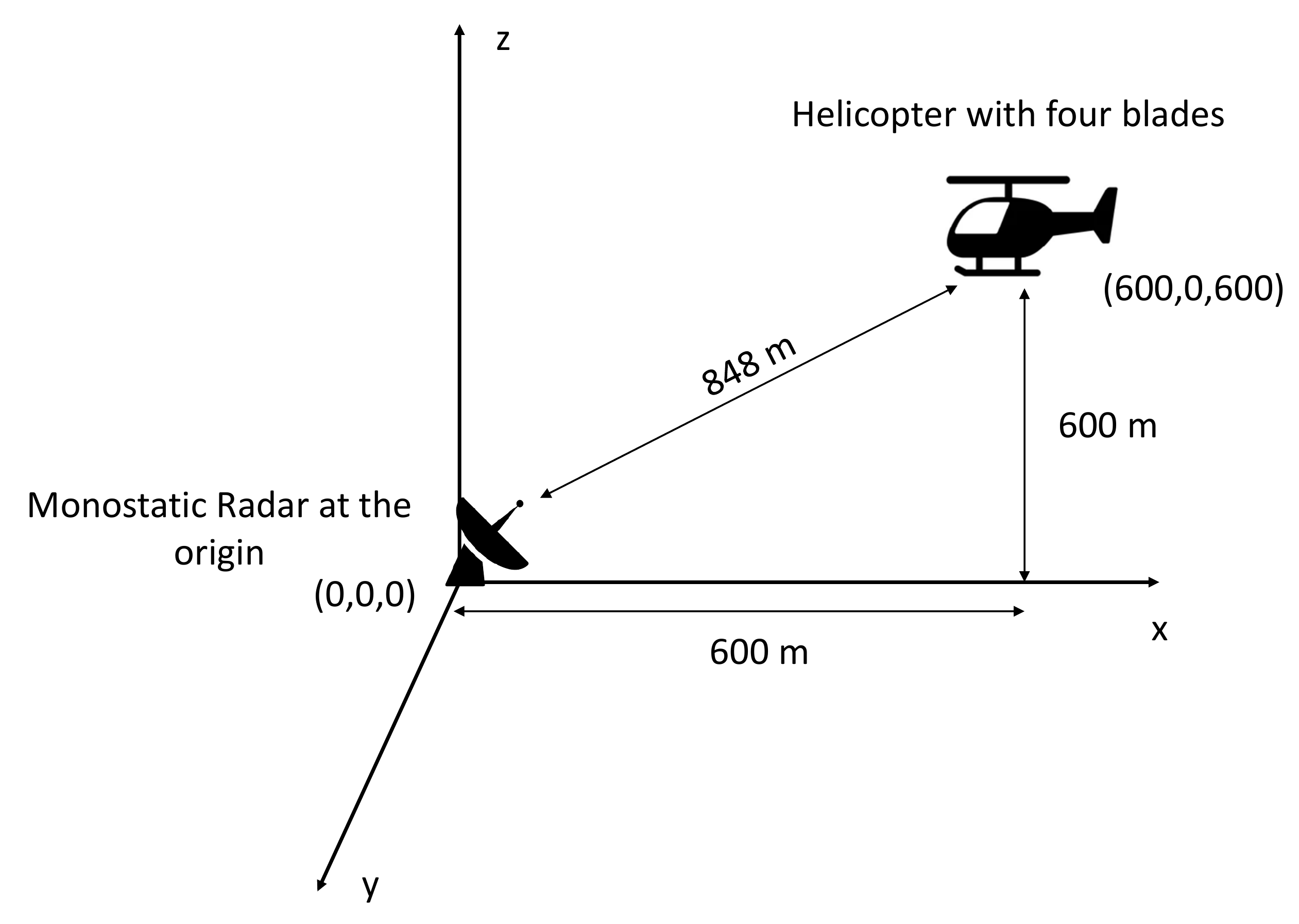}
	\caption{A simulation scenario to obtain the micro-Doppler using a monostatic radar and a four blade helicopter~\cite{Matlab_MicroDoppler}. The immovable radar is placed at the origin. The helicopter has four blades and the initial coordinates of the helicopter are $(600, 0, 600)$. The helicopter is moving at a velocity of $75$~m/s away from the radar, and the blades rotate at a constant speed of $4$ revolutions per second. }\label{Fig:microdoppler_scenario}
\end{figure}

\subsection{Classification Using Micro-Doppler Radar Signatures} 
\begin{figure}[!t]
	\centering
	\includegraphics[width=0.999\columnwidth]{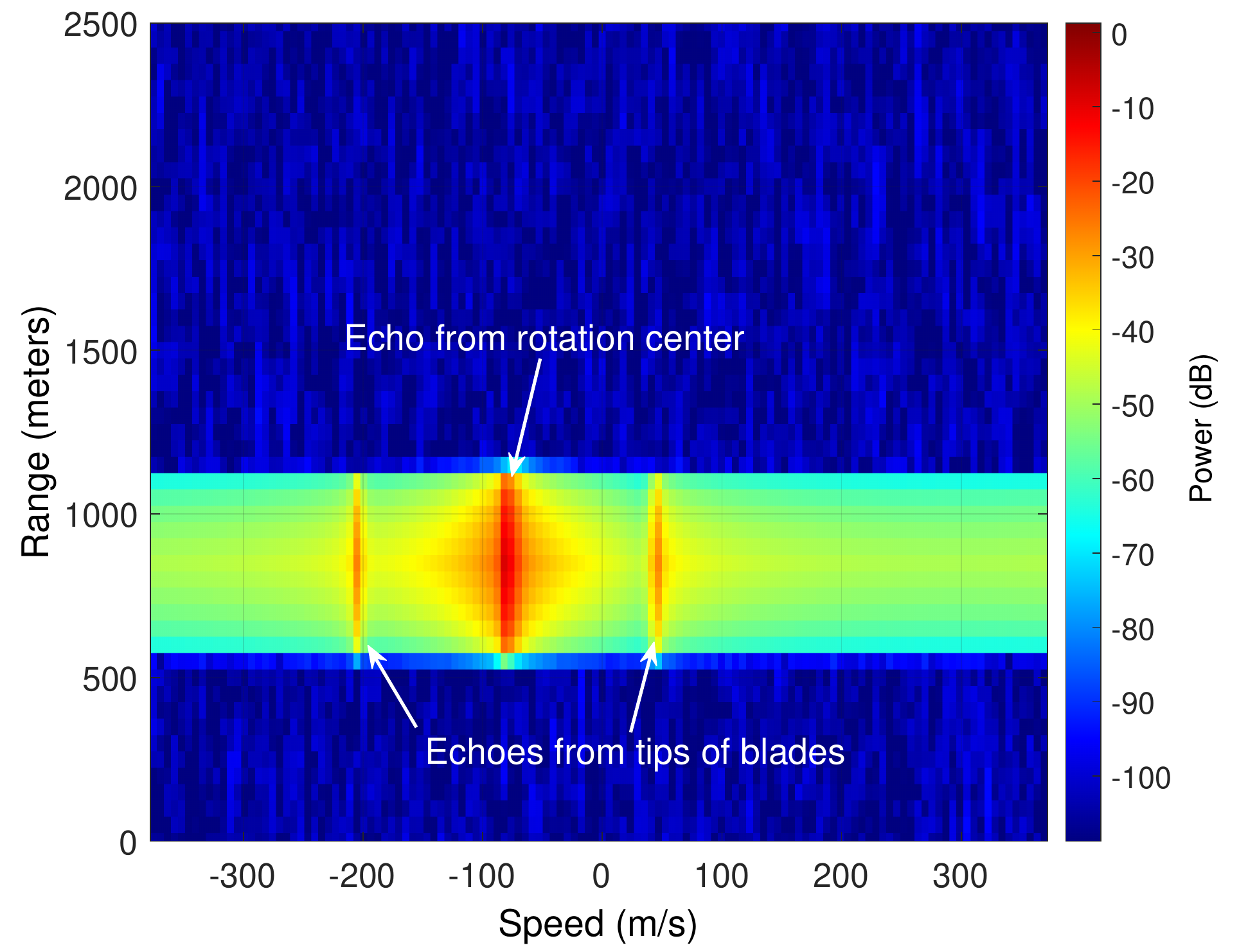}
	\caption{The range-velocity plot of the micro-Doppler scenario shown in Fig.~\ref{Fig:microdoppler_scenario}~(regenerated from \cite{Matlab_MicroDoppler}). The major illumination is centered around ($60$~m/s, $850$~m) due to the rotation center at the helicopter body. The other two illuminations are due to the radar returns from the blade tips. Therefore, there are different speed profiles from a single aerial vehicle due to different relative rotation speeds of the parts of the aerial vehicle.    }\label{Fig:speed_range_microdoppler}
\end{figure}

\begin{figure}[!t]
	\centering
	\includegraphics[width=0.999\columnwidth]{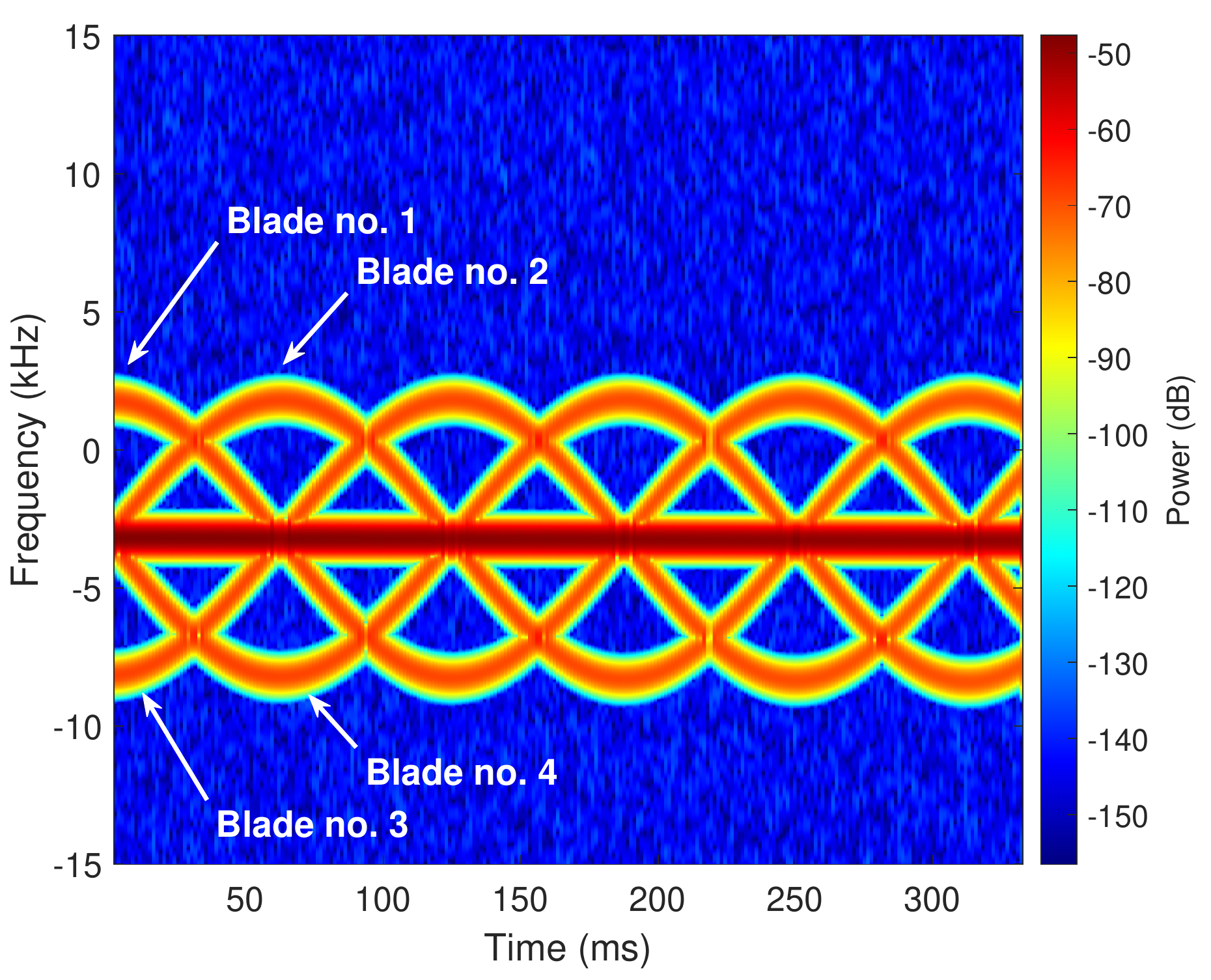}
	\caption{Illustration of micro-Doppler modulation due to blades and a constant Doppler shift~(regenerated from \cite{Matlab_MicroDoppler}). The modulation is sinusoid due to the rotation of the blades and during each sinusoid period, three other sinusoids are present at equal intervals indicating the radar returns from four blades that are equally spaced. }\label{Fig:microdoppler_pattern}
\end{figure}

There are micro-Doppler radar systems that can process the micro-Doppler returns from moving aerial vehicles and can help in the classification~\cite{micro_doppler}. For example, micro-Doppler radars can be used to differentiate between UAVs and birds. The micro-Doppler radars require high sampling rates and have a limited range. A micro-Doppler scenario is shown in Fig.~\ref{Fig:microdoppler_scenario}. In Fig.~\ref{Fig:microdoppler_scenario}, there is a single monostatic radar placed at origin and has zero velocity. A helicopter with four blades is at an initial coordinate location of ($600$~m,$0$,$600$~m) and the helicopter is moving at a velocity of $75$~m/s, away from the radar. The length of each blade is $7$~m and the blades rotate at a speed of $4$ revolutions per second. The center frequency is $6.0$~GHz, the sampling frequency is $3$~MHz, and the PRF is $30$~KHz. The simulations are carried out using Matlab~\cite{Matlab_MicroDoppler}. The simulation results of range-velocity for the setup shown in Fig.~\ref{Fig:microdoppler_scenario} are provided in Fig.~\ref{Fig:speed_range_microdoppler} and the micro-Doppler modulation results due to blades are shown in Fig.~\ref{Fig:microdoppler_pattern}. In both Fig.~\ref{Fig:speed_range_microdoppler} and Fig.~\ref{Fig:microdoppler_pattern}, the contribution of the Doppler shifts from different parts of the helicopter is shown. The contributions can be divided into two main categories. One contribution is due to the center rotation and main body of the helicopter, and the second contribution is due to the blades of the helicopter. In Fig.~\ref{Fig:microdoppler_pattern} a micro-Doppler modulation pattern due to the tips of the blades and a constant Doppler shift are shown. The micro-Doppler signatures in Fig.~\ref{Fig:speed_range_microdoppler} and Fig.~\ref{Fig:microdoppler_pattern} can be used to detect and classify the presence of an equally spaced four-blade rotor helicopter. 

In \cite{UAV_microDopp}, the micro-Doppler signature of the UAV is experimentally studied for characterization of payload and intent. In \cite{track_new10}, Hough Transform is used to aid in the detection and classification of micro-Doppler returns from multi-rotor UAVs. In \cite{track_new2}, hovering UAVs are detected in a cluttered background using range-Doppler and micro-Doppler signatures. The micro-Doppler signatures from the rotating propellers are used to distinguish the UAV from the clutter. A multistatic radar and micro-Doppler are used for detection and tracking of micro-UAVs in \cite{track_new2}. The detection of UAV is performed using micro-Doppler signatures. The micro-Doppler signatures were found to provide better results compared to the conventional Doppler shift procedure for clutter rejection. 

In \cite{UAV_unloaded_loaded}, the classification of loaded or unloaded UAVs is carried out using micro-Doppler radar signatures. The algorithm uses the spectral kurtosis technique and the principal component analysis. Machine and deep learning are used for the detection of UAVs using micro-Doppler signatures, and kinematic criteria in \cite{UAV_ML_new2}. Enhanced UAV detection and feature extraction are carried out using Doppler and micro-Doppler signatures in \cite{UAV_ML_new3}. A uniform planar array radar is used to track the small rotary-wing aerial vehicles using the micro-Doppler signatures before detection in \cite{UAV_trackbeforedetect}. The efficiency of the track before detect approach is also compared with conventional radar system approaches. Different techniques available in the literature for classification of aerial vehicles by radar systems, including sensors used, classification features, and classification algorithm used, are summarized in Table~\ref{Table:Classification_algos}. 

\begin{table*}[t]
	\begin{center}
     \footnotesize
		\caption{Classification of aerial vehicles by radar systems using different types of classification algorithms.} \label{Table:Classification_algos}
\begin{tabular}{@{}|P{ 3.3cm}|P{ 2.5cm}|P{3.5cm}|P{4.5cm}|P{0.7cm}|@{}}
 \hline
\textbf{Aerial vehicle type}&\textbf{Radar/sensor type}&\textbf{Classification features}&\textbf{Classification algorithm}&\textbf{Ref.}\\
\hline
Metafly, MavicAir2, Parrot Disco&FMCW radar&Micro-Doppler signatures from radar spectrogram dataset&Deep learning&\cite{class_new1}\\
\hline
See Table~7.1 in \cite{class_new2}&CW radar, automotive radar&Diameter of the rotor& Support vector machine, Naive Bayes&\cite{class_new2}\\
\hline
Mobile objects in air&Mobile autonomous radar stations&Image of the environment&Neural network algorithm&\cite{class_new4}\\
\hline
Multi-rotor UAVs&X-band CW radar&Features based on micro-Doppler signatures of the UAVs&Spectrographic pattern analysis of UAVs &\cite{class_new5}\\
\hline
Planes, quadrocopter, helicopters, stationary rotors, and birds&X-band CW radar&Features based on micro-Doppler signatures of aerial vehicles&Support vector machine, Naive Bayes &\cite{class_new6}\\
\hline
Fixed wing and multi-rotor UAVs&FMCW surveillance radar&Features based on micro-Doppler signatures&Total
error rate minimization based classification &\cite{class_new7}\\
\hline
Multi-rotor UAVs and birds&S-band BirdRad system&Polarimetric parameters provided in Table~I of \cite{class_new8}&Nearest neighbor classifier &\cite{class_new8}\\
\hline
DJI Matrice 600 Pro, DJI Matrice 100, Trimble zx5, DJI Mavic Pro 1, DJI Inspire 1 Pro, and DJI Phantom 4 Pro&Surveillance radar operating at 15~GHz and 25~GHz&RCS of UAVs&15 different classifiers  provided in Table~IV of ~\cite{class_new9}&\cite{class_new9}\\
\hline
Rotary UAVs&FMCW radar and acoustic sensor array&Spectral power bins&Neural network algorithms&\cite{class_new10}\\
\hline
14 UAV controllers provided in Table~I of \cite{class_new11}& High-frequency
oscilloscope&Energy transient signals and features associated with them&Naive Bayes, neighborhood component
analysis, k-nearest neighbor, discriminant analysis, state vector machine, and neural networks&\cite{class_new11}\\
\hline
\end{tabular}
		\end{center}
			\end{table*}

\begin{figure}[!t]
	\centering
	\includegraphics[width=0.999\columnwidth]{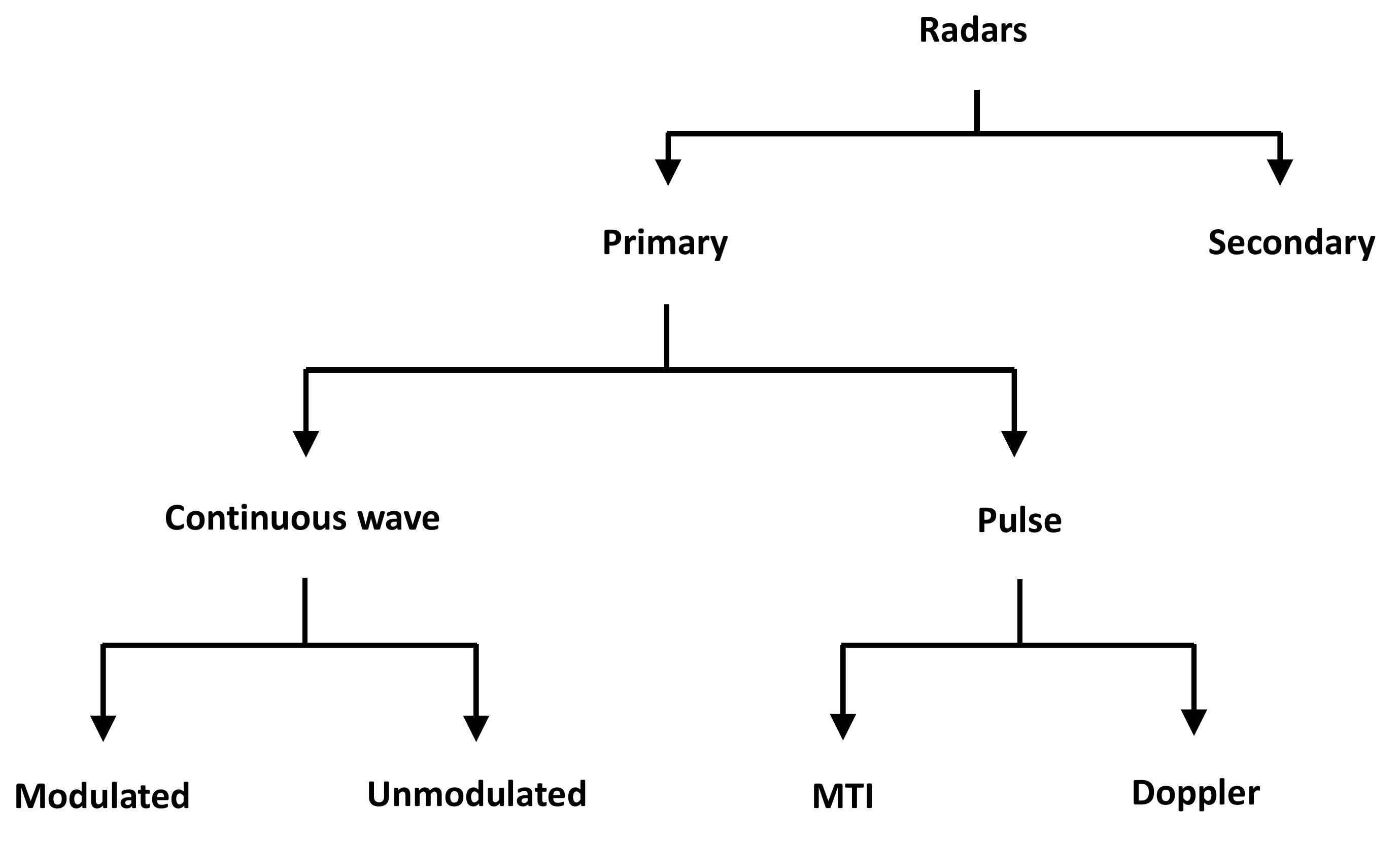}
	\caption{Different types of radars.}\label{Fig:Radar_types}
\end{figure}

\section{Types of Radar Systems} \label{Section:Radar_types}
In this section different types of radar systems for detection, tracking, and classification of aerial vehicles are reviewed. 

\subsection{Conventional Radar Systems}
There are a number of conventional radars that use conventional signal processing. Fig.~\ref{Fig:Radar_types} shows a high  level overview of the different types of conventional radars. Majority of radars for aerial vehicle detection are real aperture radars. A monostatic active real aperture radar in the search mode shown earlier in the survey at Fig.~\ref{Fig:radar_basic1}. The range resolution of a real aperture radar is dependent on the width of the transmitted pulse, whereas, the azimuth resolution is inversely proportional to the length of the antenna. The azimuth resolution can be increased by increasing the length of the radar antenna. Increasing the size of the antenna may not be possible beyond certain limits for airborne platforms. To be carried by such airborne platforms, e.g., satellites and planes, the size of the antenna needs to be compact. 

\begin{figure}[!t]
	\centering
	\includegraphics[width=\columnwidth]{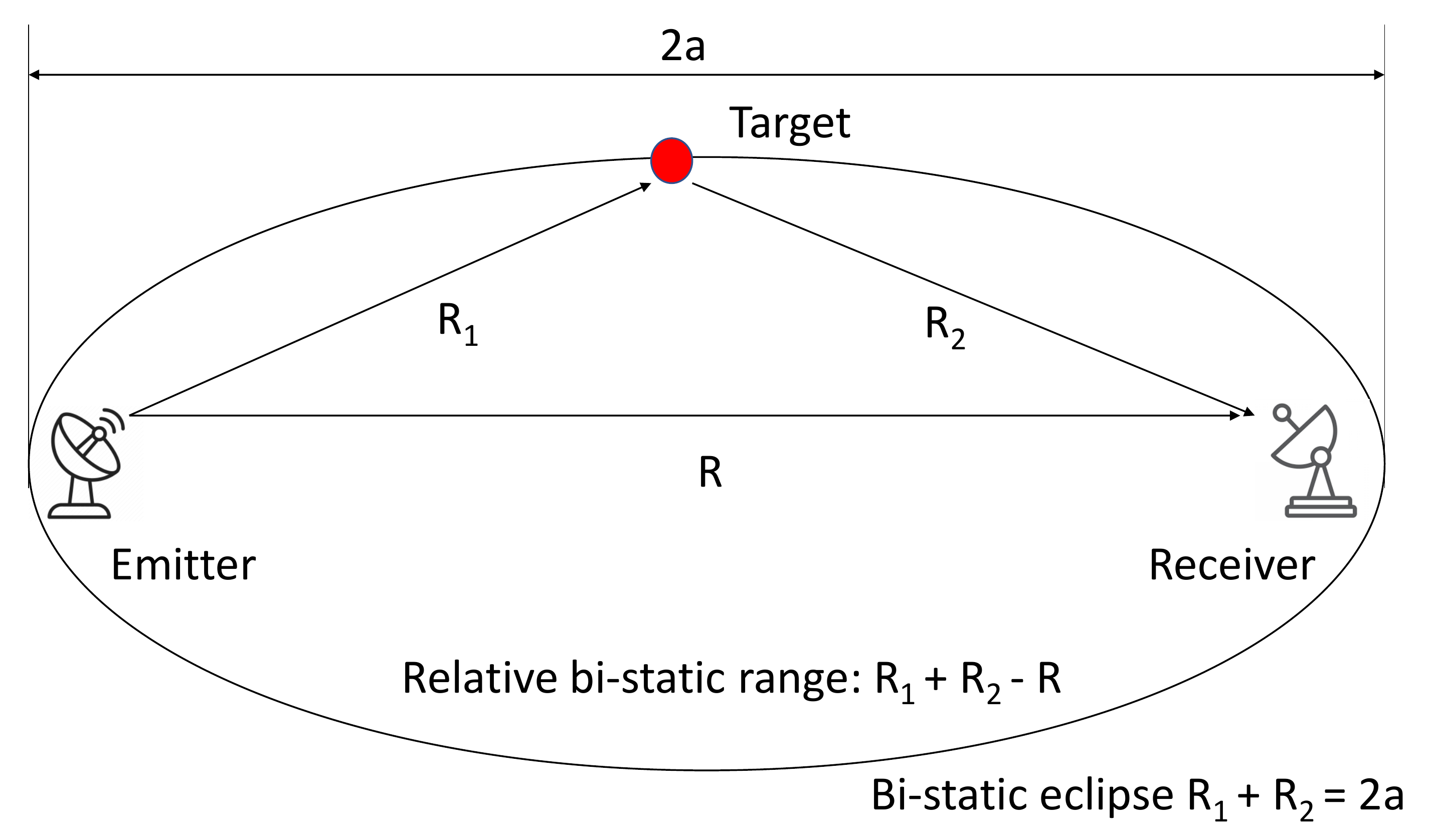}
	\caption{Geometry of a bi-static radar system. The aerial vehicle is illuminated by the emitter source forming the upper side of the triangle. }\label{Fig:bistatic_figure}
\end{figure}

The majority of the radar systems are monostatic. Monostatic radars are simpler and typically receive stronger and unambiguous signals compared to bi-static radars. The receivers of the bi-static and multi-static radars collect radar energy at different aspect angle(s) when compared to monostatic radars. In particular, the scattered energy from small UAVs can be collected by the spatially spaced receivers/antennas~\cite{dvb_uav}. Therefore, the bi-static and multi-static radars provide better situational awareness and early warning capabilities against small, and low flying aerial vehicles. The source of illumination for bi-static and multi-static radars can be either a dedicated source or a third-party source, e.g., TV/radio broadcasting. The illumination of a target aerial vehicle by a bi-static radar system is shown in Fig.~\ref{Fig:bistatic_figure}. The geometry formed due to illumination of an aerial target by a bi-static radar is also shown in Fig.~\ref{Fig:bistatic_figure}. A major limitation of bi-static and multi-static radars is the data transfer and synchronization between the TX and RX(s).

A major category of conventional radar systems consist of ground/sea-based radars. The ground/sea-based radar systems can be search, tracking, and guidance radars. Generally, search, tracking, and guidance radars are co-located. The ground/sea-based radar systems are used for aerial surveillance of aerial vehicles flying at a few meters from the ground to the upper atmosphere. The ground/sea-based radars are generally large and use large transmit power to cover long ranges. Different types of signals and processing (as discussed in Section~\ref{Section:Signal_processing}) are used by ground/sea-based radars for transmission and reception. The clutter for ground-based radars depends on the terrain. Major clutters for ground-based radar systems include mountains, trees, and man-made structures. The clutters for sea-based radar systems are mainly from the sea waves. The signal variation from clutter depends on the condition of the sea and the frequency used. Different types of popular radar systems used in the field are shown in Table~\ref{Table:Radar_types}. The radar systems in Table~\ref{Table:Radar_types} include mobile/immobile, airborne, search, track, and guidance radars. 

\begin{table*}[htbp]
	\begin{center}
     \footnotesize
		\caption{Different types of radars for search, track and guidance towards different types of aerial vehicles.}
\begin{tabular}{@{}|P{ 1.8cm}|P{2cm}|P{2cm}|P{2cm}|P{2cm}|P{1cm}|P{1.1cm}|P{0.9cm}|P{1cm}|@{}}
 \hline
 \textbf{Name}&\textbf{Type}&\textbf{Installation platform}&\textbf{Purpose}&\textbf{Possible aerial vehicles}&\textbf{Frequency}&\textbf{Coverage (Az., El.)}&\textbf{Range}&\textbf{Antenna}\\
             \hline
           AN/SPY6 \cite{radar1}& Search& Ground/shipboard& Early warning& Aerial vehicles& S-band& 360$^\circ$& Not specified&AESA\\
             \hline
AN/FPS-35 \cite{radar2}& Search& Ground (immobile)& Early warning& Aerial vehicles& 420-450 MHz& 360$^\circ$& 320 km&Parabolic reflector\\
            \hline
             AN/TPQ-53 \cite{radar4}& 3D search \& track& Ground (immobile)& Early warning& Aerial vehicles& 1215-1400~MHz& 360$^\circ$,90$^\circ$& 1000 km& AESA\\
             \hline
             AN/TPY-2 \cite{ANTPY2}& Surveillance, track, classify aerial vehicles& Ground (mobile)& Long range, very high altitude& Aerial projectiles& X-band& $\pm 60^\circ$,90$^\circ$& 4700 km& Digital antenna array\\
             \hline
             AN/APY-12 \cite{radar5}& Airborne surveillance& Airborne& Imaging& Ground, aerial vehicles& X-band& Ground imaging& Not specified& SAR\\
             \hline     
            AN/MPQ-53 \cite{MPQ53}& Control radar& Ground (mobile)& Surveillance, IFF, tracking/guidance, ECM& Aerial projectiles& G/H-bands& 360$^\circ$,65$^\circ$& 100 km& PESA\\
             \hline 
             AN/SPG-62 \cite{radar6}&Guidance& Ground/shipboard& Aerial vehicle illumination& Aerial vehicles& I/J-Bands& Not specified& 305 km& CW\\
             \hline
             91N6 \cite{radar91n6}& 3D surveillance \& track& Ground (mobile)& Early warning& Aerial vehicles& 2.9-3.3~GHz& 360$^\circ$,90$^\circ$& 600 km& Phased array antenna\\
             \hline
             TPY-4 \cite{radar4}& Search, track& Ground (mobile)& Close range& Rockets/artillery& S-band& 360$^\circ$,90$^\circ$& 60 km& AESA\\
             \hline
                 EL/M2084 \cite{ELM2084}& Guidance& Ground (mobile)& Long range& Projectiles interception& S-band& 360$^\circ$,80$^\circ$& 470 km& 3D AESA\\
             \hline
             IBIS 150 \cite{IBIS150}& Tracking and guidance& Ground (mobile)& Tracking aerial vehicles& Aerial projectiles& S-band& Not specified& 130 km& PESA\\
\hline
\end{tabular} \label{Table:Radar_types}
		\end{center}
			\end{table*}

\subsection{Miscellaneous Types of Radar Systems}
In addition to conventional radar systems there are different types of radars available in the literature for the detection and tracking of modern aerial vehicles. These radars will be grouped as miscellaneous radars. Table~\ref{Table:Miscellaneous_advance_radars} summarizes the miscellaneous modern radar systems. For example, hybrid radars (categorized as miscellaneous radar here) that combine the FMCW and Doppler radars or FMCW and interferometry radars are provided in~\cite{hybrid_radar}. Hybrid radars have the additional benefit of combining the advantages of different radars. Other miscellaneous radars include surveillance radar used for the detection and tracking of UAVs, as provided in \cite{UAV_survelliance}. The radar in \cite{UAV_survelliance} is a Doppler radar with pulse compression, MTI, and constant false alarm rate~(CFAR) algorithms. The antenna is mechanically scanned and can detect and track ground and aerial moving objects. Shape features are used in \cite{UAV_shapefeatures} to help in the detection of the micro-UAVs using a surveillance radar. The shape features help to mitigate the false alarms caused by the ground clutter. 

Lidar systems, which use light for ranging and mapping of the target area instead of radio waves, can be interpreted to be similar to radars that operate at  significantly lower wavelengths than the radio waves used by radar systems.  Due to the high range resolution of the light, small details of the terrain can be obtained precisely using lidar. Lidar can also be used for tracking aerial vehicle/s~\cite{lidar}. However, the lidar is not suitable for wide-area search in a small time duration due to small beamwidth. The performance of lidar is also effected during heavy rain, snow, and fog. Marine radars installed on small boats can be used for UAV detection~\cite{clutter6, marine_radar_UAV2}. Marine radars are off-the-shelf and easy to operate for UAV detection and tracking. Marine radars can be magnetron pulsed radars, FMCW radars, or solid-state chirped pulsed radars. A major limitation of the marine radars is their inability to handle clutter in urban areas for UAV detection and require additional clutter rejection processing.  

Another example for miscellaneous radars is Aveillant’s Gamekeeper radar  used in \cite{UAV_intent}. The Bayesian approach is used to determine the intent of a UAV around a given geographical location using the radar data in \cite{UAV_intent}. In \cite{UAV_twin_radar}, a performance analysis of twin radars is provided for civilian air traffic control. Using twin radars, a higher detection probability of small UAVs is possible. In addition, better angular resolution and increased update plot frequency can be achieved. VHF radars are used for the demonstration of the performance of twin radars. In \cite{UAV_ML_new}, UAV detection and tracking are carried out in a complex background environment using a combination of multi-modal deep learning and computer vision techniques. A novel 2D omnidirectional radar system named Omega360 is introduced in \cite{UAV_Omega360}. The Omega360 radar generates multiple staring beams simultaneously to cover the $360^{\circ}$ azimuth plane. The prolonged staring time helps to achieve long integration times for a given aerial vehicle that helps in achieving a larger SNR, and therefore, detection of small UAVs becomes possible.  

Open-ended radar architectures and SDRs, as well as digital and adaptive beamforming techniques can be used for modern radar systems~\cite{adaptive_SDR_radar}. An example of SDR-based radars is AESA TX/RX modules. If there are $N$ TX and RX elements in an AESA module, then $N$ SDRs are required. SDRs operating at different frequency bands can also be used for the detection of UAVs~\cite{SDR1,SDR2,SDR3}. The discussion about quantum radars is provided in \cite{UAV_quantum}, including pros and cons of common types of quantum radars. The ability of quantum radars to detect stealth and unconventional aerial objects is also provided. A radar warning receiver~(RWR)~\cite{RWR} is used by modern aerial vehicles to detect the illumination by a radar. The RWR allows the pilot to analyze the intercepted radar signal illuminating the aerial vehicle, find the location of the radar emitting the signals, and take countermeasures based on the extracted information. 

Multiple-input and multiple-output~(MIMO) radar systems~\cite{MIMO_radar1,MIMO_radar2} can be used for aerial vehicle detection and tracking. MIMO radars typically transmit mutually orthogonal signals from different antennas. The mutually orthogonal transmitted signals are extracted at each receive antenna. Additional features of MIMO radars include dramatic refresh rates and a good performance against UAV swarms. MIMO radar systems are also helpful in removing ground clutter. UAVs are detected using a MIMO radar in~\cite{UAV_MIMO1}, where the detection of a hexacopter UAV is carried out experimentally in a forest area using $8\times16$ MIMO radar. A MIMO radar array is proposed in \cite{UAV_MIMO_new}. Experimental results are provided in \cite{UAV_MIMO_new} that demonstrate the capability of the MIMO radar array system for detecting and tracking small UAVs. In \cite{MIMOradar_USAirforce}, a prototype of 3D MIMO radar is introduced for UAV swarming. In addition, the 3D MIMO radar is used on-board UAV in a swarm for tracking and collision avoidance. Design requirements of different sensors and up-gradation for specific performance indicators for a radar mounted on the UAV in the swarm are discussed in detail in~\cite{MIMOradar_USAirforce}

Cognitive radar systems~\cite{Cognitive_radar} can also be used for the detection and tracking of modern aerial threats. Cognitive radars can perform better compared to conventional radar systems for detection, and tracking of modern aerial threats in unknown terrains. The ability to sense the environment and change the radar parameters based on the aerial vehicle type and surroundings can greatly help in the detection and tracking of small RCS and low flying aerial vehicles in complex environments. The parameters of the cognitive radar, e.g., frequency, PW, and transmit power can be changed in real-time based on the aerial vehicle and terrain. In \cite{cognitive_deeplearning}, deep learning is used for the classification of mini-UAVs for cognitive radars. In \cite{cognitive_deeplearning} the micro-Doppler signatures of an aerial vehicle are collected cognitively that are not present in the training database. Similar to cognitive radars, reconfigurable antennas (both in software and hardware) can be used. The features of the re-configurable antenna are adjusted based on the situation. 
	
\begin{table}[!t]
	\begin{center}
    \footnotesize
		\caption{Miscellaneous advance radar systems. } \label{Table:Miscellaneous_advance_radars}
\begin{tabular}{@{}|P{ 1.5cm}|P{4.5cm}|P{0.7cm}|@{}}
\hline
Radar&Unique aspects& Ref.\\
\hline
Hybrid radar&Combination of FMCW and Doppler radar or FMCW and interferometry radar. Helps to combine the advantages of the different radars&\cite{hybrid_radar}\\
\hline
Surveillance radar & Doppler radar with pulse compression, MTI, constant false alarm rate~(CFAR) algorithms, and target's shape feature extraction &\cite{UAV_survelliance,UAV_shapefeatures}\\
\hline
3D lidar&Multiple targets tracking in complex environments by simplifying the information obtained from the sensor &\cite{lidar}\\
\hline
Marine radar&FMCW radars that are off-the-shelf and easy to operate for UAV detection and tracking&\cite{clutter6,marine_radar_UAV2}\\
\hline
Aveillant’s Gamekeeper radar&The Bayesian approach is used to determine the intent of a UAV around a given geographical location using the radar data&\cite{UAV_intent}\\
\hline
Twin radars&A higher detection probability of small UAVs is possible using twin radars. Moreover, better angular resolution and increased update plot frequency can be achieved&\cite{UAV_twin_radar} \\
\hline
Omega360& A 2D omnidirectional radar that generates multiple staring beams simultaneously to cover the $360^{\circ}$ azimuth plane. A high SNR is achieved by prolonged staring and long integration times that enables the detection of small UAVs&\cite{UAV_Omega360}\\
\hline
SDR-based radar&It offers open-ended architecture, and digital and adaptive beamforming techniques can be used. SDRs can be used for UAV detection&\cite{adaptive_SDR_radar,SDR1,SDR2,SDR3}\\
\hline
Quantum radar& The ability of quantum radars to detect stealth and unconventional aerial objects is provided&\cite{UAV_quantum}. \\
\hline 
MIMO radar& The mutually orthogonal transmitted signals by MIMO radar are extracted at each receive antenna. Features of MIMO radars include dramatic refresh rates, good performance against UAV swarms, and removing ground clutter& \cite{MIMO_radar1,MIMO_radar2,UAV_MIMO1,UAV_MIMO_new,MIMOradar_USAirforce}\\
\hline
Cognitive radar&Better performance compared to conventional radars in unknown/complex terrains for detection and tracking of small RCS and low-flying aerial vehicles. The parameters of the cognitive radar, e.g., frequency, PW, and transmit power can be changed in real-time based on the aerial vehicle and terrain&\cite{Cognitive_radar,cognitive_deeplearning}\\
\hline
\end{tabular}
		\end{center}
			\end{table}

\subsection{Airborne Radar Systems}
Airborne radar systems are popular due to their extended range, and better situational awareness due to the height of the airborne platform. However, airborne radar systems require complex signal processing due to motion of the airborne platform and variable clutter compared to ground/sea-based radar systems. The size and transmit power of radars mounted on airborne platform are also small due to space constraints. Low frequencies cannot be used for airborne radar systems due to the large size requirement of the antenna and the limited scanning angles. Popular modern radar systems onboard aircraft use: 1) pulse-Doppler; 2) agile beam; and 3) phased arrays. Compared to ground radar systems, the clutter in the airborne radar systems cannot be filtered out using Doppler frequency shifts. This is mainly because the Doppler component of the clutter~(mainly from the ground) observed by an airborne radar system is not negligible and the Doppler component of the clutter is also angle-dependent~\cite{clutter_airborne}. There are numerous airborne radar systems available in the literature for the detection and tracking of UAVs. Popular airborne radar systems are summarized in Table~\ref{Table:Airborne_radars}. STAP is used by airborne radar systems to suppress clutter and jammer interference~\cite{STAP_processing}. STAP is popular to filter out the received signal in both the Doppler and angular domains. This can result in the cancellation of the radar returns from clutter and jammers.

The aerial Dragnet program supervised by DARPA aims to provide detection and tracking of UAVs in an urban area using UAVs~\cite{DARPA_dragnet}. The airborne UAVs required for surveillance can either be tethered or long endurance platforms. A mmWave radar mounted over a UAV is used to detect and track another target UAV in \cite{UAV_UAV_mmWaveradar}. The 2D radar measurements are processed to obtain the 3D position of the target UAV in \cite{UAV_UAV_mmWaveradar}. UAV swarms can also be used to carry airborne radars, where each element of the swarm can carry components of the radar system. For example, each UAV in the swarm can carry an antenna and form a dynamic antenna array. In \cite{UAV_stealth_airborne}, passive and active radar systems are placed on a swarm of UAVs for detecting stealth aerial vehicles. Simulations were carried out in \cite{UAV_stealth_airborne} to detect and track stealth aerial vehicles using bi-static radars mounted on UAV swarms. 

There are also satellite-based radar systems for remote sensing and imaging~\cite{satellite_imaging}. SAR systems are mainly used on-board satellites~\cite{satellite_SAR}, and aerial vehicles. The SAR can simulate a large aperture through the motion of the satellite/aerial vehicle assisted by signal processing~\cite{SAR}. The SAR data can be further processed using different algorithms to obtain high-resolution images~\cite{SAR_algorithms}. Compared to image-based remote sensing, the performance of SAR is independent of weather and light conditions. The SAR can be used for moving aerial vehicle detection and tracking~\cite{SAR_target}. In \cite{UAV_ISAR} an inverse SAR~(ISAR) is used for detection at a long range of different types of modern threats. A major benefit of the ISAR is quick detection of the aerial vehicle with low PFA and at long ranges.

Airborne sense and avoid (or detect and avoid~(DAA)) systems sense and avoid a possible future collision between the flying aircraft and other airborne platforms or high-rise ground obstacles. Sense and avoid systems are critical for advanced aerial mobility and UAV traffic management use cases~\cite{S_and_Avoid1,S_and_avoid2}. Collision avoidance radars come under the category of sense and avoid systems. Collision avoidance radars are used onboard aerial vehicles to avoid collisions with other aerial vehicles during flight. In \cite{UAV_collosion_avoidance}, a monopulse radar at $24$~GHz is mounted on a UAV. The radar uses $1\times 4$ TX and RX antennas. The angular information of the aerial vehicle objects in the 3D space is obtained using the setup. The detection of the obstacle in real-time in \cite{UAV_collosion_avoidance} allows autonomous flight possible. Collision avoidance between two UAVs is achieved using video cameras and computationally efficient algorithms in \cite{UAV_airborne_video}. The detection and tracking of a moving UAV target from a moving UAV source in \cite{UAV_airborne_video} are carried out using a camera mounted on one of the UAV and a UAV-to-UAV detection and tracking algorithm is used. 

Airborne collision avoidance system for small UAVs~(ACAS sXu) developed by FAA provides decentralized DAA capabilities for small UAVs~\cite{ACAS1,ACAS2}. ACAS sXu is a robust system that can operate autonomously and can help small UAVs to avoid manned and other UAVs during autonomous or beyond visual range flights. The primary task of the ACAS sXu is to provide alerts and advisories to small UAVs while in flight. The alerts and advisories by ACAS sXu are generated by first detecting an aerial vehicle/s in the vicinity using sensors~(usually radar) or a shared data link between the aerial vehicles. The system then estimates the position and speed of the aerial vehicles through a tracking algorithm and suggest if there is a possibility of a collision. Overall, ACAS sXu provides a tunable, modular and scalable solution for small UAVs and ensures a safe air space.

\begin{table}[!t]
	\begin{center}
    \footnotesize
		\caption{Airborne radar systems. } \label{Table:Airborne_radars}
\begin{tabular}{@{}|P{ 1.5cm}|P{4.5cm}|P{0.7cm}|@{}}
\hline
Radar & Unique aspects & Ref.\\
\hline
Airborne radar using STAP & STAP is used by airborne radar systems to suppress clutter and jammer interference. STAP also filters out the received signal in both the Doppler and angular domains & \cite{STAP_processing}\\
\hline
Aerial Dragnet system & The aim of the Aerial Dragnet radar system supervised by DARPA is to provide detection and tracking of UAVs in urban areas using UAVs. The airborne UAVs required for surveillance can either be tethered or long endurance platforms & \cite{DARPA_dragnet} \\
\hline 
mmWave radar & A mmWave radar mounted over a UAV is used to detect and track another target UAV. The 2D radar measurements are processed to obtain the 3D position of the target UAV & \cite{UAV_UAV_mmWaveradar}\\
\hline
Active/passive radars on-board UAVs in a swarm & UAV swarms can carry airborne radars. Passive and active radar systems placed on a swarm of UAVs can be used for the detection of stealth aerial vehicles & \cite{UAV_stealth_airborne}\\
\hline
Satellite-based radar & Satellite-based radar systems for remote sensing and imaging & \cite{satellite_imaging, satellite_SAR}\\
\hline 
SAR on-board aerial vehicles&The SAR can be used for imaging and moving aerial vehicle detection and tracking at long ranges. ISAR provides quick detection of the aerial vehicle with low PFA and at long ranges & \cite{satellite_SAR,SAR_target,UAV_ISAR}\\
\hline
Detect and avoid~(DAA) Radar & Avoid a possible future collision between the flying aircraft and other airborne platforms or high-rise ground obstacles using sensor returns. DAA radar systems allow safe autonomous flights possible. In addition to radars, cameras are also popular sensors for DAA & \cite{S_and_Avoid1,S_and_avoid2, UAV_collosion_avoidance,UAV_airborne_video}\\ 
\hline
Airborne collision avoidance system for small UAVs~(ACAS sXu) & ACAS sXu is developed by FAA and provides decentralized DAA capabilities for small UAVs. ACAS sXu can operate autonomously. The primary task of the ACAS sXu is to provide alerts and advisories to small UAVs while in flight & \cite{ACAS1,ACAS2}\\
\hline
\end{tabular}
		\end{center}
			\end{table}


\subsection{Radars Using Popular Frequency Bands}
The radars use different frequency bands for their operation, and they are  sometimes named based on the frequency band used. An active C-band multistatic surveillance radar system is used for the detection and positioning of a quadcopter UAV in \cite{UAV_Cband}. In \cite{UAV_Wband} it is demonstrated that W-band can be used for detection, localization, and classification of small UAVs. A L-band 3D staring radar is used for the detection of low-RCS UAVs in \cite{UAV_Lband}. An S-band non-coherent radar and neural network are used for a UAV target recognition in \cite{UAV_Sband}. Radar data from birds, ships, and UAVs are used for classification. The classification is carried out using track information and video classification methods in \cite{UAV_Sband}. The image-based classification is found to be better for recognizing small UAVs. 

A C-band multistatic radar system is introduced in \cite{UAV_Cband_new} for the detection of small UAVs. The radar provides scanning capabilities by using a phased array at the TX and digital beamforming at the RX. A simulator is also built based on the system's capabilities. In \cite{X_L_band}, a NeXtRAD radar system is used for simultaneous monostatic and bi-static measurements. The measurements were carried out at X-band and L-bands. In \cite{X_L_band}, it is shown that the NeXtRAD radar system is capable of detecting multiple small UAVs. A holographic 3D radar capable of wide-area multi-beam surveillance for detection of micro UAVs is provided in \cite{UAV_holographic_radar}. Experiments are conducted using a $32\times 8$ element RX array operating in the L-band. A high processing gain and detection sensitivity of holographic radar system help to detect small RCS UAVs.

\subsection{A Network of Radar Systems}
The emerging aerial threats require better situational awareness and early warning. Using a network of distributed radars at different locations across the globe can provide early warning and better monitoring compared to stand-alone radar systems. Radar systems onboard satellites, airborne platforms, ships, and ground can form a network for improved surveillance against a wide range of threats. This network can be centralized or decentralized. If any node in the network detects a threat it can share the information with the other nodes. The mobile nodes can use their mobility to verify the threat and position themselves for tracking. Once a threat is identified and tracked, each node can suggest countermeasures based on the observed features of the aerial vehicle, and the best countermeasure is adopted.

In \cite{UAV_network_passive}, a network of passive airborne radar sensors is scheduled to detect and track multiple airborne UAV targets carrying active radars. The passive sensors are controlled and scheduled through a proposed algorithm in \cite{UAV_network_passive}. The algorithm is based on long and short-term rewards and the Markov decision process. In \cite{track_new3}, experiments were performed using two radar systems networked to detect and track UAVs over a wide area. UAV tracks were created using a recursive random sample consensus algorithm from the combined detection paths of the two radar systems. In \cite{UAV_staringradar}, UAV detection is carried out using a staring cooperative radar network. A wide-beam antenna and a directive digital array are used by the staring radars. The setup in \cite{UAV_staringradar} allows simultaneous multiple fixed beams at the RX allowing long integration times that help in the detection of slow-moving small UAVs.

\section{Miscellaneous Factors Affecting Radar Performance} \label{Section:Misc_factors}

In this section, limitations of radar systems and miscellaneous factors related to radar systems for the detection, tracking, and classification of aerial vehicles are discussed. 

\subsection{Limitations of Radar Systems}
The limitations of different types of radar systems are already discussed briefly in different sections of the paper. Here, we will summarize the limitations of the radar systems for countering modern aerial threats. The modern aerial threats are mainly from UAVs, stealth, different flight formations of UAVs, e.g., UAV swarms, and RF jamming. The limitations of the radar systems for handling modern aerial threats are as follows:
\begin{itemize}
    \item In general, the radar systems are inefficient because a large amount of energy is transmitted, whereas, only a fraction of the transmitted energy is received back from clutter and from a potential aerial vehicle. Also, the energy emission from radar systems~(active) exposes them to potential of detection. 
    \item The detection of unconventional aerial threats~(small RCS) requires adaptive adjustments of radar thresholds and PFA generally increases.
    \item A major limitation of any radar system is in the detection and classification of aerial vehicles with stealth at long ranges. The stealth aerial vehicles are designed to absorb,  deflect, and scatter incoming radar energy in directions other than the source.
    \item The radar systems also have limitations in detecting small, low flying UAVs in cluttered environments. The ground hugging UAVs are renowned to evade radar systems. 
    \item The radar systems have limitations in tracking very high altitude, maneuverable, and high speed aerial vehicles, e.g., hypersonic glide vehicles. 
    \item The detection and tracking of multiple aerial vehicles in a UAV swarm requires high range and angular resolution radars, and complex algorithms for tracking and classification. In case of a dense UAV swarm, the accuracy of detection, tracking, and classification will be limited for any modern radar system. 
    \item In a scenario of a swarm of aerial vehicles, the ambiguity in range (discussed in Section~\ref{Section:PW_dutycycle}) increases for a given PRI and PW. Similarly, if the aerial vehicles in a swarm are moving at different velocities, the ambiguity in velocity is expected to increase for a given PRI and PW~(as discussed in Section~\ref{Section:Doppler_ambiguity}).
    \item Modern offensive ECM can be used to disable radar systems. ECM devices mounted on aerial vehicles can be used to jam and spoof radar systems.
    \end{itemize}

\subsection{Countering Multiple Aerial Vehicles}
The detection and tracking of multiple aerial vehicles is challenging~(as discussed in Section~\ref{Section:multipleUAVs}). There are different techniques used by radar systems for the successful detection, and tracking of multiple aerial vehicles. These techniques include the use of UWB radars, multiple narrow beams, and classification algorithms for differentiating close by aerial vehicles. In \cite{uwb_multiple} multiple aerial vehicle detection and tracking were performed using UWB radars. The large bandwidth offers high range resolution and helps to resolve multiple aerial vehicles flying closely. Multiple aerial vehicles can also be detected and tracked using FMCW frequency division multiplexing MIMO radar operated at $60$~GHz as shown in \cite{MIMOradar_vasilii}. The radar is capable of detecting and tracking multiple small UAVs simultaneously at a moderate distance.

Single/multiple electronically steered beams by PESA and AESA radars can be used to track multiple aerial vehicles simultaneously~\cite{aesa_pesa}. Complex algorithms are also available in the literature to detect and track multiple aerial vehicles~\cite{multiple_tracking1,multiple_tracking2}. AI algorithms are mainly used for the detection and tracking of multiple aerial vehicles~\cite{multiple_ML1, multiple_ML2}. A major emerging threat of multiple aerial vehicles is UAV swarms. Small UAVs flying in a swarm are difficult to track by radar systems and a challenging scenario for interdiction~\cite{UAVSwarms_Threat}.

\subsection{Noise in Radar Systems}
The noise in a radar system is similar to any other remote sensing system and hinders the detection of an aerial vehicle. The noise can increase the PFA, localization error, and classification inaccuracy. The noise can be narrow-band or broadband, frequency-dependent or independent, and the statistics of the noise can be stationary or non-stationary. There can be different sources of noise for a radar system discussed in~\cite{noise}. The sources of noise are broadly classified as internal and ambient. The internal sources of noise in a radar system are due to: 1) thermal emissions and random currents in the components of TX and RX; 2) data quantization noise; 3) signal processing limitations; and 4) noisy data from sidelobes. The ambient sources of noise lie in the channel between the TX and RX. The sources of ambient noise can be either on the ground or in the atmosphere and space. In addition to noise, there are different sources of interference to the radar signal propagation. The interference can be from a radar using the same frequency or due to out-of-band emissions from nearby radio devices. 

During radar measurements, the noise floor is important in deciding the threshold for detection. There are also noise radars that use random signals for propagation, where, either a noise source or modulation using a white noise source at low frequency is used for transmission~\cite{noise}. The transmission of a truly random signal for radar has many advantages. These include immunity to noise, unambiguous range and Doppler measurements, robust ECM, and high EM compatibility.

\begin{figure}[!t]
	\centering
	\includegraphics[width=\columnwidth]{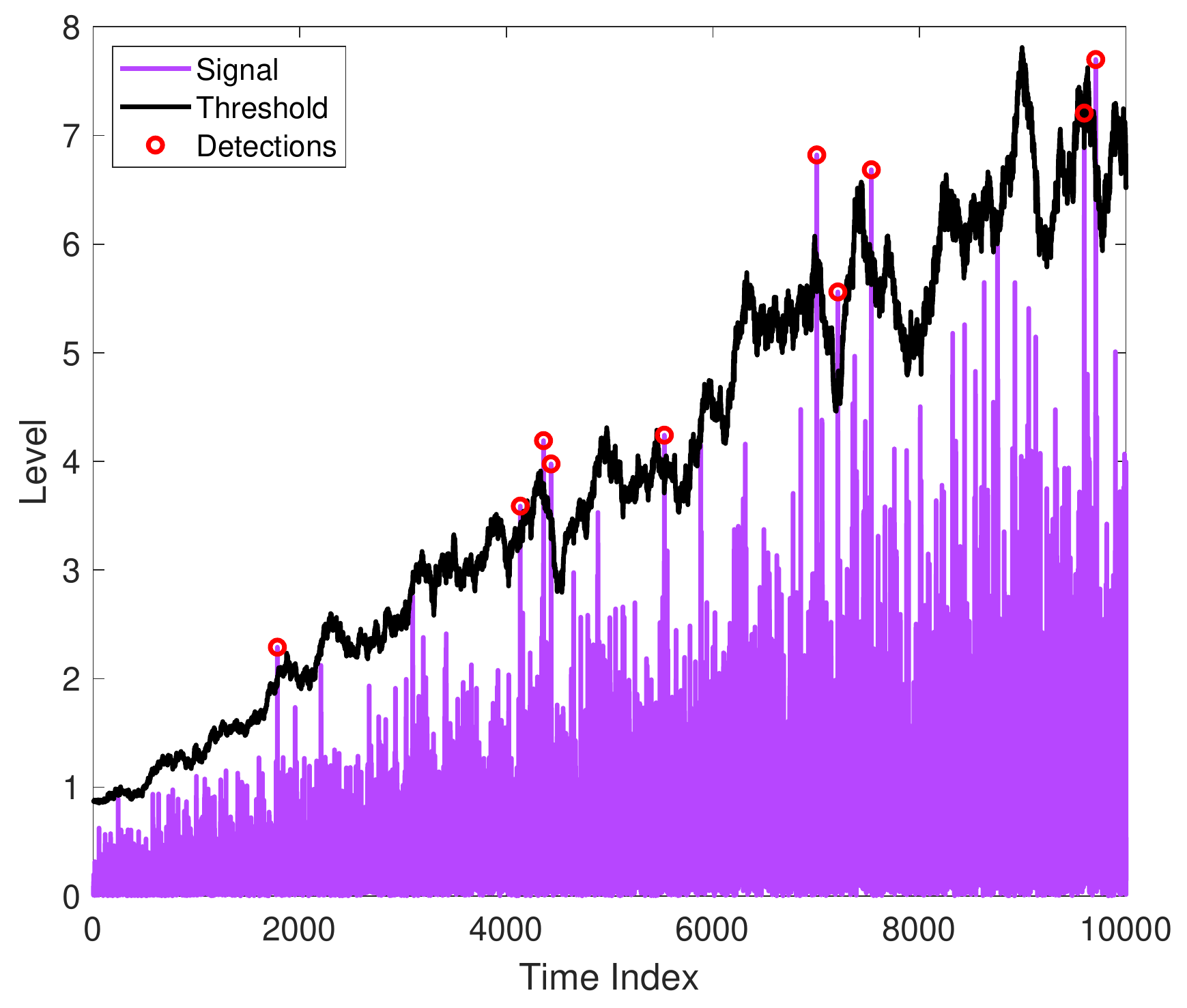}
	\caption{An example of automatic threshold factor where the threshold increases with the noise power to keep the false alarm rate constant. Whenever the signal level is above the threshold, a detection is made~(regenerated from \cite{Matlab_threshold}). } \label{Fig:Threshold_pfa}
\end{figure}

\subsection{Probability of False Alarms}
The probability of a false~(target present) event, when the true~(no target present) event has occurred will constitute the PFA. The performance of a radar detector is measured based on how low the PFA is for a given SNR and detection threshold. The PFA for a radar system should be very small, typically less than $10^{-6}$~\cite{pfa_value}. The threshold for detection can be selected adaptively to keep the PFA low~\cite{pfa_thr}. The PFA for detection of small and stealthy aerial vehicles is high due to the low SNR of received echoes~(from small RCS).

In \cite{pfa_thr2}, AI is used in conjunction with other available databases to adaptively decide detection threshold, sense the environmental conditions, and select the appropriate CFAR algorithms~\cite{pfa_thr2}. An example of CFAR is shown in Fig.~\ref{Fig:Threshold_pfa} using an adaptive threshold~\cite{Matlab_threshold}. The threshold is adjusted adaptively based on the noise level to keep the false alarm rate at a constant level as shown in Fig.~\ref{Fig:Threshold_pfa}.  

\subsection{Scan Volume}
A major feature of a radar system is its ability to cover a given scan volume in a given duration of time. There is a trade-off between the scan volume, the beamwidth of the radar beam (in the azimuth and elevation planes), the number of simultaneous beams available, and the total scan duration to cover the volume. The beamwidth of the radar beam can be increased to cover a given volume in a short duration of time, however, the angular resolution will reduce~(see Section~\ref{Section:Range_Angular_resolution}). The scan duration for a given volume can also be reduced by increasing the number of independent beams e.g., in an AESA radar. Another method of managing the scan volume is by adjusting the radar resources according to the situation. For example, if there is (are) aerial vehicle(s) detected at a given volume of space, then the majority of scan energy can be directed towards the volume where the aerial vehicle(s) is (are) detected. 

\subsection{Position of the Radar System}
 The range and coverage area of a radar system depend on the position of the radar antenna. The range and coverage area can be increased by increasing the height of the radar antenna above the ground/sea. The antenna height above the ground can be achieved by using airborne platforms. For example, a radar placed at the geostationary orbit can cover one-third of the earth's geographical area. However, increasing the height of the radar antenna above the ground/sea increases the overall echo time and the range/angular resolution is low. The position of a radar antenna also depends on the terrain. For example, in hilly and urban terrain, the radar antenna can be placed on the hill and rooftops of buildings for better coverage. Furthermore, the detection and tracking of small aerial vehicles such as UAVs can be performed using distributed antennas installed at different heights above the ground for better situational awareness.

\subsection{Interfacing of Radars with Other Systems}
A complete radar system is a combination of different sub-systems controlled by a command/control center. For example, search radar, tracking radar, and guidance radar are sub-systems of an overall radar system. The different sub-systems are networked for a combined response. Similarly, radar data~(from active and passive radars) and other sensor data from different locations are networked for accurate detection and tracking of an aerial vehicle. The real-time aerial vehicle data obtained from the front-end radar system is relayed to nearby or remote data centers for post-processing and analysis. For example, data collected by radars onboard UAVs are fed back to GSs via satellite links for analysis. Similarly, in a track, while scan radar operation mode, tracking, and scanning are performed simultaneously and the results are continuously analyzed by a control center. The remote data/control links between the field radars and control centers are susceptible to jamming and spoofing. 

\subsection{Electronic Countermeasures}
All wireless links are vulnerable to ECM which can be categorized as offensive and defensive approaches. The radar systems can be used for both offensive and defensive ECM. Popular offensive ECM include jamming the hostile radio communications, spoofing the radio communications with false data~\cite{jamming}, or transmitting multiple copies of the echo signals to get an impression of multiple aerial vehicles. The defensive ECM include robust frequency hopping over broadband, multilayered authentication and encryption of data, detection, and reporting of jammed strobe, and dedicated RX for monitoring of the ECM~\cite{ECCM}.

The simplest ECM acting as a jammer can be a noise generator. The noise generator raises the noise floor at a given operating frequency of the intended radar system. There are different types of jamming~\cite{jamming_techniques}. In spot jamming a particular frequency of the radar is targeted, whereas, in sweep jamming, sweeping across a band of frequencies is carried out. Another jamming is barrage jamming, where a large frequency band is covered, e.g., UWB. Deceptive jamming can also be used, where spoofed signals are sent to the radar. CW radars are more prone to jamming compared to pulse radar systems. Passive or semi-active~(active on request) pulse radar systems are resistant to jamming. The UWB radars also offer resistance to jamming mainly due to their low PSD. 


\subsection{Friend or Foe Identification}
Friend or foe determination known as identification friend or foe~(IFF) technology is a part of the detection and classification process~\cite{IFF_radar}. Secondary surveillance radars and radar transponders are generally used for IFF. A friendly aerial vehicle is equipped with a radar transponder that replies to interrogation by a secondary surveillance radar. The hostile aerial vehicle, on the other hand, will not be able to decode the information sent from the secondary radar. Therefore, a hostile aerial vehicle will not be able to provide an appropriate reply to the secondary radar in due time. IFF system can also be installed on the UAVs to possibly identify malicious UAVs. The IFF can only identify a friend when the transponder on the aerial vehicle is turned on. In the presence of an IFF jammer, the identification of a friend may not be possible.

\section{Communication Systems for the Detection, Tracking, and Classification of UAVs} \label{Section:Comm_Sys}

In this section, the importance of communication systems, and factors affecting the detection, tracking, and classification of UAVs using communication systems are discussed. The popular communication systems used for UAV detection, tracking, and classification, comparison of communication systems with radar systems, and future directions for the communication systems in the UAV detection, tracking, and classification are also provided.

\subsection{Importance of Communication Systems for UAV Detection, Tracking, and Classification}
A major limitation of the radar systems is finding small RCS aerial vehicles in highly cluttered environments. Furthermore, if the clutter is not static, rejecting the dynamic clutter becomes challenging. In complex environments, e.g., dense urban, the clutter is high-rise~(often obstructs the LOS of the radar), diverse~(different shapes, sizes, heights, and surface materials of the buildings), and dynamic~(due to the motion of pedestrians and ground and aerial vehicles). Moreover, the shape and flight characteristics of many commonly available UAVs resemble birds. However, providing a seamless detection cover against UAVs~(that have small RCS, and fly low and slow) in a dense urban environment by radar systems alone is not always possible. On the other hand, communication systems offer thorough coverage in the majority of dense urban areas. Similar to radar systems, the communication systems use RF-based propagation and can either directly detect UAVs or aid radar systems.

Major advantages of communications systems for  detection, tracking, and classification of UAVs can be listed as follows: 
\begin{itemize}
    \item Communication systems offer a thorough coverage in the majority of the urban areas compared to radar systems.
    \item Communication systems can operate in the NLOS conditions, and use different algorithms for aerial vehicle detection. On the other hand, the majority of the radar systems require a LOS path with the aerial vehicle for precise detection.
    \item Multiple types of communication systems are available in an urban area, e.g., FM, amplitude modulation~(AM), television broadcast, and mobile and satellite communications. Each communication system uses a distinct frequency band and communication methodology. Moreover, different types of communication systems offer long, medium, and short-range communications. 
    \item The height of the antenna/s of communication towers is selected to maximize coverage in a given area. The height of the majority of the communication tower antennas is such that small UAVs flying at typical heights~(e.g., lower than $400$~feet in the United States under FAA Part 107 Rules) can be detected and tracked.  
    \item The communications using satellites offer ubiquitous and LOS coverage on the ground mainly due to the height of the satellites above the ground. The ubiquitous LOS coverage by satellites can help in the detection and tracking of UAVs.
    \item Modern communication systems, e.g., 5G and beyond uses phased arrays and can electronically steer antenna beams similar to phased array radar systems.  
    \item Different communication systems update the channel state information~(CSI) at regular intervals of time, e.g., massive MIMO systems. These periodic CSI updates can help in UAV detection.
    \item Majority of the communication systems used for the detection and tracking of UAVs operate in the passive mode and hence avoid detection and subsequent countermeasures. 
    \item Communication systems can be used to obtain the Doppler signature of the UAV, which can help to classify a UAV. 
    \item A significant research literature and standardization details are available for communication systems in the public domain that can help research for UAV-based sensing and tracking. 
    \end{itemize}

\subsection{Factors Affecting Detection, Tracking, and Classification using Communication Systems}

There are many factors that affect the detection, tracking, and classification of UAVs using communication systems. The main factors include the type of the transmitted signal, modulation and coding, frequency, bandwidth, transmit power, receiver sensitivity, antenna characteristics, and position of communication nodes. Different types of communication systems use different types of signals. The selection of a signal for transmission depends mainly on the application and propagation channel. Orthogonal frequency division multiplexing~(OFDM) is a popular signal for the majority of communication systems where sub-carriers experience flat fading. Similarly, the selection of modulation and coding is based on the propagation channel and given application. Modulation and coding are selected to suppress noise, and interference, increase communications reliability and provide variable data rates. For the detection, tracking, and classification of UAVs, low noise/interference and reliable propagation can be achieved using appropriate modulation and coding schemes.  

The center frequency and communication range are directly related. However, the available spectrum for a given communication system is limited, e.g., GSM and LTE. Communication systems working at different portions of the spectrum can provide a diverse set of frequencies and bandwidhts that can collectively assist in the detection of UAVs, due to different propagation characteristics in such frequencies and bandwidths. For example, larger bandwidths (e.g., ultra-wideband) can better resolve the MPCs from the scatterers that can improve the detection, tracking, and classification of UAVs. Complex amplitude, delay, and angle of arrival~(AoA) of each MPC returned from the UAV can provide an estimate of the size, material, shape, and position of the UAV. Alternatively, if the bandwidth is not sufficient to resolve MPCs from the UAV and scatterers, then changes in the received signal strength indicator~(RSSI), either directly or through their statistical distribution over time, can be used for the detection of UAVs. 

The transmit power and antenna gain can help to compensate for the signal attenuation in the propagation channel. A large antenna array, adaptive beam steering, and beamwidth~\cite{adaptive_beamwidth} can provide high gain and reduce clutter. A good receiver sensitivity can also help to recover weak signals at the receiver. Due to the small RCS of the UAV, high transmit power, good receiver sensitivity, and high gain antennas are required for detection. Other factors such as the position and height of the communication tower are important for seamless coverage to a given area. The position and height of the communication tower also help to reduce interference with other zones using the same frequency band(s). For UAV detection and tracking, a preferable position and height of the communication tower is the one that ensures a LOS path with the flying UAV. 

A communication channel can be a slow or fast-fading channel. A fast fading channel due to the motion of the TX/RX or scatterers in the vicinity presents many challenges. In a fast-fading channel, the channel changes during the transmitted symbol period and introduces rapid amplitude fluctuations. The fast varying channel can result in high Doppler spread and aliasing. In a fast fading channel, frequent CSI estimates are required. Moreover, fading in a communication channel can be flat or frequency selective, depending on the bandwidth of the transmitted symbols and the channel conditions. Frequency selective fading results in MPCs, and for UAV detection purposes, a slow and frequency selective fading channel is preferable. For a flat fading channel, a high SNR can help in the detection based on the RSSI approach.

\subsection{Popular Communication Systems for the Detection, Tracking, and Classification of UAVs}
The detection, tracking, and classification of UAVs using communication systems can be carried out either in the active mode, e.g., using JCR systems or analyzing changes in the common communication signals e.g., FM, LTE, and sensing the UAV RF communications passively. The majority of the communication systems used for the detection, tracking, and classification of UAVs operate in the passive mode. For example, links between communicating nodes for radio, television broadcast services, and mobile and satellite communications services can be considered virtual threads. Any change in the communication link due to the presence of UAV(s) can be analyzed passively.

\begin{table}[!t]
	\begin{center}
    \footnotesize
		\caption{Radar operation aided by Communication systems.} \label{Table:Comm_radars}
\begin{tabular}{@{}|P{ 1.7cm}|P{5cm}|P{0.7cm}|@{}}
\hline
Communication System& Unique aspects & Ref.\\
\hline
GSM/LTE/5G & Detection, tracking, and classification using GSM/LTE communication networks in the passive mode. The weak reflections of GSM/LTE signals from a UAV are analyzed. Phased arrays for 5G communications are used for detecting small aerial vehicles in complex urban environments& \cite{UAV_passive_radar,UAV_passive_bistatic, GSM_new1,GSM_new2,LTE1,LTE_classify,LTE3,LTE4,LTE5,LTE6,UAV_comm_radar,UAV_comm_radar2,UAV_5G,UAV_5G_3,3GPP_classify}\\
\hline 
Satellite&Satellite communications that cover large spaces between upper atmosphere and ground can be used for the detection, tracking, and classification of UAVs&\cite{Satellite1,Satellite1,Satellite2,Satellite3,Satellite4,Satellite4}\\
\hline
WiFi&The interference in the WiFi signals outdoors due to presence of UAV or WiFi signals from UAVs can be analyzed for detecting UAVs&\cite{Wifi1,Wifi2,Wifi3,Wifi4}\\
\hline
DAB&Passive radar using DAB transmissions can be used for the detection and tracking of UAVs&\cite{DAB1,DAB2,DAB3,DAB4,DAB5,DAB6,DAB7,DAB8}\\
\hline
DVB-T&The DVB-T transmissions can be used to detect and classify UAVs in passive mode at long ranges. The tracking is also possible using RX antenna arrays&\cite{UAV_PBR,UAV_DVBT,DVB_classify2,DVB_T_classify,UAV_DVBT2,UAV_4D,UAV_DVBT2_Clean}\\ 
\hline 
JCR& A JCR system can simultaneously provide communication services and radar-based sensing. To provide radar-based sensing, additional hardware, algorithm and processes, and specified waveform are used by the communication systems&\cite{JCR1, JCR2, JCR3, JCR4, JCR5}\\
\hline
RF analyzers&The RF communications between the UAV and other nodes are sensed and analyzed to detect presence of a UAV. The analysis can provide classification of the UAV also &\cite{RF_analysis,RF_analysis_clas,UAV_signal_analysis,RF_analysis_new}\\
\hline
\end{tabular}
		\end{center}
			\end{table}

In the active mode, certain resources from the communication system(s) are allocated for the active radar operation, e.g., JCR systems. A common waveform and the same hardware are used for both communications and radar operation in a JCR system. A third possibility is to analyze the communications between the UAV and other UAVs or between the UAV and satellites/GS to detect the presence of a UAV. The patterns of the RF communications~(e.g., frequency, frequency hopping pattern, bandwidth, modulation, packet size, the delay between the packets, and sequence of burst transmissions) between UAVs and other aerial vehicles or GS can be used to distinguish a UAV from other mobile communication nodes~(discussed in Section~\ref{Section:RF_analyzers}). Table~\ref{Table:Comm_radars} lists popular communication systems and RF analyzers used for the detection, tracking, and classification of UAVs.

\subsubsection{GSM}
Major advantages of using a GSM network for the detection and tracking of UAVs include the collection of detection measurements from different aspect angles using base stations, and combining measurements from different base stations to overcome weak measurements from a single base station. In~\cite{UAV_passive_radar}, a GSM-based radar in the passive mode is used to collect weak reflections from a UAV. Experiments were conducted to detect and track a quadcopter using passive GSM-based radar achieving a high detection rate. A track-before-detect method is used to detect and track small UAVs using GSM signals. A GSM-based detection system for small UAVs operating in the passive mode is introduced in \cite{GSM_new1}. Passive bistatic radar measurements were carried out using GSM signals in \cite{UAV_passive_bistatic} for the detection of very low-RCS aerial vehicles. Experiments were carried out in \cite{GSM_new2} using multiple mobile telecommunication illuminators including GSM in the passive mode for the detection of aerial targets.  


\subsubsection{LTE}
A major benefit of the LTE is variable bandwidth available from $1.4$~MHz to $20$~MHz. LTE offers a high range and velocity resolution compared to GSM. Also, the ambiguity function in LTE results in lower sidelobes, and the global coverage allows coordinated passive sensing. Similar to GSM, an LTE network can be used for the detection and tracking of UAVs in the passive mode. A USRP-based setup in \cite{LTE1} uses LTE signals from multiple transmitters and a single receiver is used for the detection of UAV in the passive mode. In \cite{LTE2_classify} a flying UAV is classified as common user equipment~(UE) node using LTE services. The classification helps to identify rogue UAVs from common UEs. The classification is carried out using measurements and statistics of UE and UAV nodes and machine learning algorithms. In \cite{LTE3}, LTE-based measurements based on interference and using multiple machine learning algorithms were carried out to identify UAVs carrying UEs. A passive LTE-based radar system is used in \cite{LTE5} for the detection of UAVs. 
Experiments were conducted to demonstrate the capability of the passive LTE radar to detect low-flying UAVs. A multi-static passive radar based on LTE is studied in \cite{LTE6}. Multiple eNodeBs provide bi-static range measurements and the hyperbolic positioning method is used for the localization of the UAV. Passive SAR imaging is demonstrated in \cite{LTE4} using LTE signals. 

\subsubsection{5G}
To fulfill high data rate requirements, mmWave band is utilized for 5G communications. Salient features of 5G communications include dense network coverage through a large number of access points and connected devices, highly directional antenna beams, and beam steering capabilities. The phased arrays used for 5G communications can also help in the detection of small aerial vehicles. In \cite{UAV_comm_radar} a 5G prototype using FMCW signals and phase arrays at $28$~GHz is used for the detection of UAVs in an urban environment. The phased arrays offered better angular tracking and precise range measurements in different directions. Similarly, in \cite{UAV_comm_radar2}, 5G phased arrays at $64$~GHz are used for UAV detection. An FMCW-based system offered a high resolution and medium-range detection capability. In \cite{UAV_5G}, the possibility of using 5G infrastructure for the detection of UAVs is provided. 5G deployments using mmWaves can help in the efficient detection of small UAVs. Different design perspectives of the system based on 5G, e.g., the density of base stations, bandwidth, and directional antennas are provided for the detection of small UAVs. In \cite{UAV_5G_3}, detection and identification of rotor-based UAVs is carried out using 5G networks. Remote identification of UAVs can also be made using 5G networks as provided in the 3GPP report~\cite{3GPP_classify}.

\subsubsection{Satellites}
Satellite communications can be used for the detection and tracking of UAVs near the ground. However, the long distance of the satellites from the ground, low data rates, and small RCS of common UAVs make detection and tracking challenging. In \cite{Satellite1}, a digital video broadcasting satellite~(DVB-S) is used for the detection of UAVs. The micro-Doppler signature from rotating blades is used for UAV identification. To overcome the challenges of UAV detection by satellite networks, forward scattering radar~(FSR) configuration is used in the passive mode. The passive DVB-S and FSR are also used in \cite{Satellite2} for the detection of multi-rotor UAVs and performing micro-Doppler analysis of the copter UAVs. Similarly, in \cite{Satellite3}, Malaysian Satellite Measat3a/3/3b, and DVB-S FSR are used for UAV detection. In \cite{Satellite4}, DVB-S is used for the detection of multi-rotor UAVs and micro-Doppler signature extraction. The micro-Doppler signature helps in the UAV classification. A reference channel and a surveillance channel are used. The reference channel points to the satellite for reference signaling, whereas the surveillance channel points to the UAV. The two channels are used to calculate the cross-ambiguity function which assists in UAV detection.  

\subsubsection{WiFi}
WiFi signals outdoors can be used to detect the presence of a UAV. In \cite{Wifi1}, statistical fingerprint analysis of the WiFi signals is carried out to detect the presence of unauthorized UAVs. The characteristics of the UAV control signals and first-person view transmissions, features of WiFi traffic generated by UAVs, and machine learning algorithms help to identify a UAV in a WiFi coverage area. Similarly, in \cite{Wifi2}, a WiFi based statistical fingerprinting method is used for the detection of UAVs. The approach claims to achieve an identification rate of $96$\%. In \cite{Wifi3}, the detection and identification of UAVs are carried out using WiFi signals and RF fingerprinting. Firstly, a UAV is detected based on the captured signals, then machine learning algorithms are used to identify the category or the UAV. The identification accuracy is higher outdoors compared to indoors. In \cite{Wifi4}, machine learning algorithms are used to analyze the WiFi signals originating from UAVs for control and video streaming. The analysis helps to detect and identify UAVs. The proposed approach can work on encrypted WiFi traffic also, using the packet size and WiFi traffic inter-arrival time features for the detection and identification of UAVs over the encrypted WiFi traffic. WiFi spectrum analyzers, e.g., Yellowjacket-Tablet~\cite{Wifi5} operating in the frequency range of $2$~GHz to $5.9$~GHz can also be used for UAV detection.   

\subsubsection{Digital and Analog Audio Broadcast}
Digital audio broadcast~(DAB) has been a popular source of communications to large populations. DAB can also be used for the detection of UAVs in the passive mode. A passive radar in the VHF band and using DAB transmissions is used for the detection of a fixed-wing UAV in~\cite{DAB1}. The detection range of the fixed-wing UAV reported was $1.2$~km. A DAB and FM~(using analog signals) radio based passive radar sensor is used in~\cite{DAB2}. The sensor can be used for air-surveillance. Similarly, passive multi-static aerial surveillance using DAB signals is carried out in~\cite{DAB3}. The Doppler and bi-static time difference of arrival measurements are used and a multi-hypothesis tracking algorithm is developed to aid at different stages of tracking. Similarly, in~\cite{DAB4}, DAB based passive multistatic radar is used to detect stealth UAVs. Moreover, a passive multistatic radar based on the FM radio signal transmissions in~\cite{DAB6} is used to detect aerial vehicles. The tracking is carried out using three passive radars placed at different locations. In~\cite{DAB5}, a passive coherent location~(PCL) system using DAB, FM, and cellular communication signals is used. The PCL can be used to detect UAVs. A passive radar based on AM~(analog) signals that performs track before detect operation is provided in~\cite{DAB7}. The time-azimuth and time-Doppler traces of multiple targets are tracked and analyzed under low SNR conditions. FM radio signals can also be used for the detection of stealth aerial vehicles~\cite{DAB8}.

\subsubsection{Digital Video Broadcast Television}
Similar to DAB, DVB terrestrial~(DVB-T) signals can be used for the detection and tracking of UAVs. The potential of a passive radar using DVB-T is shown in \cite{UAV_DVBT}. Experimental data for the detection and identification of multi-rotor and fixed-wing UAVs is provided. It is claimed that passive radars can play an important role in the detection of different types of UAVs at long ranges using DVB-T transmissions. The potential uses of DVB passive radar for the detection and identification of different small UAVs are provided in \cite{DVB_classify2}. The detection and identification using DVB passive radar help to counter illegal UAV flights. The detection of UAVs using DVB-T2 passive coherent radar is provided in \cite{UAV_DVBT2}. A passive radar using DVB-T illuminators of opportunity is used for 4D~(range, Doppler, azimuth, and elevation) detection and tracking of small UAVs in~\cite{UAV_4D}. The approach uses a uniform linear array and two separate antennas for estimation of range, Doppler, azimuth, and elevation angles of the aerial vehicle. A passive multistatic digital TV-based radar for UAV detection is provided in \cite{UAV_SDPR}. Passive radar is used for the detection and tracking of UAVs in a highly cluttered scenario near airports in~\cite{UAV_SDPR}. In \cite{DVB_T_classify}, DVB-T2 signals are used to illuminate small UAVs experimentally. The illumination observed by the passive radar and the specific Doppler-signatures is used to classify UAVs from non-UAVs targets. In \cite{UAV_DVBT2_Clean} it is observed that the strong reflections from large RCS of commercial planes make software-based detection and tracking of small RCS UAVs difficult. Long CPIs and multi-stage CLEAN algorithms are, therefore, used to filter strong reflections from commercial planes in \cite{UAV_DVBT2_Clean}. A measurement scenario is created with DVB-T-based passive radar. In \cite{UAV_PBR}, a passive multi-channel bi-static radar system is used for small UAV detection using TV broadcast transmissions. Experiments were performed to validate the system.

\subsubsection{Joint Communications-Radar Systems}
\begin{figure}[!t] 
    \centering
	\begin{subfigure}{\columnwidth}
    \centering
	\includegraphics[width=\columnwidth]{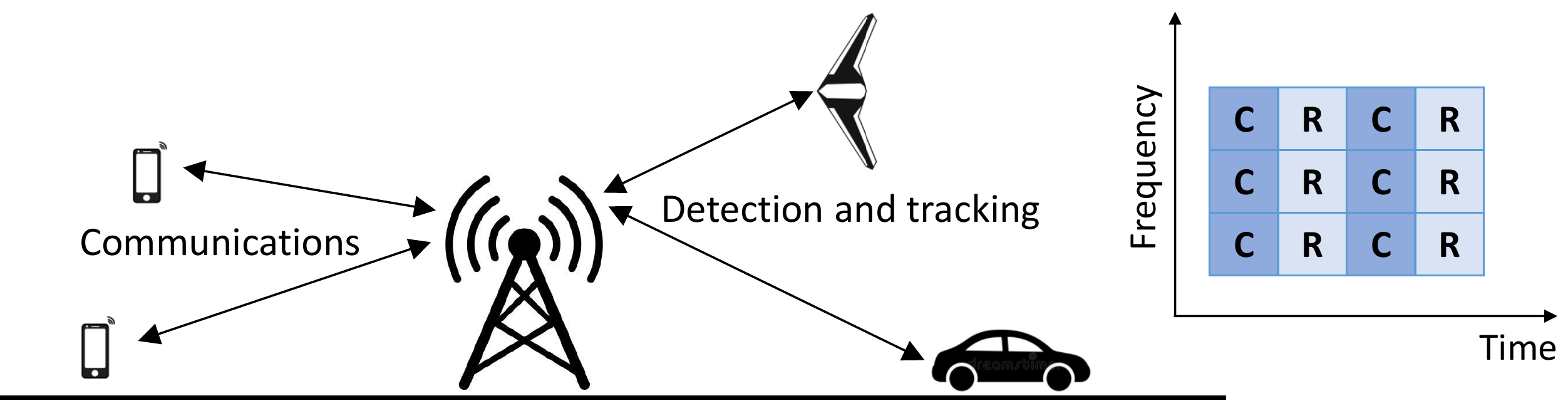}
	  \caption{}  
    \end{subfigure}    
    \begin{subfigure}{\columnwidth}
    \centering
	\includegraphics[width=\columnwidth]{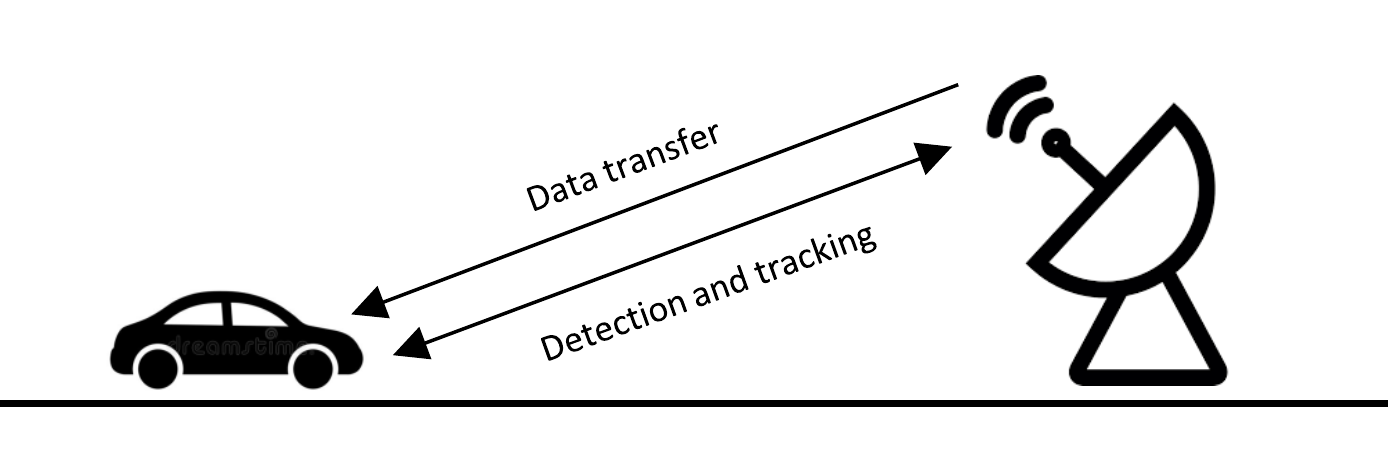}
	  \caption{}  
    \end{subfigure}
    \caption{(a) A JCR system example in which time-frequency resources are equally distributed between the communication and radar operations~(C represents communications and R represents radar); (b) a RadCom system example where, in addition to radar operation, data is also transferred.} \label{Fig:JCR}
\end{figure}

There are numerous possibilities for using communication systems for the detection and tracking of UAVs. One such possibility is the JCR/RadCom systems. Both communications and radar systems use radio waves for transmission and reception, and therefore, can be combined into a single JCR system. The benefit of the JCR system is efficient use of the available spectrum and compact hardware. In a JCR system, either a radar functionality can be added to a communication system or for a radar operation, communication data can be transmitted as shown in Fig.~\ref{Fig:JCR}. Furthermore, in a JCR system, either the resources are shared equally for communication and radar operations as shown in Fig.~\ref{Fig:JCR}(a) or dynamically based on the requirement. In a RadCom system, in addition to radar operation, the communication data can also be transferred as shown in Fig.~\ref{Fig:JCR}(b). Generally, in a JCR system, additional hardware and processes are introduced to the existing communication system for radar operation. In a JCR system, a common waveform is used for both communication and radar operation. Different waveforms, and channel and system models used for JCR systems are reported in~\cite{JCR4,JCR5}. JCR systems can be used in different emerging applications, e.g., self-driving vehicles, traffic monitoring and management, and ambient assistance for elderly and disabled people~\cite{JCR2, JCR4}. The JCR system can also assist in future UAV traffic management and detection and tracking of UAVs~\cite{JCR2,JCR6}. 

\subsection{Comparison of Communication Systems and Radar Systems}
As discussed earlier, there are many similarities between radar systems and communication systems. The similarities and differences between the communication systems and radar systems are summarized as follows:
\begin{itemize}
     \item Both communication systems and radar systems use EM waves at different frequencies.
      \item The radio wave propagation through free space occurs in the same way for communication systems and radar systems. 
    \item Both radar systems and communication systems use a signal generation stage and RF stage. The signal generation involves pulse selection, coding, encryption, shaping, and modulation. The RF stage includes amplification and steering~(by changing the signal phase across antenna elements) from the antenna array. 
    \item The receiver of a communication system and a radar system includes a low noise amplifier, matched filter, decoder, decrypter, demodulator, and detector.
    \item Both communication systems and radar systems are time-critical systems.
    \item The major difference between the communication systems and the radar systems is the transmit power. The radar system transmits EM signals at much higher power compared to common RF communications. The high transmit power is achieved by using amplifiers or high-gain antennas.
    \item Complex frequency hopping patterns are frequently used in radar systems to avoid jamming compared to communication systems.
    \item The selection of frequencies for the communication systems takes into account the transmission~(or penetration losses) through the building materials. However, for radar systems, the transmission through materials is not mainly considered during frequency selection and the main focus is on the reflection properties. 
    \item Complex encoding and signal processing techniques are used for communication systems. The techniques help to attain a high data rate and serve multiple users simultaneously. On the other hand, for radar systems, simple pulse signals are commonly used because the main focus is reliability. 
    \item The radar systems operate mainly in the LOS, whereas, the communication systems can operate for  LOS and NLOS links.
    \item The receiver sensitivity is higher for radar systems compared to common communication systems. 
    \item The TX and RX are generally co-located in a radar system~(monostatic radar), whereas TX and RX are a distance apart in the communication systems. 
\end{itemize}

\subsection{Future Possibilities of using Communication Systems for the Detection and Tracking of UAVs}
The future possibilities of using communication systems for the detection and tracking of UAVs are as follows:

\begin{itemize}
\item JCR/RadCom systems can offer joint communication and sensing capabilities on the ground and in the air for beyond 5G systems.
\item Cognitive radios in addition to providing opportunistic communications can also be used to detect UAVs simultaneously. 
\item HAM radio networks can be used to create a mesh of interconnected devices that can provide sensing capabilities for UAVs.
\item Marine radio communications is a worldwide system that uses the VHF band for communications between ships or between ships and shore. Marine radio communications can be analyzed to detect UAVs flying over sea or near shores.
\item The microwave links between telecommunication back-end towers can provide sensing capabilities for UAVs in addition to communications.
\item Public safety communication networks can be used for the detection and tracking of UAVs.
\item There is significant research going on for the detection of stealth aerial vehicles using passive radars and the source of illumination is common communication networks, e.g., FM/AM radio and television broadcasts, and mobile telecommunications~\cite{DAB8}.
\item A combination of different communication systems~(as discussed above) providing layered detection, tracking, and classification capabilities against UAVs can be used.
\end{itemize}

\section{Methods other than Radar Systems}      \label{Section:others_than_radar}
In this section, detection, tracking, and classification of aerial vehicles using methods other than radar systems are discussed. All the systems developed/acquired for the detection and mitigation of malicious UAVs should follow the joint advisory issued by the Federal Aviation Administration, Federal Communications Commission (FCC), Department of Justice (DOJ), and Department of Homeland Security~\cite{advisory_CUAS}. A comparison of radar systems and other sensors over different performance indicators~\cite{comparison} is shown in Table~\ref{Table:comparison_radar}. From Table~\ref{Table:comparison_radar}, overall, it can be concluded that radar systems perform better compared to other sensors for countering UAVs. Furthermore, the pros and cons of popular non-radar systems available in the literature are summarized in Table~\ref{Table:Non_radar}. 

\begin{table*}[t]
	\begin{center}
     \footnotesize
		\caption{Comparison of different sensors over a range of performance indicators (adapted from ~\cite{comparison}).} \label{Table:comparison_radar}
\begin{tabular}{@{}|P{ 2.5cm}|P{ 1.8cm}|P{ 1.8cm}|P{1.3cm}|P{1.3cm}|P{1.3cm}|P{1.3cm}|P{1.3cm}|@{}}
\hline
&\textbf{Long range radar with micro-Doppler}&\textbf{Short range high frequency radar with micro-Doppler}&\textbf{Airborne short range radar}&\textbf{RF triangulation}&\textbf{Clustered audio sensors}&\textbf{EO/IR sensor}&\textbf{Airborne EO/IR sensor}\\
\hline
\textbf{Range}&Extreme&Extreme&Medium&Good&Short&Medium&Medium\\
\hline
\textbf{Dynamic environment coverage}&Bad&Bad&Good&Good&Good&Bad&Good\\
\hline
\textbf{Detection in urban environment}&Medium&Medium&Medium&Medium&Medium&Bad&Good\\
\hline
\textbf{Detection in darkness}&Good&Good&Good&Good&Good&Good&Good\\
\hline
\textbf{Distance to target}&Good&Good&Good&Good&Bad&Bad&Bad\\
\hline
\textbf{Speed of target}&Good&Good&Good&Good&Bad&Bad&Good\\
\hline
\textbf{Mobility}&Bad&Medium&Good&Bad&Medium&Good&Good \\
\hline
\textbf{Detection in fog}&Good&Good&Good&Good&Good&Bad&Bad\\
\hline
\textbf{Detection in rain}&Good&Good&Medium&Good&Medium&Medium&Medium\\
\hline
\textbf{Weight}&High&Medium&Low&High&Medium&Low&Low\\
\hline
\textbf{Rogue UAV detection}&Yes&Yes&Yes&No&Yes&Yes&Yes\\
\hline
\textbf{UAV swarm detection}&Medium&Medium&Good&Good&Good&Good&Good\\
\hline
\textbf{Processing power}&High&High&High&High&Low&Medium&Medium\\
\hline
\textbf{Power consumption}&High&Low&Low&Medium&Medium&Low&Low\\
\hline
\textbf{Low speed UAV detection}&Good&Good&Good&Good&Good&Good&Good\\
\hline
\textbf{Low cost}&Bad&Medium&Medium&Medium&Medium&Good&Good\\
\hline
\textbf{Maintenance}&Good&Good&Bad&Good&Medium&Medium&Bad\\
\hline
\textbf{Installation ease}&Bad&Medium&Good&Medium&Medium&Medium&Good\\
\hline
\end{tabular}
		\end{center}
			\end{table*}

\subsection{Electro-Optical/Infrared}
Passive imaging sensor EO/IR can be used to detect, track, and classify small RCS and stealthy aerial vehicles with high precision. EO/IR sensors generally operate in the infrared and visible regions of the EM spectrum. The major benefits of EO/IR sensors are precise 3D detection, tracking, and classification capabilities in day/night conditions~\cite{EOIR,EOIR2}. However, the complexity of the aerial vehicle's background and thermal image saturation affects the performance of the EO/IR sensors. The range of the EO/IR is also limited by the horizon. The range can be increased by increasing the height of the EO/IR sensors above the ground, e.g., installing EO/IR sensors on airborne platforms. Moreover, the atmospheric effects, e.g., haze, fog, rain, and snow can affect the performance of EO/IR sensors. In \cite{UAV_vision}, the detection and tracking of UAVs are carried out using a vision-based approach and deep learning. The approach offers an economical medium-range solution for UAV detection and tracking alternative to radar. The detection of small and low flying UAVs is provided in \cite{UAV_imaging_new} using images from off-the-shelf cameras. A YOLO model and two neural networks namely DenseNet and ResNet are used. 

In \cite{UAV_CA}, a single camera onboard a UAV is used for aerial object detection during flight along the trajectory. The detection helps in collision avoidance. First, an image of a possible obstacle is captured from a UAV camera. The image is analyzed in real-time and a decision regarding change of UAV trajectory is made based on the features of the obstacle. In \cite{UAV_vision_motion}, imaging and a motion-based hybrid approach are used for 2D detection of objects, 3D localization, and tracking. The hybrid approach helps to detect and track small and fast-moving UAVs in cluttered environments. In \cite{UAV_image_new}, a UAV tracker using Global Positioning System~(GPS) sensor and a 2D camera are used. The camera is used to detect the presence of an intruding UAV, and then using a GPS sensor the position of the intruding UAV is estimated. A mathematical model based on the area expansion principle is used. In \cite{UAV_video}, UAV detection and tracking are carried out using video observations. The paper provides optical system processing for the detection and tracking of small UAVs. 

\subsection{Acoustic Methods}
Every aerial vehicle produces noise or acoustic signals while flying. The noise from a flying UAV can be detected using a microphone. The acoustic signals captured by a microphone and AI algorithms can also be used for the detection and classification of small, low, and slow-flying aerial vehicles, e.g., UAVs~\cite{acoustic_AI}. Among the UAVs, the noise is higher for multi-rotor UAVs compared to fixed-wing UAVs. The higher noise from multi-rotor UAVs is mainly from the rotating propellers. In \cite{UAV_acoustic}, passive acoustic radar systems are used for the detection of aircraft and UAVs. The passive radar system setup consists of omnidirectional microphones, a signal processing unit, and peripherals. Digital audio broadcast signals are used in \cite{UAV_acoustic2} for the detection and localization of micro UAVs. Using audio signals, the UAVs can be detected at a distance of $2.6$~km. The bi-static range and Doppler shifts, and direction of arrival of audio echoes are used for localization of the UAV. In \cite{UAV_acoustic3}, machine learning and multiple acoustic nodes are used for the detection of UAVs. The training is performed using short-time Fourier transform and Mel-frequency cepstral coefficients. Short-time Fourier transform and support vector machine provided the best detection results in \cite{UAV_acoustic3}. 

There are many limitations of the acoustic methods. The acoustic methods for the detection of aerial vehicles perform poorly in high noise environments and the range is limited. The acoustic methods work in the passive mode and a single acoustic sensor cannot provide precise localization and tracking of the UAV. The acoustic methods have small significance against high speed and high altitude flying aerial vehicles. 

\subsection{Radio Frequency Analyzers} \label{Section:RF_analyzers}
All aerial vehicles are equipped to communicate with their peers in the air and ground controllers through radio links. The radio communications between an aerial vehicle and peers or ground controllers can be detected and analyzed using RF analyzers. A simple spectrum analyzer or SDR can be used as an RF analyzer. Energy detection at specific frequency bands and patterns of the energy burst of the radio link can be used for detection and classification. AI algorithms can be used to differentiate the emissions of an aerial vehicle from other communication devices~\cite{RF_analysis}. The energy spectrum of the radio emissions using a multistage detector is used to detect and classify UAVs in the presence of noise and interference from other communication nodes~\cite{RF_analysis_clas}. In \cite{UAV_signal_analysis}, multi-dimensional signal features are used for the detection and localization of UAVs. The communication signal and channel state information between the UAV and the controller and subsequent features are analyzed. The spatial features of azimuth and elevation angles are used for UAV localization. 

Similar to acoustic analysis, there are limitations of the RF analysis method. The range of the RF analysis method is dependent on the strength of the radio link, and propagation channel between the aerial vehicle and the peer in the air or ground controller. RF analysis is effective in detecting UAVs that often require a continuous radio control link between the UAV and the remote controller. However, if a UAV is flying autonomously without any active radio link, then RF analysis may not be effective. RF analysis method operates in a passive mode and the precise location of the UAV cannot be precisely determined using a single antenna. However using a single MIMO RF sensor to determine the AoA of the RF signal from the UAV and if the range estimate of the UAV is available, then the location of the UAV can be estimated. 

\subsection{Laser based Technique}
\begin{figure}[!t]
	\centering
	\includegraphics[width=\columnwidth]{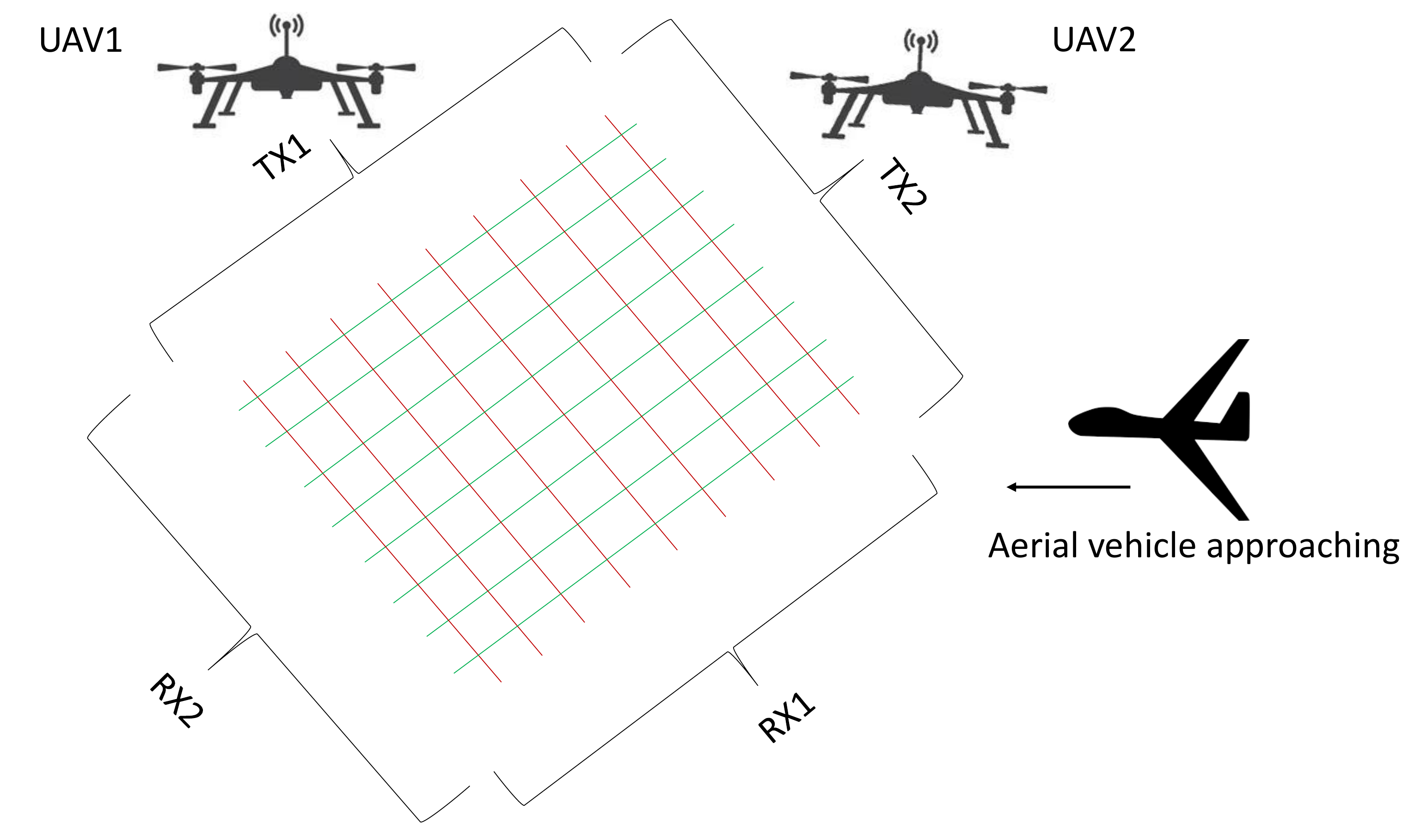}
	\caption{A laser mesh created in the air using two airborne UAVs for detection, tracking, and classification of aerial vehicles~(regenerated from \cite{UAV_wahab_laser}).}\label{Fig:laser_wahab}
\end{figure}

Laser is used for different civilian and defense applications. The laser can also be used for the detection, tracking, and classification of aerial vehicles. A novel approach using a mesh of laser beams to detect, track, and classify small UAVs and stealth aerial vehicles is provided in \cite{UAV_wahab_laser}. At least two airborne platforms at different spatial positions are required. The two airborne platforms transmit laser beams towards the ground RXs such that a mesh is formed as shown in Fig.~\ref{Fig:laser_wahab}. If an aerial vehicle blocks the path of the laser beams, it is detected and subsequently tracked and classified. Laser beam steering is also suggested to increase the coverage range and for better tracking and classification in \cite{UAV_wahab_laser}.  

\begin{table*}[!t]
	\begin{center}
    \footnotesize
		\caption{Non-radar approaches for detection, tracking and classification of UAVs. } \label{Table:Non_radar}
\begin{tabular}{@{}|P{2cm}|P{13cm}|P{1.8cm}|@{}}
\hline
\textbf{Non-radar approach}& \textbf{Pros and cons} & \textbf{Representative References.}\\
\hline
\textbf{Electro-Optical/Infrared (EO/IR)} & Passive imaging sensors and can detect, track, and classify small RCS aerial vehicles with high precision in day/night conditions. Major limitations include the complexity of the aerial vehicle's background and thermal image saturation. Also, the range of the EO/IR sensors is limited by the horizon, and atmospheric changes affect the performance& \cite{EOIR,EOIR2,UAV_vision,UAV_imaging_new}\\
\hline
\textbf{Acoustic} & Operate in the passive mode and can be used to detect, localize, and classify small, low, and slow-flying aerial vehicles. Major limitations include poor performance in high-noise environments and the range is limited. Also, the acoustic methods have small significance against high-speed and high-altitude aerial vehicles & \cite{acoustic_AI,UAV_acoustic,UAV_acoustic2,UAV_acoustic3}\\
\hline
\textbf{Radio frequency analysis} & Operate in the passive mode. RF control/communication links can be used to detect, track, and classify UAVs. Limitations include small detection range, and dependence of range on the strength of the radio link, and propagation channel. Not effective against UAVs flying autonomously without any active radio link & \cite{RF_analysis,RF_analysis_clas,UAV_signal_analysis}\\
\hline
\textbf{Laser} & Can be used for the detection, tracking, and classification of different types of aerial vehicles including stealth aerial vehicles. The scanning is fast, the equipment is compact, simple, and off-the-shelf. Major limitations include small scanning area of the beam compared to radar, limited range, and atmospheric effects on the laser propagation  
& \cite{UAV_wahab_laser,UAV_lidar_new,UAV_lidar_new2,track_new6}\\
\hline
\textbf{Distributed sensors} & Multiple types of off-the-shelf sensors can be used at different locations to detect, track, and classify UAVs. Distributed sensors are economical, provide better situational awareness, and can cover a large area. A major challenge is to connect distributed sensors to a central location. Maintenance of distributed sensors is also challenging & \cite{UAV_multisensor, UAV_distributed1,track_new4,UAV_distributed3}\\
\hline
\textbf{Space-based sensors}& Provide early warning and have a long-range and wide coverage area. However, the space sensors need to take into account the atmospheric and space effects into account. There is also inherent delay due to long distance & \cite{UAV_space1,UAV_space2}\\
\hline
\textbf{Combat UAVs} & The initial and maintenance cost is small and the design can be modular and mission-specific. Other advantages include the protection of the pilot's life, cost, and quick replaceability in the field. Major limitations include small payload, the delay in the data reception, the vulnerability of the RF communication/control link to jamming and spoofing & \cite{UAV_combat1,UAV_combat2,UAV_combat3}\\
\hline
\end{tabular}
		\end{center}
\end{table*}

A laser scanner is used for distinguishing birds and UAVs in \cite{UAV_laser_new}. A cross-polarization ratio analysis is carried out using the received optical echoes. The approach uses simple and off-the-shelf equipment. A lidar system is used for the detection and tracking of UAVs in \cite{UAV_lidar_new}. A classification sensor and image processing are used for the identification of UAVs. A scanning lidar is used for long-range detection and tracking of UAVs in \cite{UAV_lidar_new2}. Time of flight is used to calculate the range of the aerial vehicle, and imaging is performed by subsequent scanning of the scene. Tracking is based on imaging of the scene at a rapid rate. Lidar-based detection and tracking of mini/micro UAVs are also provided in \cite{track_new6}. 

\subsection{Distributed Sensors}
A network of different types of off-the-shelf sensors, e.g., imaging, and acoustic sensors, and SDRs can be placed at distributed locations in the possible flight path of aerial vehicle(s). The location of the sensors can be electricity and telecommunication towers, tall buildings, and other infrastructure. The sensors can be connected to a central network for monitoring. The distributed sensors provide the first layer of detection in addition to conventional layer of detection. In \cite{UAV_multisensor}, the concept of using different types of heterogeneous sensors for the detection and tracking of UAVs is discussed. In \cite{UAV_distributed1}, visual detection and tracking of cooperative UAVs is carried out using cameras onboard UAVs, and deep learning methods. A distributed dynamic radar network carried by UAVs is presented in \cite{track_new4}. Instead of using terrestrial radars, the radars carried by UAVs can be dynamically positioned to optimally detect and track malicious UAVs. The detection and tracking are improved by the cooperative working of the UAVs. In \cite{UAV_distributed3}, airborne UAV assets for surveillance, intelligence gathering, and reconnaissance for the US air force are discussed.

\subsection{Regulating the UAV Air Traffic}
A novel method to counter malicious and hostile UAVs is by regulating the UAV air traffic. The UAV traffic can be regulated by installing control towers similar to air traffic control towers. The control towers can be installed on already available infrastructure, e.g., telecommunication towers. A control tower will autonomously regulate the UAV traffic in the vicinity by assigning tags to registered UAVs. The tags will be updated when the UAV approaches the next control tower limit, which can help to identify unregistered and malicious UAVs. 

\subsection{Space Assets}
The monitoring of aerial vehicles can also be performed using sensors in space, such as radar, imaging, and infrared sensors. The space-based sensors provide early warning against several different aerial threats e.g. UAV swarms. The sensors placed in space have a long-range and wide coverage area, and  have added advantage of mobility. However, the space sensors need to take into account the atmospheric and space effects into account. A literature review of synergies between UAVs and satellites is provided in \cite{UAV_space1} for remote sensing applications. A UAV detection system using satellites was awarded a grand prize in European Satellite Navigation Competition~\cite{UAV_space2}. 

\subsection{Combat UAVs}
The initial and maintenance cost of modern fighter jets is large. Also, the modern fighter jets may not be equipped to counter a wide spectrum of aerial threats, e.g., autonomous UAV swarms. A possible solution is to use combat UAVs, which can be used separately or in combination with fighter jets. The combat UAVs can be either remotely controlled by a human operator,  can work autonomously using AI, or both. The combat UAVs can have a modular and mission-specific design. For example, a swarm of combat UAVs can be used to counter an attacking UAV swarm. The major advantage of combat UAVs compared to manned fighter jets is the protection of pilot's life, cost, flexible design modifications, and quick replaceability in the field. 

An autonomous UAV equipped with a camera and autopilot driven by a micro-controller to track and shoot down another UAV is provided in \cite{UAV_combat1}. An image of the target UAV is obtained by the camera on-board UAV. The image processing is performed by the Raspberry Pi 2 microcontroller for detection and subsequent tracking. In \cite{UAV_combat2}, a single camera~(as payload) onboard a UAV is used for detection and tracking of other UAVs in the vicinity. First, the background motions are estimated using a perspective transformation model, then spatio-temporal traits are applied for each moving object, and motion patterns are used to identify each moving object. Kalman filtering is also applied to boost the performance. A visual detection and tracking system on-board a UAV is presented in \cite{UAV_combat3}. A 3-axis gimbal-supported camera and kernelized correlation filter are used for real-time UAV detection and tracking. 

\begin{table*}[htbp]
	\begin{center}
     \footnotesize
		\caption{Popular counter-UAV radar systems~\cite{CSD_UAV,CSD_UAV2}. } \label{Table:CSD_UAV}
\begin{tabular}{@{}|P{ 5cm}|P{5.3cm}|P{2.5cm}|P{0.5cm}|@{}}
 \hline
\textbf{Product name}&\textbf{Manufacturer}&\textbf{Platform}&\textbf{Ref.}\\
\hline
NM1-8A drone radar system&Accipter, Canada&Ground-based&\cite{CSD1}\\
\hline
Spartiath&ALX Systems, Belgium&UAV/Ground-based&\cite{CSD2}\\
\hline
Gamekeeper 16U&Aveillant, United Kingdom&Ground-based&\cite{CSD3}\\
\hline
Laser Avenger&Boeing, United States&Ground-based&\cite{CSD4}\\
\hline
Harrier Drone
Surveillance Radar&DeTect, Inc, United States/United Kingdom&Ground-based&\cite{CSD5}\\
\hline
RadarOne&DroneShield, Australia&Ground-based&\cite{CSD6}\\
\hline
RadarZero&DroneShield, Australia&Ground-based&\cite{CSD7}\\
\hline
SABRE&DRS/Moog, United States&Ground-based&\cite{CSD8}\\
\hline
GroundAware&Dynetics, United States&Ground-based&\cite{CSD9}\\
\hline
Red Sky 2 Drone Defender System&IMI Systems, Israel&Ground-based&\cite{CSD10}\\
\hline
SharpEye&Kelvin Hughes, United Kingdom&Ground-based&\cite{CSD11}\\
\hline
SR-9000S&Meritis, Switzerland&Ground-based&\cite{CSD12}\\
\hline
Drone Detection Radar&Miltronix, United Kingdom&Ground-based&\cite{CSD13}\\
\hline
EAGLE&MyDefence Communication ApS, Denmark&Ground-based&\cite{CSD14}\\
\hline
1L121-E&NNIIRT, Russia&Ground-based&\cite{CSD15}\\
\hline
3D Air Surveillance UAV Detection Radar&OIS-AT, India&Ground-based&\cite{CSD16}\\
\hline
OBSIDIAN&QinetiQ, United Kingdom&Ground-based&\cite{CSD17}\\
\hline
Multi-Mission Hemispheric Radars&RADA Electronic
Industries, Israel&Ground-based&\cite{CSD18}\\
\hline
HEL Effector Wheel XX&Rheinmetall AG, Germany&Ground-based&\cite{CSD19}\\
\hline
Elvira&Robin Radar Systems, Netherlands&Ground-based&\cite{CSD20}\\
\hline
Drone Sentinel&Advanced Radar Technologies, Spain&Ground-based&\cite{CSD21}\\
\hline
RR Drone/radar detection system&Aaronia, Germany&Ground-based&\cite{CSD22}\\
\hline
Fortem TrueVIEW radar&Fortem Technologies&Ground-based and airborne&\cite{Fortem}\\
\hline
EchoFlight&Echodyne corp.&Airborne&\cite{echodyne}\\
\hline
LSTAR (V)2&SRC Inc.&Ground-based&\cite{lstar}\\
\hline
$60$~GHz FMCW MIMO&Custom built&Ground-based&\cite{MIMOradar_vasilii}\\
\hline
\end{tabular}
		\end{center}
			\end{table*}
			
\subsection{Popular Counter-UAV Systems}
There are many research activities available in the literature for countering UAVs~\cite{C-UAV}. Major counter UAV efforts in \cite{C-UAV} include using multistatic radars, passive radars, cognitive radars, MIMO radars, and air-to-air radars. The majority of the radar systems currently used for countering UAVs are either battlefield radars, short-range air defense radars, marine radars, or bird detection radars. In \cite{C-UAV2}, numerous challenges in detecting UAVs using different types of radar systems are discussed. An X-band radar equipped with electronic scanning is claimed to be a reliable and economical solution for detecting UAV threats in \cite{C-UAV2}. Table~\ref{Table:CSD_UAV} provides popular counter-UAV radar systems. The name of the radar system, installation platform, and manufacturer information is also provided in Table~\ref{Table:CSD_UAV}. 

Various methods of detecting and tracking malicious UAVs are provided in \cite{UAV_detect_survey}. Different methods of UAV detection and classification using radar systems and imaging systems are discussed in \cite{UAV_detect_survey}. A survey on the detection, tracking, and interdiction of small UAVs is provided in \cite{UAV_Survey_Farshad}. In the survey, various physical and cyber threats from amateur UAVs are discussed, and then various detection tracking and interdiction methods using radar, RF analysis, acoustic sensors, and computer vision techniques are provided. 

A study on different anti-UAV technologies for detection and tracking of modern UAV threats is discussed in \cite{C-UAV_new}. The advantages and disadvantages of different types of anti-UAV architectures for low altitude aerial security are provided in \cite{UAV_antidrone, UAV_antidrone2}. An anti-UAV system named ADS-ZJU is provided in \cite{UAV_antidrone2}, which uses different surveillance technologies in the passive mode, for detection, localization, and RF jamming of UAVs. Different technologies for the detection and localization of amateur UAVs are provided in \cite{UAV_technol}. The technologies for modulation classification, localization of UAVs based on received signal strength are discussed. A discussion about current and future challenges for surveillance of UAVs is also provided. Different counter UAV systems, state of art, current and future challenges are provided in \cite{UAV_technol2}. 

\subsection{Anti-Stealth Methods}
The procedures to reduce the visibility of aerial vehicles from radar systems~(small RCS) can be carried out in different ways. Specific devices, shapes, software, and materials in the form of surface coatings, paints, and parts are used to tailor the reflection, refraction, diffraction, and scattering of EM waves from an aerial vehicle~\cite{stealth_tech}. These include conductive fibers, foam rubber sheets, transparent radio absorbent material similar in appearance to polycarbonate sheets, loaded sprayed ceramic tiles, absorbing honeycomb consisting of a lightweight composite, and open cells. Furthermore, other surface coverings in the form of conductive paints or inks, painted small cell foams, magnetic RAM applied as surface coverings, resistive cards consisting of fiber paper sheets covered with conductive paint, and infrared-based treatments in the form of paints and coatings on the surface can be used. Molded edges, avoiding edges and corners at $90^\circ$~(that can strongly reflect EM waves to the source), and cavities during the design of the aerial vehicles can also help to achieve small RCS. The collective stealth techniques help to achieve low/disguised RCS at certain distances, angles, and frequencies, therefore, avoiding/confusing detection. 

There are many techniques available to counter stealth in aerial vehicles. High frequency~(HF) radars can be used for the detection of stealth aerial vehicles at long range. In particular OTHR radars using high frequency band are known to detect stealth aircraft at long ranges~\cite{OTH_stealth}. VHF/UHF radars are also used for anti-stealth purposes. The size of the majority of the equipment installed on stealth aerial vehicles is smaller than the wavelength of the VHF radars. Therefore, the incident radar waves are scattered from different parts of the stealth aerial vehicle. Bi-static or multi-static radars or other passive radio listeners can be used to collect the scattered energy. Centimeter radars can also be used for the detection of stealth aircraft as the size of the imperfections of the aircraft e.g., joints, rivets are the same as the wavelength. Sparse signal processing is extensively used by passive and bi-static or multi-static radars.

Bi-static and multi-static passive radars are helpful in the detection and tracking of stealth aerial vehicles~\cite{bistatic,multistatic}. The position of the TXs and RXs is important for stealth aerial vehicle detection using bi-static/multi-static passive radars. 
Broadcast signals from frequency modulation, amplitude modulation, TV-DVB, and satellite communications can also be used by passive radars for the detection of stealthy aerial vehicles~(see Section~\ref{Section:Comm_Sys}). Cognitive radars can be yet another alternative as discussed in~\cite{cognitive_stealth}. In addition, multiple airborne UAVs can be used for collecting information of the stealthy aerial vehicles in the air as passive radar listeners, RF analyzers, EO/IR sensors, or acoustic sensors.

\section{Concluding Remarks}     \label{Section:conclusions}
In this survey paper, a comprehensive overview of the radar systems for detecting, tracking, and classification of modern aerial threats is provided. The salient features of modern radar systems helpful in countering modern aerial threats are presented. The detection, tracking, and classification of modern aerial threats using communication systems are also discussed. Finally, methods other than radar systems for detection, tracking, and classification of unmanned vehicles are reviewed briefly. 

\bibliographystyle{IEEEtran}


\end{document}